\documentclass[a4paper,12pt]{article}
\usepackage{amsfonts,amsthm,amsmath,amssymb, graphicx
,hyperref}

\newcommand{\bea}{\begin{eqnarray}}
\newcommand{\eea}{\end{eqnarray}}
\newcommand{\nnm}{\nonumber}
\begin{document}
{}~
{}~
\hfill\vbox{\hbox{}\hbox{QMUL-PH-06-11} 
\hbox{WITS-CTP-028}
}\break

\vskip .6cm

\centerline{{ \bf \Large  Correlators, Probabilities
 and Topologies in $\mathcal{N}=4$ SYM   }}


\medskip

\vspace*{4.0ex}

\centerline{\large \rm 
T.~ Brown${}^{1}$, R.~ de Mello Koch${}^{2}$, 
S.~Ramgoolam${}^{1}$, N.~Toumbas${}^{3}$}

\vspace*{4.0ex}
\begin{center}
{\large ${}^1$Department of Physics\\
Queen Mary, University of London\\
Mile End Road\\
London E1 4NS UK\\
}
\vskip.1in 
{\large ${}^{2}$ Department of Physics\\
University of Witwatersrand  \\ 
Wits, 2050  \\
South Africa  \\ 
}
\vskip.1in 
{\large ${}^3$ Department of Physics\\
University of Cyprus\\
Nicosia 1678, Cyprus \\}
\end{center}

\vspace*{5.0ex}

\centerline{\bf Abstract} \bigskip

We calculate transition probabilities for various 
processes involving giant gravitons and 
small gravitons in AdS space, using the dual $ \mathcal{N}=4$ SYM theory. 
 The normalization factors for these probabilities  involve, in general, 
 correlators for manifolds of  non-trivial topology which are obtained by gluing simpler
 four-manifolds.  This follows from the factorization properties which relate 
 CFT correlators for different topologies. These points are illustrated, 
in the first instance,  
in the simpler example  of a two dimensional Matrix CFT.
We give the bulk five dimensional interpretation, involving neighborhoods
of Witten graphs, of these gluing properties of the four 
dimensional boundary CFT. 
As a corollary we give a simple description, based on Witten graphs, 
of a multiplicity of bulk topologies corresponding to 
a fixed boundary topology.     
We also propose to 
interpret the correlators as topology-changing transition amplitudes 
between LLM geometries.

\thispagestyle{empty}

\vfill
\noindent{\it  {t.w.brown@qmul.ac.uk, robert@neo.phys.wits.ac.za,
s.ramgoolam@qmul.ac.uk, nick@ucy.ac.cy}}
\eject

\vspace*{5.0ex}

\def\Sec{\S}
\def\vac{|0\rangle}
\def\bra#1{\left\langle #1\right|}
\def\ket#1{\left| #1\right\rangle}
\newcommand{\braket}[2]{\langle #1 | #2 \rangle}
\def\corr#1{\left\langle #1 \right\rangle}
\def\tr{{\rm tr}\, }
\def\Tr{{\rm Tr\, }}
\def\adj{\textrm{adj}}
\def\Id{\textrm{Id}}
\def\id{\textrm{id}}
\def\ind{\textrm{ind}}
\newcommand{\Sym}{\operatorname{Sym}}
\def\dim{\textrm{dim}}
\def\Dim{\textrm{Dim}}
\def\Res{\textrm{Res}}
\def\Ind{\textrm{Ind}}
\def\ker{\textrm{ker}}
\def\im{\textrm{im}}
\def\sgn{\textrm{sgn}}
\def\ch{\textrm{ch}}
\def\STr{\textrm{STr}}
\def\ha{{1\over 2}}
\def\wt{\widetilde}
\def\ra{\rangle}
\def\la{\langle}
\def\a{\alpha}
\def\b{\beta}
\def\c{\chi}
\def\g{\gamma}
\def\G{\Gamma}
\def\e{\epsilon}
\def\f{\phi}
\def\F{\Phi}
\def\fb{{\ov \phi}}
\def\vf{\varphi}
\def\m{\mu}
\def\mub{\ov \mu}
\def\n{\nu}
\def\nub{\ov \nu}
\def\o{\omega}
\def\O{\Omega}
\def\r{\rho}
\def\k{\kappa}
\def\kab{\ov \kappa}
\def\s{\sigma}
\def\sb{\ov\sigma}
\def\S{\Sigma}
\def\l{\lambda}
\def\L{\Lambda}
\def\p{\psi}
\def\pb{\ov\psi}
\def\cb{\ov\chi}
\def\d{\partial}
\def\dbar{\ov\partial}
\def\dag{\dagger}
\def\dalpha{{\dot\alpha}}
\def\dbeta{{\dot\beta}}
\def\dgamma{{\dot\gamma}}
\def\ddelta{{\dot\delta}}
\def\da{{\dot\alpha}}
\def\db{{\dot\beta}}
\def\dg{{\dot\gamma}}
\def\dd{{\dot\delta}}
\def\t{\theta}
\def\T{\Theta}
\def\tb{{\ov \theta}}
\def\lb{{\ov \lambda}}
\def\eb{{\ov \epsilon}}
\def\gb{{\ov \gamma}}
\def\zb{{\ov z}}
\def\wb{{\ov w}}
\def\ib{{\ov i}}
\def\jb{{\ov j}}
\def\kb{{\ov k}}
\def\mb{{\ov m}}
\def\nb{{\ov n}}
\def\qb{{\ov q}}
\def\xh{\hat{x}}
\def\D{\Delta}
\def\DD{\Delta^\dag}
\def\Db{\ov D}
\def\M{{\cal M}}
\def\rd{\sqrt{2}}
\def\ov{\overline}
\def\C{\mathbb{C}}
\def\R{\mathbb{R}}
\def\Z{\mathbb{Z}}
\def\N{\mathbb{N}}
\def\H{\mathbb{H}}
\def\Slash{\, / \! \! \! \!}
\def\slash #1{{\not \hspace{-.5mm}#1}}

\newcommand{\cA}{\mathcal A}
\newcommand{\cB}{\mathcal B}
\newcommand{\cC}{\mathcal C}
\newcommand{\cD}{\mathcal D}
\newcommand{\cE}{\mathcal E}
\newcommand{\cF}{\mathcal F}
\newcommand{\cG}{\mathcal G}
\newcommand{\cH}{\mathcal H}
\newcommand{\cI}{\mathcal I}
\newcommand{\cJ}{\mathcal J}
\newcommand{\cK}{\mathcal K}
\newcommand{\cL}{\mathcal L}
\newcommand{\cM}{\mathcal M}
\newcommand{\cN}{\mathcal N}
\newcommand{\cO}{\mathcal O}
\newcommand{\cP}{\mathcal P}
\newcommand{\cQ}{\mathcal Q}
\newcommand{\cR}{\mathcal R}
\newcommand{\cS}{\mathcal S}
\newcommand{\cT}{\mathcal T}
\newcommand{\cU}{\mathcal U}
\newcommand{\cV}{\mathcal V}
\newcommand{\cW}{\mathcal W}
\newcommand{\cX}{\mathcal X}
\newcommand{\cY}{\mathcal Y}
\newcommand{\cZ}{\mathcal Z}

\newcommand{\be}{\begin{equation}}

\newcommand{\ee}{\end{equation}}

\newcommand{\ret}{\nonumber \\}
\newcommand{\nn}{\nonumber}

\tableofcontents

\section{Introduction} 

 AdS/CFT duality \cite{malda}\cite{gkp}\cite{witten} 
 provides a framework to study hard questions of quantum
  gravity, using tractable calculations in gauge theory. 
 The discovery of giant gravitons \cite{mst}\cite{GMT}\cite{HHI} and the identification of 
 their dual gauge theory operators \cite{bbns}\cite{cjr} open the way to 
 exploring transitions among these brane-like objects, as well as transitions from giant gravitons into 
 small, ordinary gravitons. From the point of view of the bulk gravity theory, these processes are
non-perturbative in nature and difficult to analyze quantitatively.

In this paper, we explain how to calculate the corresponding transition probabilities. 
These can be obtained by appropriately normalizing the relevant gauge theory correlators 
describing the bulk
interactions. We show that, in general, the normalization factors involve correlators on  
manifolds of non-trivial topologies.
The result is a direct consequence of CFT factorization equations, 
which relate correlators on manifolds of different topologies. Factorization is expected to be a generic
property of conformal field theories, which follows from the operator/state correspondence 
and sewing properties 
of path integrals. Here, we explore some of its implications for the case of the four dimensional 
$\cN=4$ Super Yang Mills theory. We prove explicit inequalities 
that follow after we discard some intermediate states
from four dimensional factorization equations.  
As we shall demonstrate with specific examples, 
 factorization relations among correlators on spaces of different
 topologies constrain the relative growth of the correlators as 
the number of colors is increased, in a manner consistent with the probability 
interpretation.  These probabilities are
the generic observables of string theory in asymptotically $AdS$ backgrounds. 

This paper is organized as follows.
In Section \ref{sec:puzzle}, we consider two normalization prescriptions, one
which we call the overlap-of-states normalization, and the other which we call
the multi-particle normalization, and use them to compute transition probabilities.  
Both normalization schemes have been used in various contexts in the literature
\cite{bbns,cjr,dms}.
We find that there is a problem with
the use of the multi-particle normalization prescription. 
We give several examples for which ``multi-particle normalized'' 
amplitudes
grow with $N$, and so they do not yield well defined probabilities.  
The resolution of this
puzzle is the first main result of this paper. In general, to get
well defined probabilities, we need to divide by correlators on manifolds of more
complicated topologies, as implied by factorization. The main ideas relating 
factorization and probabilities are explained in Section \ref{sec:facttoprob}.  

In Section \ref{sec:fact2d}, we
review the main aspects of factorization and apply them to derive inequalities and probabilities 
in a simpler two dimensional model involving
free complex matrix fields.    
In Section \ref{sec:fact4d}, we extend
the discussion to the more relevant case of the four dimensional $\cN=4$ Super Yang Mills theory. In Section
\ref{sec:correctresults}, we summarize  results of explicit transition
probability computations for processes involving giant and small gravitons in $AdS$ space.

Motivated by the need for the gluing properties of the boundary CFT in the correct
formulation of probabilities for bulk spacetime processes, we investigate
how to lift the geometrical boundary gluing properties to the bulk five
dimensional Euclidean space. The results are presented in Section
\ref{sec:bulkinterpreta}. Witten graphs, i.e. graphs with end
points corresponding to CFT operator insertions on the boundary of AdS 
and vertices in the bulk, and their neighborhoods, are found to provide a simple framework for the bulk
lifting of the boundary gluings. Finally, we propose to 
interpret CFT correlators involving operators of large $R$ charge as topology-changing 
transition amplitudes between LLM geometries \cite{llm}. 
Section \ref{sec:bulkinterpreta} may
be read independently of the rest of the paper.
Technical computations are described in the Appendices. 
A summary of notation used is given in Appendix \ref{sec:idnotcon}.

\section{Transitions from giants to KK gravitons: a puzzle}\label{sec:puzzle}

\subsection{1/2 BPS states in the AdS/CFT correspondence: a brief review}
We are interested in various interactions among particles and branes
in the  $AdS_5\times
S^5$ geometry, and for cases in which the interacting states have non-zero angular momentum on $S^5$. 
These interactions can be studied in the context of the
AdS/CFT correspondence by making use of the dictionary relating bulk
states with conformal field theory operators.  We focus on $1/2$ BPS
states characterized by a single angular momentum charge $J$ under a
$U(1)$ subgroup of the $SO(6)$ rotation group, and for which exact,
non-perturbative results can be obtained.  Such states correspond to
chiral primary operators of conformal weight $\Delta=|J|$ in the dual
 $U(N)$ $\cN=4$ Super Yang Mills theory.

For small angular momentum, $J \ll N$, the states describe
Kaluza-Klein (KK) bulk gravitons.  Single particle KK states
correspond to single trace operators of the form $\tr (\Phi^J)$ in the
boundary conformal field theory \cite{gkp}\cite{witten}.  Here $\Phi$
is a complex field in the adjoint representation of $U(N)$, and this
field has unit charge under the particular $U(1)$ $R$-symmetry subgroup
we are considering.  Perturbative supergravity interactions among KK
graviton states have been studied in \cite{fmmr}\cite{lmrs}, and the
results have been matched with boundary conformal field theory
computations.

When we increase the angular momentum so that $J \sim \sqrt{N}$, the states describe
strings in plane wave backgrounds \cite{bmn}. More precisely, the operator
$\tr (\Phi^J)$ for $J \sim \sqrt{N}$ can be associated to the ground state of the string in light-cone
gauge. Excited string states can be obtained by replacing some of the $\Phi$'s in the trace with
other transverse scalars. These states are nearly BPS and their interactions have been studied in 
\cite{cfh}\cite{kpss}\cite{sv}
from the gauge theory and the bulk point of view.

When the angular momentum is a finite fraction of $N$, $J \sim N$,
some of the states describe large spherical $D3$ branes inside the
$S^5$ component or spherical branes inside the $AdS_5$ component of
the bulk geometry, the so called giant gravitons \cite{mst}\cite{GMT}\cite{HHI}.  To describe giant
graviton states in the boundary Super Yang Mills theory, we use a
basis for the space of $1/2$ BPS operators that consists of Schur
polynomials of the matrix $\Phi$.  The space of Schur polynomials is
in one-to-one correspondence with the set of Young diagrams
characterizing irreducible representations of $U(N)$.  Thus we denote
the Schur polynomials by $\chi_R(\Phi)$, with $R$ denoting the
corresponding $U(N)$ representation. Now if the Young diagram corresponding to
the $U(N)$ representation $R$ has $n$ boxes, then it also characterizes an irreducible representation of 
the symmetric group $S_n$.   
Explicitly $\chi_R(\Phi)$ is given by
\begin{equation}
\chi_R(\Phi)={1 \over n!}\sum_{\sigma \in S_n}\chi_R(\sigma)\tr(\sigma \Phi)
={1 \over n!}\sum_{\sigma \in S_n}\chi_R(\sigma)
\left[\sum_{i_1,i_2,\dots, i_n}\Phi^{i_1}_{i_{\sigma(1)}}\Phi^{i_2}_{i_{\sigma(2)}}\dots\Phi^{i_n}_{i_{\sigma(n)}}\right]
\end{equation}
Using the dictionary developed
in \cite{cjr}\cite{cr} (see also \cite{bbns}), the Schur operator corresponding to an $AdS$
giant with $L$ units of angular momentum on $S^5$ is given by
$\chi_{[L]}(\Phi)$, where we denote by $[L]$ a Young diagram with a
single row of length $L$. This Young diagram describes a symmetric
representation of $U(N)$.  Similarly, the operator corresponding to a
sphere giant is given by $\chi_{[1^L]}(\Phi)$, where $[1^L]$ denotes a
Young diagram with a single column of length $L$ describing an
antisymmetric representation \footnote{Kaluza-Klein gravitons can also
  be described in the Schur polynomial basis: for small angular
  momenta, the single trace operators $\tr (\Phi^J)$ can be expressed
  in terms of combinations of Schur polynomials corresponding to small
  Young diagrams. The choice of the single trace basis allows one to match 
directly the Fourier modes of the operators 
with the particle creation and annihilation operators of perturbative bulk supergravity \cite{bdhm}}.
Operators describing open string excitations on giant gravitons have been discussed in 
\cite{Berenstein:2005fa}\cite{Berenstein:2006qk}.

At even larger values for the angular momentum, $J\sim N^2$, one finds
bulk geometries \cite{llm}.  These geometries have an $SO(4)\times
SO(4)\times R$ isometry group and preserve $16$ of the original $32$
supersymmetries.  In the boundary theory, they are described by free
fermion droplets in the two dimensional phase space occupied by the
fermions.  These fermions are the eigenvalues of
 the matrix $\Phi$ \cite{Berenstein}.

 Interactions involving small KK
gravitons, giant gravitons and LLM geometries 
 should be encoded in the correlation functions of the
corresponding dual CFT operators\footnote{For some discussions of such gauge theory 
correlators see 
\cite{deMelloKoch,taktak,okuyama,dms, taktsu}.}.
 Our basic remarks on normalizations and probabilities are general, valid for any coupling 
in the gauge theory, but our explicit computations are done in the free gauge  theory limit. 
When they involve the special class of correlators called 
 extremal, the explicit results are valid for any coupling. 
  These are correlators in which the spacetime
coordinates of all anti-holomorphic operators involving
$\Phi^{\dagger}$ coincide while the positions of holomorphic operators
are arbitrary (and vice versa).  Non-renormalization theorems protect
extremal correlators of $1/2$-BPS chiral primaries so that the weak
coupling computation of the correlators can be extrapolated to strong
coupling without change \cite{lmrs,intril,ehsw1,ehsw2}.

Some extremal correlators describe transitions from a giant graviton
state into multi-particle KK graviton states.  For example, the
correlator
\begin{equation}\label{gtokk} 
\langle \chi_{[N]}(\Phi^{\dagger})(y)\tr(\Phi^{J_1})(x_1)\tr(\Phi^{J_2})(x_2)\dots \tr(\Phi^{J_n})(x_n) \rangle
\end{equation}
such that $\sum_i^n J_i=N$ and $J_i \ll N$, encodes information 
about the transition 
from an $AdS$ giant with $N$ units of angular momentum into several KK gravitons.
Note that these processes involve ``in'' and ``out'' states that are half-BPS 
and stable. Their existence does not indicate an instability of the initial
state, since the survival probability does not fall off exponentially with time.   
We may view the transitions in terms of a choice of detectors. In the above case  
for example, the detectors are chosen to detect KK gravitons. The strong dependence 
of transition probabilities on the choice of measurement was emphasized in 
\cite{balbabel1}\cite{balbabel}.   
The reverse process, where several KK gravitons give rise to a giant graviton is also of interest. 
We wish to calculate the probabilities for such transitions to occur. Most of 
these probabilities will be exponentially suppressed in $N \sim 1/g_s$ 
indicating the non-perturbative nature of such transitions.

\subsection{Statement of the puzzle}

We want to work out the normalized amplitudes for the transition from
$AdS$ and sphere giant graviton states either into other giant gravitons or
into many Kaluza-Klein gravitons.  We make use of two different normalizations:
the multi-particle normalization and the overlap-of-states
normalization.  For the multi-particle normalization we divide the
correlator by the norms of each of the  products separately; for
the overlap-of-states normalization we divide by the norm of all the
outgoing states together.  In this section,  we ignore the spatial structure of the
correlators and only consider the matrix-index structure. In our exact treatment
later we cannot ignore the spatial dependencies of the correlators.

The multi-particle-normalized transition from an $AdS$ giant graviton state with 
angular momentum $N$ into several
Kaluza-Klein gravitons, all of which have angular momentum $J$, is given by
\begin{equation}
  \label{AdStoidenticalKK}
  \frac{\left|\langle \chi_{[N]} (\Phi^{\dagger}) (\tr (\Phi^J) )^{N/J}
    \rangle\right|^2}{\langle\chi_{[N]} ( \Phi^\dagger )\chi_{[N]}
    ( \Phi )\rangle\, \,\langle \tr ( \Phi^{\dagger J} )\tr ( \Phi^J ) \rangle^{N/J}}
\end{equation}
and the overlap-of-states-normalized $S$ giant transition is given by
\begin{equation}
  \frac{\left|\langle \chi_{[1^N]} (\Phi^{\dagger}) (\tr (\Phi^J) )^{N/J}
    \rangle\right|^2}{\langle\chi_{[1^N]} ( \Phi^\dagger )\chi_{[1^N]}
    ( \Phi )\rangle\, \,\langle( \tr  (\Phi^{\dagger J} ))^{N/J}(\tr ( \Phi^J ))^{N/J} \rangle}
\end{equation}

The first part of the puzzle is that, in general, the multi-particle normalization
does not yield well-defined probabilities.
For example if we calculate the $AdS$ giant graviton process (\ref{AdStoidenticalKK})
for $J = N/2$, we get the answer
\begin{equation}
 \frac{\left|\big\langle
       \chi_{[N]}(\Phi^\dagger)  \tr(\Phi^{\frac{N}{2}})\tr(\Phi^{\frac{N}{2}})\big\rangle
    \right|^2}{\langle\chi_{[N]} ( \Phi^\dagger )\chi_{[N]}
    ( \Phi )\rangle  \big\langle
    \tr(\Phi^{\dagger\frac{N}{2}})\tr(\Phi^{\frac{N}{2}})\big\rangle\big\langle
    \tr(\Phi^{\dagger\frac{N}{2}})\tr(\Phi^{\frac{N}{2}})\big\rangle}  \sim \frac{1}{6\sqrt{2}}  \left(\frac{32}{27}\right)^{N}
\end{equation}
which is bigger than 1 and therefore does not yield a well-defined probability.

Similarly the multi-particle-normalized transition (\ref{AdStoidenticalKK}) for $J << N$ is
given by
\begin{equation}
  \frac{\left|\langle \chi_{[N]} (\Phi^{\dagger}) (\tr (\Phi^J) )^{N/J}
    \rangle\right|^2}{\langle\chi_{[N]} ( \Phi^\dagger )\chi_{[N]}
    ( \Phi )\rangle\, \,\langle \tr ( \Phi^{\dagger J} )\tr ( \Phi^J ) \rangle^{N/J}}\sim  2^{-\frac{1}{2}} e^{-N + 2N\log(2) -(N/J)\log(J)}
\end{equation}
The factor multiplying $N$ in the exponential is $-1/2 + \log(2) -
(1/2J)\log(J)$, which is positive for all $J$ (because  $\log(2)$
dominates).  Thus this amplitude exponentially increases with $N$ for
all $J$.  This is also inconsistent with a probability interpretation.

When we consider the multi-particle normalized transition from an $AdS$ giant
into two smaller $AdS$ giants, we get similar divergent results
\begin{equation}
 \frac{\left|\big\langle
       \chi_{[N]}(\Phi^\dagger) \chi_{[\frac{N}{2}]}(\Phi)\chi_{[\frac{N}{2}]}(\Phi) \big\rangle
    \right|^2}{\big\langle
   \chi_{[N]}(\Phi^\dagger)\chi_{[N]}(\Phi)\big\rangle  \big\langle
    \chi_{[\frac{N}{2}]}(\Phi^\dagger)\chi_{[\frac{N}{2}]}(\Phi)\big\rangle\big\langle
    \chi_{[\frac{N}{2}]}(\Phi^\dagger) \chi_{[\frac{N}{2}]}(\Phi)\big\rangle}  \sim \frac{3}{\sqrt{8}}  \left(\frac{32}{27}\right)^{N}
\end{equation}  

Note however that the multi-particle normalization does not always
give divergent results.  For example the transition from a \emph{sphere} giant state
into KK gravitons with  $J << N$ is given by 
\begin{eqnarray}
  \frac{\left|\langle \chi_{[1^N]} (\Phi^{\dagger}) (\tr (\Phi^J) )^{N/J}
    \rangle\right|^2}{\langle\chi_{[1^N]} ( \Phi^\dagger )\chi_{[1^N]}
    ( \Phi )\rangle\, \,\langle \tr ( \Phi^{\dagger J} )\tr ( \Phi^J )
    \rangle^{N/J}} \sim (2\pi)^{\frac{1}{2}} e^{-N + \frac{1}{2}
  \log(N) -(N/J)\log(J) }
\end{eqnarray}
which is exponentially decreasing for all $J$.

The second part of the puzzle is that there is no clear way to decide
which normalization to use.
In this paper we solve both puzzles.  We will show that the
multi-particle normalization requires us to divide by the two-point
function on a higher genus manifold.  This will yield well-defined probabilities
for transitions from a single giant graviton state into a collection of smaller objects. We
will also find that different transition probability interpretations
require different normalizations.

A final subtlety is that for transitions from a giant state to states described by single trace operators, 
we cannot just naively take
the square of the absolute value of the overlap amplitude of the giant graviton operator with a bunch
of traces.  Instead we should take the overlap of the giant graviton operator with
traces and multiply with the overlap amplitude involving the duals of the trace
operators. The dual is defined in terms of the metric on the space of
traces: $\cG^{ij}\cO_j$.

Details of the calculations presented in this section, as well as several other computations, are given in Appendix
\ref{appendixcalc}.  The correctly normalized results for the processes
discussed here are given in Section \ref{sec:correctresults}. These are exponentially suppressed in $N$ as expected.

\section{From factorization to 
 probability interpretation of correlators}\label{sec:facttoprob}

\subsection{Factorization on $S^4$ and probabilities}\label{factprob243}

Factorization in conformal field theory relates n-point correlators on the sphere to lower point 
correlators. 
Consider 
\bea 
 |\langle A^{\dagger} ( x^* ) B  ( Q )  \rangle|^2  =\langle A^{\dagger} ( x^* ) B  ( Q)\rangle\, \langle B^{\dagger} ( Q^* ) A  ( x )  \rangle
\eea 
Factorization implies that  
we can interpret a normalized version of this as a probability 
for the state created by the operator $A$ at $x$ to evolve into 
the state created 
by the operator $B$ at $Q^*$. The action of conjugation acts by
 reversing the sign of the 
Euclidean time coordinate. 

\begin{figure}[t]
\begin{center}
\includegraphics{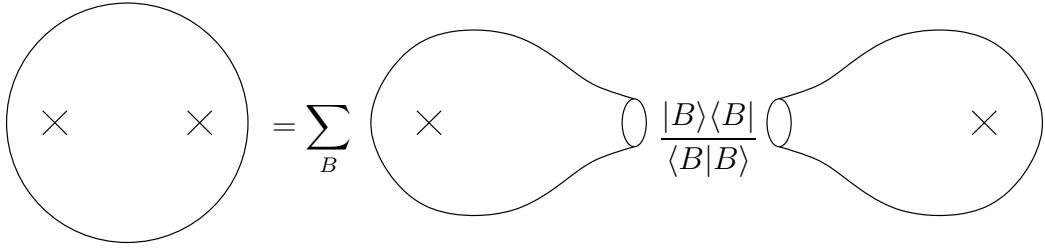}
\caption{A sphere correlator by gluing two spheres} \label{fig:sphere}
\end{center}
\end{figure}

Using a basis $B$ for the set of all possible operators, which we
choose to diagonalize the metric on the space of local operators,    
the  factorization equation takes the form  
\bea 
\langle A^{\dagger}(x^*) A ( x )  \rangle = 
\sum_B  \frac{ \langle A^{\dagger} (x^*) B  ( Q )  \rangle 
\,\langle B^{\dagger} ( Q^* ) A  (x)   \rangle }{ \langle B^{\dagger} ( Q^* )  B ( Q )       \rangle }
\eea 
See Figure \ref{fig:sphere}.
Dividing by the term on LHS we have 
\bea 
1 = \sum_{B} P ( A(x) \to  B(Q) ) 
\eea
where $P$ is interpreted as the probability for $A$ to evolve into $B$, given by 
\bea\label{Probs}  
P ( A(x) \to B(Q)  ) =   \frac{ \langle A^{\dagger} (x^*) B  ( Q )  \rangle 
\,\langle B^{\dagger} ( Q^* ) A  (x)   \rangle}{\langle A^{\dagger}(x^*) A ( x )  \rangle \langle B^{\dagger} ( Q^* )  B ( Q )       \rangle }
\eea

In the context of the  2D Matrix CFT model
 ( see section  \ref{sec:fact2d} ) 
 where the fields  have matrix oscillators in their mode expansion, 
the  operation of conjugation acts as 
$ \alpha_{-n} \rightarrow \alpha_{n}^{ \dagger} $ in the holomorphic sector. 
This is a symmetry of $ L_0   $.   
Similarly in the antiholomorphic sector, we have that
 $  \bar \alpha_{-n} \rightarrow \bar \alpha_{n}^{ \dagger} $, which is a symmetry of 
$ \bar L_0  $. Hence the conjugation is a symmetry of the Hamiltonian 
 $ H = L_0 + \bar L_0 $ which generates translations in time. 

 In  Euclidean theories, the proper definition of the adjoint of an operator involves the usual
 conjugation as well as the reversal of the Euclidean
 time. This operation guarantees that
 self-adjoint operators remain self-adjoint under Euclidean time
 evolution: $ A ( \tau ) = e^{ H \tau } A ( 0) e^{ - H \tau } $.  It
 also means that for a physical theory $\langle A^{ \dagger} (- \tau , \theta )
 A (\tau , \theta ) \rangle$ must be positive, a
 condition called reflection positivity \cite{ost}. 
 Thus the RHS of eq. (\ref{Probs}) is positive as it must
be the case for a proper probability interpretation.

The same thing can be said about extremal correlators which involve
 holomorphic operators at a number of different points: 
\bea 
&& \langle A_1^{\dagger}(x_1^*) A_2^{\dagger}(x_2^*) \dots A_k^{\dagger}(x_k^* ) A_1( x_1 )A_2( x_2) \dots A_k ( x_k )   \rangle \nnm \\ 
&& = \sum_{B } 
{  \langle A_1^{\dagger}(x_1^*) A_2^{\dagger}(x_2^*) \dots A_k^{\dagger}(x_k^* ) B (  Q )  \rangle \,\langle 
 B^\dagger(Q^* )  A_1( x_1 )A_2( x_2 ) \dots A_k ( x_k)  \rangle  \over 
 \langle B^\dagger( Q^* ) B ( Q)  \rangle } 
\nnm \\
\eea 
Then we can still derive a sum of probabilities equal to $1$ 
with 
\bea\label{ovlpnorm}  
&& P ( A_1(x_1) , A_2(x_2) \dots A(x_k) \to B(   Q ) ) \nnm \\
&& = \frac{ |  \langle A_1^{\dagger}(x_1^*) A_2^{\dagger}(x_2^*) \dots A_k^{\dagger}(x_k^* ) B (  Q ) \rangle |^2 }{
  \langle A_1^{\dagger}(x_1^*) A_2^{\dagger}(x_2^*) \dots A_k^{\dagger}(x_k^* ) A_1( x_1 )A_2( x_2) \dots A_k ( x_k )   \rangle \, 
 \langle B^\dagger( Q^* ) B ( Q) \rangle}
\eea 
Note that these arguments involve the overlap-of-states normalization, not 
the multi-particle normalization. If we replace $|  B( Q ) \rangle $ by 
a state created by more than one operator e.g
 $ | B_1  (y_1) B_2 (y_2) \rangle $ the formula (\ref{ovlpnorm})
can be used but it will not give an answer corresponding to a probability for 
separate detectors measuring $B_1  (y_1) $ and
 $B_2  (y_2) $. We will describe the case of multiple detectors and 
multi-particle normalization in the next subsection.

We will describe the detailed factorization equations later on, which
follow from conformal invariance and the sewing properties of path integrals. These
equations involve sums over all operators. 
There is a limit of large
separations where the factorization can be restricted to BPS states,
and gives the combinatoric (position independent) factorization
equations in terms of the Littlewood-Richardson coefficients obtained in \cite{cr}.

If we use the trace basis for the $B$'s in (\ref{Probs}), we still have
a factorization equation.
In this basis, the probability is defined by  
\begin{equation}
  P ( A  \to B ) = \frac{\langle A^{\dagger} B \rangle \, \langle \tilde {B }^{\dagger} 
A \rangle}{ \langle A^{\dagger} A \rangle }   
\end{equation}
where $ \tilde B $ is the dual operator to $B$, with duality being given 
by the inner product defined by the 
2-point function  (see the Appendix Section \ref{dualbasisec} for more details).

\subsection{Higher topology and multi-particle normalization}
 
We can extend these arguments to derive the probability
interpretation for the case of multiple outgoing particles.

We need to consider correlators of higher topology. Take the $\R^4$ manifold with
two $B^4$'s cut out and an operator insertion. 
 This gives a manifold with two $S^3$ boundaries and a puncture. 
 Take a second copy of $\R^4$ with the $B^4$'s cut
out and an operator inserted. 
 Glue each $S^3$ boundary with a corresponding $S^3$ boundary on
 the other $\R^4$.
Call this manifold $X$ and consider a two-point function on $X$:
\bea 
\langle A^{\dagger}  ( x ^*) A ( x)  \rangle_{G=1}  
\eea
This procedure is analogous to that of gluing two cylinders in 2d CFT to get a
genus one surface with two punctures.  Here we are doing the gluing in a 4d
CFT, but we have used the notation $G=1$ by analogy. 
We introduce the notation $ \Sigma_4 ( G ) $, to denote the four dimensional analog of 
a genus $G$ surface in two dimensions. It can be obtained by taking two copies of 
$S^4$ with  $G+1$ non-intersecting balls removed, and gluing the two along the $S^3$ boundaries. 
To define probabilities for some set of states to go into 
$G+1$ states we need to normalize with correlators on  $ \Sigma_4 ( G ) $. 

\begin{figure}[t]
\begin{center}
\includegraphics{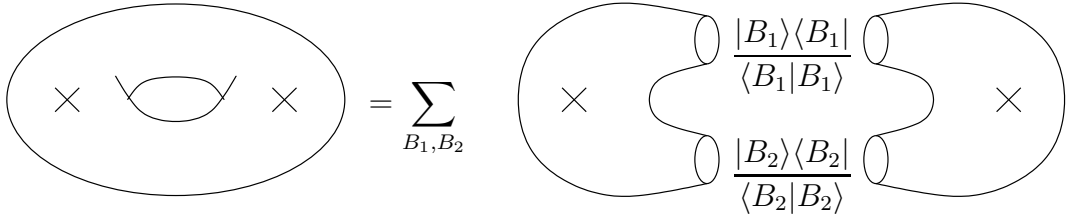}
\caption{A torus correlator by gluing two spheres} \label{fig:torus}
\end{center}
\end{figure}

We can argue for this as follows. By the factorization argument we have 
\bea\label{gonefac} 
\langle A^{\dagger}(x^*)  A ( x )  \rangle_{  G=1  }  = \sum_{B_1, B_2 }
\frac{ \langle  A^{\dagger}  ( x^* ) B_1 (C_1 )   B_2 ( C_2 )   \rangle \, 
\langle B_2^{\dagger} ( C_2^*  )  B_1^{\dagger} ( C_1^* ) 
  A  ( x ) \rangle }{ 
 \langle B_1^{\dagger} ( C_1^*  )  B_1  ( C_1 ) \rangle\, \langle B_2^{\dagger} ( C_2^* )  B_2( C_2) 
 \rangle }
\eea
See Figure \ref{fig:torus}.
$C_1$ and $C_2$ are circles along which we cut the torus. The operators $B_i ( C_i) $ 
create states localized on these circles. By scaling, these are related to the more familiar 
states which, in the operator-state correspondence, are obtained 
 by local operators acting on the
 vacuum. Hence the equation above can be related to correlation functions 
 of usual local operators.  Eq. (\ref{gonefac}) is explained in more detail in section \ref{sec:fact2d} in the two dimensional case, 
and in section \ref{sec:fact4d} in the four dimensional case. 

It follows from (\ref{gonefac}) that 
\bea 
 1 = \sum_{B_1, B_2 }
\frac{ \langle  A^{\dagger}  ( x^* ) B_1 (C_1)   B_2 ( C_2 )   \rangle \, \langle B_2^{\dagger} ( C_2^* )  B_1^{\dagger} (C_1^*)   A  ( x ) \rangle }{ 
 \langle A^{\dagger}(x^*)  A ( x )  \rangle_{G=1}\;\langle B_1^{\dagger} ( C_1^*  )  B_1  ( C_1) \rangle\, \langle B_2^{\dagger} ( C_2^* )  B_2( C_2) 
 \rangle }
\eea 
More generally 
\bea\label{gonefaci}  
1 = \sum_{B_1, B_2 }
\frac{  \langle A_1^{\dagger} (x_1^*)   \cdots  A_{k}^{\dagger} ( x_k^* ) 
   B_1 (C_1)    B_2 (C_2)   \rangle \, \langle B_2^{\dagger} ( C_2^*)  B_1^{\dagger} (C_1^*)
A_k ( x_k) \cdots    A_1 (x_1)  \rangle }{ 
 \langle A_1^{\dagger}(x_1^*)   \cdots A_{k}^{\dagger} (x_k^* ) 
 A_{k} ( x_k ) \cdots     A_1(x_1)   \rangle_{G=1  }\, \langle B_1^{\dagger} (C_1^*) 
 B_1(C_1) \rangle \,\langle B_2^{\dagger} ( C_2^*) B_2(C_2) \rangle  
 } 
\eea 
Since every summand is real and positive, it can be interpreted as a probability.  
We conclude that to normalize  correlators  
in order to get a probability for the case of multiple outgoing objects
we need to divide by factors 
involving higher genus correlators.
This corrects the naive multi-particle prescription used in the previous section.

We conclude this section with some comments:
\begin{itemize}
\item  
Notice that the probabilities we describe  
are defined subject to the constraint that the  number of final states is fixed.  
Multi-particle states in this context are obtained by the action 
of products of well separated operators on the vacuum.
 A brief discussion of conditional probabilities subject to additional conditions, 
such as fixing one of the outgoing states, is given  in  Appendix Section \ref{condprob}.
\item
In this paper we focus on Euclidean correlators on $\R^4$ (or $S^4$)
and higher genus spaces. 
A Lorentzian interpretation can be developed by choosing 
an appropriate time direction so that the out-states appear at a later time. 
When the factorization equations are appropriately 
continued to Lorentzian signature,  they still provide 
 relations between correlators.
We have not described the normalization procedure in a purely Lorentzian 
set-up, but we expect that the probabilities continue to be  relevant.  
Certainly in the large distance limits where the probabilities 
are independent of separations (see section \ref{sec:correctresults}), this is the case. 
 A more thorough 
investigation of the Lorentzian picture is desirable, where 
issues of bulk causality of the results can be explored along 
the lines of \cite{pst}.
\item
We work in a  basis where the states are characterized by the action of a 
local operator on the CFT vacuum. These states are natural to consider from the CFT point of view. 
In general, such states
are linear superpositions of states carrying arbitrary four-momentum. Definite momentum states must be constructed
so as to recover the S-matrix of type IIB string theory in the flat space limit, 
as described in
\cite{susskindflat}\cite{polchinskiflat}\cite{giddingsflat}.  It would
be interesting to express the factorization equation in the momentum
basis and study which features survive in the flat space limit.

\end{itemize}

\section{Factorization and gluing amplitudes in two dimensions}\label{sec:fact2d}

\subsection{Matrix Model CFT}

Consider the 2-dimensional CFT with action 
\begin{equation}
 S={1\over 4\pi}\int d^2 z \Tr \left(\partial X\bar{\partial}X+\partial Y\bar{\partial}Y\right)
\end{equation}
where $X$ and $Y$ are Hermitian $N\times N$ matrices.
For these fields the two point functions on the sphere are given by
\begin{equation}
\langle X^i_{j}(z_1,\bar{z}_1)X^k_{l}(z_2,\bar{z}_2)\rangle =-\delta^i_{l}\delta^k_{j}\log |z_1-z_2|^2 
=\langle Y^i_{j}(z_1,\bar{z}_1)Y^k_{l}(z_2,\bar{z}_2)\rangle 
\end{equation}
We also introduce the complex fields
\begin{equation}
 Z={X+iY\over\sqrt{2}},\qquad Z^{\dagger}={X-iY\over\sqrt{2}}
\end{equation}
for which the two-point functions on the sphere are given by
\begin{eqnarray}
&\langle {Z^\dagger}^i_{j}(z_1,\bar{z}_1)Z^k_{l}(z_2,\bar{z}_2)\rangle =-\delta^i_{l}\delta^k_{j}
\log |z_1-z_2|^2 \nn \\
&\langle Z^i_{j}(z_1,\bar{z}_1)Z^k_{l}(z_2,\bar{z}_2)\rangle =
\langle {Z^\dagger}^i_{j}(z_1,\bar{z}_1){Z^{\dagger i} }_{l}(z_2,\bar{z}_2)\rangle = 0
\end{eqnarray}

Our goal is to study factorization properties for correlators of this matrix model CFT. 
The sums that enter in the factorization identities run over the space of local operators
 of the  conformal field theory. It is natural to consider 
polynomials in the derivatives 
$\partial  Z , \d^2 Z ...  $, and $ \d {Z^{\dagger}}, \d^2 Z^{\dagger} ... $, 
along with exponentials of the matrices $ Z , Z^{\dagger} $ which generate non-zero momentum 
sectors. The non-zero momentum states decouple in most sectors of interest. 
 As discussed in Section \ref{zeroversusfree}, 
in some cases of interest it is  also consistent to truncate to the space 
of local operators invariant under global $U(N)$
 transformations\footnote{In the case of the four dimensional Super Yang Mills theory
it is always consistent to truncate to the subspace of local gauge invariant operators (see Section \ref{zeroversusfree}).}
\begin{equation}
 Z\to U^\dagger ZU
\end{equation}
This subspace of local operators is given by traces of all matrix words built using $Z$, $\partial^n Z$, ${Z^\dagger}$
and $\partial^n{Z^\dagger}$ ($n>0)$) as letters. If we consider factorization equations for $U(N)$ invariant operators
in the limit of large separations, it is possible to further restrict to just those words built using letters $\partial Z$
or $\partial {Z^\dagger}$ only. This is discussed further in Section \ref{treefact}. A basis for this subspace is provided by the loops
\begin{equation}
 {\cal A}_n (z)=\Tr \left( (\partial Z)^n\right),\qquad {\cal A}_n^\dagger (z)=\Tr \left( (\partial {Z^\dagger})^n\right)
\end{equation}
along with their products, i.e. multi-traces. 
Although this basis is complete and so perfectly acceptable, it is awkward.
In particular, the two point function on the space
 of local operators is not diagonal with respect to this basis.
This is a significant complication because the matrix 
inverse of this two point function enters the factorization equations.
Since the two point functions 
\bea 
 \langle \partial {Z^\dagger}^{i}_j  ( z_1 ) \d Z^{k}_l ( z_2 )  \rangle 
 =  { - \delta^i_l \delta^k_j   \over ( z_1-z_2 )^2 } 
\eea 
have the same index structure as those corresponding to the elementary free field $\Phi^i_j $ in the four dimensional 
super Yang Mills case, we can use results of \cite{cjr}\cite{cr} for the color combinatorics. Thus as in the four dimensional case, 
a far more convenient basis is provided by the Schur polynomials.  This basis
 is complete and further,
the two point function on the space of Schur polynomials is diagonal.

\subsection{An inner product on the states}\label{sec:innerprod243}

In unitary two-dimensional conformal field theories, we can express the
Hermitian inner product on the set of states as a product on the
space of local operators $\{\cA_i(z,\zb) \}$:
\begin{equation}
\label{innerproductdef}
  \cG_{ij} = \braket{i}{j} = \corr{\cA_i^{\dagger \prime} (z',\zb'=0)  \cA_j(z,\zb=0)}_{S^2} 
\end{equation}
where $z$ and $z'$ are related by $zz' =1$, and we denote by
$|i\rangle$ the state corresponding to the operator $\cA_i(z, \bar
z)$. Note that the prime on the first operator indicates the
$z'$-frame, and the operation of conjugation on it conjugates all {\it
  explicit} factors of i and transposes matrix indices, but leaves the
$z$ and $\bar z$ indices unchanged. This is essentially the operation
of Euclidean conjugation, which we review in the Appendix \ref{euclidconj}.

When the inner product of states is defined as an operator product, 
the hermiticity property
$\braket{i}{j} = \braket{j}{i}^*$ follows from the properties of
 conformal invariance and operator conjugation
 \cite{Ginsparg:1988ui}\cite{Polchinski}.  In a
unitary theory, the inner product of states is nonnegative, $\braket{i}{i} \geq
0$ for all $i$, implying a positivity property for the metric
$\cG_{ij}$. 

As an example, consider 
correlators involving  holomorphic derivatives of $Z$, of arbitrary order, 
 in the complex matrix model CFT. By direct computation, 
these are given by (see Appendix \ref{gluesph})
\begin{equation}
  \corr{{\partial}^{\prime m} Z^{\dagger i}_{\phantom{\dagger}j}(z'=0)
    \partial^n Z^k_l(z=0)}_{S^2}=m((m-1)!)^2\delta_{nm}\delta^i_l\delta^k_j \label{corderiv}
\end{equation}
The same result also follows if we use the operator-state map
$ \partial^k  Z^i_j  \leftrightarrow -i (k-1)!{\alpha_{-k;j}}^i$ 
and the inner product on states
\begin{equation}
\langle 0|(i (p-1)!\alpha_{p;j}^{\dagger ~ i } )(-i (k-1)!{\alpha_{-k ;q }}^l |0\rangle =k\big[ (k-1)!\big]^2\delta_{pk}
 \delta^i_q \delta^l_j
\end{equation}
Correlators involving products of derivatives of $Z$ and $Z^\dagger$ split up into sums and products 
of correlators given by eq. (\ref{corderiv}). 

Since the metric $\cG_{ij} = \braket{i}{j}$ is Hermitian and positive
definite it is diagonalizable with positive real eigenvalues. In an appropriate basis we then have
\begin{equation}
  \cG_{ij} = \braket{i}{i} \delta_{ij}
\end{equation}
with inverse
\begin{equation}
  \cG^{ij} = \frac{1}{\braket{i}{i}}\delta^{ij}
\end{equation}

From the form of eq. (\ref{corderiv}), we know that part of this
diagonal basis is given by derivatives of $Z$.  Gauge invariant
Schur polynomials of the primary field $\d Z$ are also part of this 
diagonal basis. 
This follows since the structure of the two-point
function of $ \langle\partial Z^{\dagger} \partial Z \rangle $ is the
same as that for free four dimensional fields studied in \cite{cjr}.

\subsection{Sphere factorization}\label{treefact}

 For  multi-point functions with a simple choice of operator 
positions, the spacial dependence factors out very simply 
 and all the interesting 
structure is in the dependence on $N$ and the choice of operators. For example 
\begin{equation}
 \langle \prod_{i=1}^l \chi_{R_i}( \partial Z^\dagger) ( z ) \prod_{j=1}^k 
\chi_{S_j} ( \partial Z) ( 0 ) \rangle = z^{ -2 \Delta } \sum_S
g(R_1,...,R_l;S){n_S! Dim_N(S)\over d_S}g(S_1,...,S_k;S),
\end{equation}
Here $ \Delta $ is the sum of  the number of boxes in the Young diagrams 
$R_1, R_2 \cdots R_n $. Further,   $n_S = \Delta $
 is the number of boxes in Young diagram $S$, $Dim_N(S)$ is 
the dimension of $S$ taken as a
representation of $U(N)$, $d_S$ is the dimension of $S$ taken as a representation of the symmetric group $S_n$
and $g(R_1,R_2,...,R_l;S)$ is a Littlewood-Richardson (LR) coefficient.
 It is possible to derive fusion and
 factorization identities for  appropriate ratios 
 of such correlators \cite{cr}. These
identities are a direct consequence of the sum rule
\begin{equation}
g(R_1,R_2,...,R_n;S)=\sum_{S_1,S_2,\dots,S_{n-2}}g(R_1,R_2;S_1)g(S_1,R_3;S_2)\cdots g(S_{n-2},R_n;S)
\end{equation}
satisfied by the LR coefficients. It is natural to expect
that the CFT factorization will reduce to these
 combinatoric (position independent) factorization 
identities in some limit. In this section we describe  this
 in the simplest possible
setting, where two $S^2$ correlators are glued to give another $S^2$ correlator. The local
coordinates on the first $S^2$ are denoted by $z$; the local coordinates on the second $S^2$
are denoted by $w$. The two spheres are glued around $z,\zb=0$ and $w,\wb=0$ with $zw=1$.

The CFT factorization equation states
\begin{equation}
   \corr{\cO_1(p_1)\cO_2(p_2)}_{S^2} = 
  \sum_{ij} \cG^{ij}
   \langle \cO_1(p_1) \cA_i(z,\zb=0)\rangle_{S^2}\langle\cA_j^\dagger(w,\wb=0)\cO_2(p_2)\rangle_{S^2} 
\end{equation}
This equation involves a sum over all operators. We will now argue that there is a limit of large
separations where the factorization can be restricted to ``BPS states'',
and gives the combinatoric (position independent) factorization
equations in terms of the LR coefficients. To see this,
focus on the leading contribution to the factorization equations in the large separation limit. 
By a large separation limit, we mean that
we take the distance between the operators, and the distance between the puncture and operators 
in the correlators to be large. 
From now on we will assume that we are in this limit and check the $N$ dependence that follows from 
factorization.
Since we are considering a large separation limit,
it is clear that operators that dominate the sum will be those with the smallest conformal dimension.
In addition, the only non-zero correlators have an equal number of $Z$s and $Z^\dagger$s. Taken together,
these facts imply that we can restrict to Schur polynomials in $\partial Z$ (or in $\partial Z^\dagger$). 
The operators that are dropped 
from the factorization sum, are polynomials that include at least one letter of the form $\partial^n Z$, with $n>1$.
These higher derivative terms lead to a faster fall off of the correlator as one increases 
the separation between operator locations, so that they don't contribute in the leading order.
With this restriction,
the color combinatorics for the CFT are identical to the zero dimensional model so that the only difference
between the two is extra spacial dependence in the CFT correlators.
For concreteness, consider the correlator
\begin{equation}
\langle\prod_i \chi_{R_i} (\partial Z^\dagger(z_i))\prod_j\chi_{S_j} (\partial{Z}(w_j))\rangle 
\end{equation}
%
%
Using the two point functions, in the large separation limit,
\begin{align}
\langle \partial {Z^\dagger}^{i}_j(z_1)\partial Z^{k}_l(w_1)\rangle & =-(w_1')^2\langle \partial {Z^\dagger}^{i}_j(z_1)\partial Z^{k}_l(w_1')\rangle 
 = {(w_1')^2\over (w_1'-z_1)^2}\delta^i_l \delta^k_j  \nn \\
 & = {1\over (1-z_1 w_1)^2}\delta^i_l \delta^k_j \approx {1\over (z_1)^2}{1\over (w_1)^2}\delta^i_l \delta^k_j  \nn \\
\langle \partial {Z^\dagger}^{i}_j(z_1)\partial Z^{k}_l(0)\rangle & = -{1\over (z_1)^2}\delta^i_l \delta^k_j \nn \\
\langle \partial {Z^\dagger}^{i}_j(0)\partial Z^{k}_l(w_1)\rangle & = -{1\over (w_1)^2}\delta^i_l \delta^k_j 
\end{align}
the spacial dependence factors out on both sides leaving the combinatoric (position independent) factorization
equations of \cite{cr}. The role of the higher derivatives that have been dropped is to modify the spacial
dependences so that the two sides match for any separation. This is illustrated explicitly, in a simple
setting, in Appendix \ref{gluesph}.

\subsection{Geometrical gluing and factorization of higher genus correlators }

\subsubsection{The torus gluing}\label{torusgluing}

We will now describe a  procedure
for getting correlators on a torus by
gluing together 3-punctured spheres.
The procedure can be
 generalized to the case of higher topology Riemann surfaces
following the description in 
\cite{Polchinski,Vafa:1987ea, Sonoda:1988mf,Sonoda:1988fq,Sonoda:1987ra}.

Suppose the first sphere is covered with coordinate patches
 $z$ and $z'$ glued together by
$zz'=1$, while the second sphere is covered with
 coordinate patches $w$ and $w'$ glued together by $ww'=1$.  
Let the
first sphere have punctures at $z, \zb=0$, $z',\zb'= 0$ and $z,\zb=e^{2\pi s}$
($s>0$) and the second at $w,\wb=0$, $w',\wb'= 0$ and $w,\wb=e^{2\pi s}$.

\begin{figure}[ht]
\begin{center}
\includegraphics{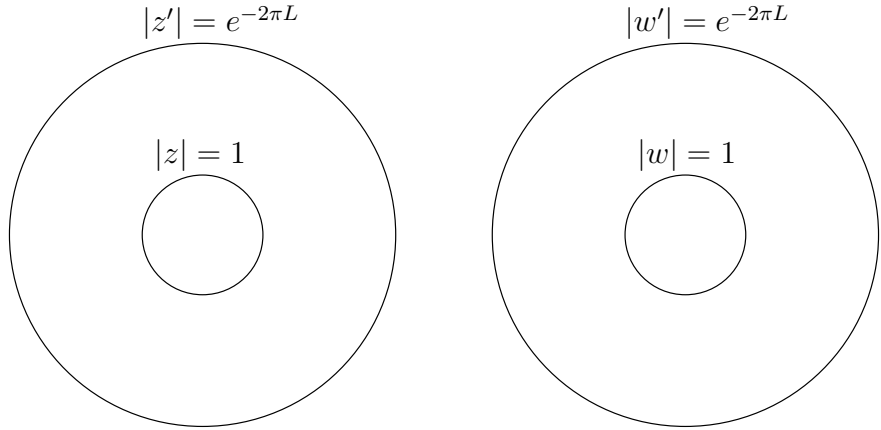}
\caption{The $z$-annulus and the $w$-annulus } \label{fig:2dgen1f1}
\end{center}
\end{figure}

We cut out from the first sphere a region around $z, \zb=0$
and another region around $ z' , \bar z' = 0 $ to give the first annulus.
  We cut out from the 
second sphere a region around $  w , \bar w  = 0 $ and a region around 
$ w' , \bar w' = 0 $ to give a second annulus. 
The cuts around $ z ,  \bar  z = 0 $ and $ w , \bar w = 0 $ can be described  
as  cutting out $|z| < e^{-2\pi \delta}$ 
and $|w| < 1$ for $\delta > 0$. We glue the two annuli by identifying points in
the regions 
 $e^{-2\pi \delta} \le |z| \le 1$ and  $1 \le |w| \le e^{2\pi
  \delta}$  with $zw = 1$.
  Thus if we are approaching $|z|=1$ from $|z| >1 $,
once we enter  the overlap region,
 we are now moving away from $|w|=1$ in the
direction of increasing $ |w| $.
The cut regions around  $z', \zb'=0$ and $w', \wb'=0$ 
can be described by 
 $|z'| <  e^{- 2\pi ({L}+\delta)}$ and $|w'| < e^{ - 2\pi {L}}$. 
Points on the two annuli regions 
$e^{ - 2\pi  ( L + \delta  ) } \le |z'| \le e^{ - 2\pi {L}}$
 and  $e^{ - 2\pi L  } \le |w'| \le e^{ - 2\pi (L - \delta )}$ respectively  
are identified by the equation $z'w' =
e^{ - 4\pi {L}}$. The gluing procedure produces a torus. If we continuously 
increase $ |z| $ from  the region near 
$|z|=  1 $ we move into the  $z^{\prime } $ patch
 with decreasing $|z'|$, via $ zz' =1$. This  
  maps to increasing $|w'|  $ in the $w'$ patch via $ z'w' = e^{-4\pi L}$. 
 This maps in turn into decreasing $|w| $ in the region near $ |w| =1 $ 
 on the second annulus, via $ ww'=1$,   which 
 maps back  to the region near $ |z| = 1 $ on the first annulus, thus 
 completing the periodic Euclidean time cycle of the torus. 
The $\delta $ factors can be taken to zero.

\subsubsection{Factorization of torus correlators }\label{opstatecorr}
We know from general arguments
\cite{Polchinski,Vafa:1987ea, Sonoda:1988mf,Sonoda:1988fq,Sonoda:1987ra}
 that for operators
$\cO_1$ and $\cO_2$ on a torus with modular parameter $\tau$, $q=
e^{2\pi i \tau}$,
\begin{eqnarray}
   \corr{\cO_1(p_1)\cO_2(p_2)}_{T^2} = &
(q\qb)^{-c/24} 
   \sum_{ij} \sum_{kl}q^{h_j} \ov{q}^{\tilde{h}_j} \cG^{ij} \cG^{kl}
   \corr{ \cO_1(p_1)\cA_j^{ \dagger \prime}(z',\zb'=0) 
 \cA_k(z,\zb=0)}_{S^2} \nn \\
  & \times\corr{\cA_l^\dagger(w,\wb=0) 
\cA_i^{ \prime}(w',\wb'=0)\cO_2(p_2)}_{S^2}\label{finalfact0}
\end{eqnarray}
where $z$ and $z'$, related by $zz'=1$, are coordinates on one sphere
and $w$ and $w'$, related by $ww'=1$, are coordinates on another
sphere.  The two spheres are sewn together around $z, \zb=0$ and
$w, \wb=0$ with $zw = 1$ and
 around $z', \zb'=0$ ($z,\zb=\infty$) and $w',\wb'=0$
($w, \wb=\infty$) with $z'w' = q = e^{-4 \pi L } $ to get a torus
with $ \tau = 2 i L $.  

We shall work with a basis of operators 
for which the metric is diagonal so that $\cG^{ij} = (1/\braket{i}{i})
\delta^{ij}$.
 Then the expression above can be written as
\begin{align}
  \corr{\cO_1(p_1)\cO_2(p_2)}_{T^2} = &
(q\qb)^{-c/24}  \sum_{i} \sum_{k} q^{h_i} \ov{q}^{\tilde{h}_i}
  \frac{1}{\braket{i}{i}\braket{k}{k}}\corr{ \cO_1(p_1) 
\cA_i^{\dagger \prime}(z',\zb'=0)
    \cA_k(z,\zb=0)}_{S^2} \nn \\
  & \times\corr{\cA_k^\dagger(w,\wb=0)\cA_i^{
      \prime} (w',\wb'=0)\cO_2(p_2)}_{S^2} \label{finalfact1}
\end{align}
Since the metric only mixes operators of 
the same dimension, the operators $\cA_i , \cA_k$ 
can be chosen to be eigenstates of the scaling operator. 
Notice that when both the operators $\cO_1,\cO_2$ are set equal to the unit operator, we
recover the modular invariant torus partition function. 

\begin{figure}[ht]
\begin{center}
\includegraphics{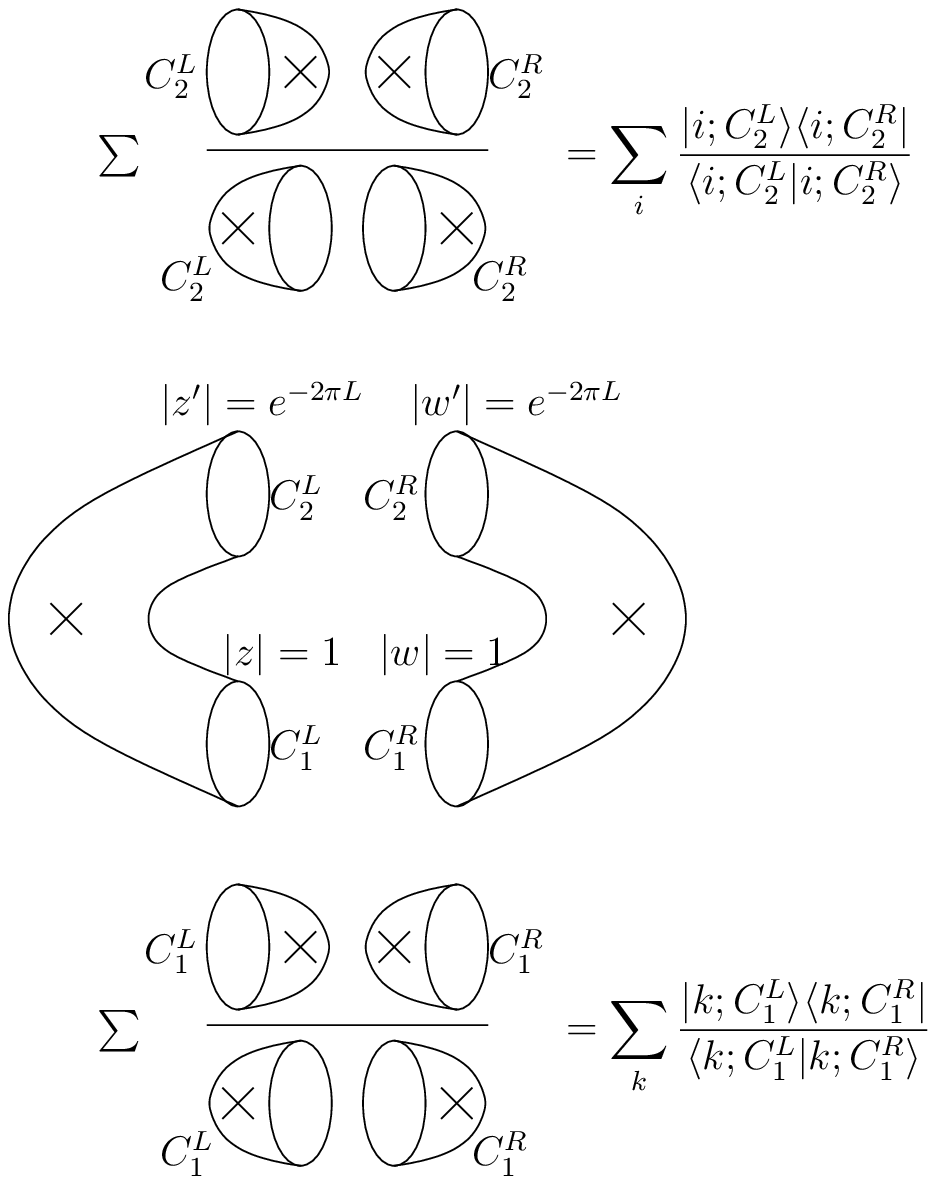}
\caption{The torus correlator obtained after gluing - see equation 
  \eqref{geomfact} } \label{fig:detailtorus}
\end{center}
\end{figure}

The geometrical gluing picture described in the previous
 section provides a set up to demonstrate how a
factorization equation such as (\ref{finalfact1}) arises.
The basic features of the following manipulations are in Figure 
\ref{fig:detailtorus}.
The result can be understood naturally 
in terms of the operator-state
correspondence of conformal field theories. 
We let the operator $\cO_1$
to be located at $z,\zb=e^{2\pi s}$
and the operator $\cO_2$ to be located at $w,\wb=e^{2\pi s}$ with $0<s<L$.  

Consider first the $z$-correlator appearing in eq. (\ref{finalfact1})
\begin{equation}
  \corr{\cA_i^{\dagger \prime} (z',\zb'=0)\cO_1(z,\zb=e^{2\pi s})
\cA_k (z,\zb=0)}_{S^2} \label{samplecorr}
\end{equation}
where we choose to order the operator insertions radially with
 respect to $|z|$.
To construct the $z$-annulus we remove the interior of the unit
 disk, $|z|<1$, and replace the operator at
$z, \zb=0$, $\cA_k (z,\zb=0)$, by a state on the boundary at
$|z| =1$.  We also remove the patch $|z'| < e^{-2\pi {L}}$ and replace the
operator at $z', \zb'=0$ ($z, \zb = \infty$),
 $\cA_i^{\dagger \prime}(z',\zb'=0)$, by a
state on the boundary at $|z'|=e^{-2\pi L}$.

These states arise as follows. Using the operator/state correspondence
 we associate to the operator 
$\cA_k (z,\zb=0)$ an ``in-state'' $|\cA_k \rangle$ defined by
$\lim_{z,\zb \to 0}\cA_k (z,\zb)|0\rangle$.
We can think of this state as living on a small circle of radius
 $|z|=\epsilon$ surrounding the origin $z,\zb=0$, and 
consider  
the limit $\epsilon\to 0$. Notice that there are no other
 operator insertions in the unit disk. Then 
the path integral over the unit disk amounts to radially
 propagating (or scaling) this state to a state at 
$|z|=1$.  This operation is
 equivalent to acting with the radial evolution operator 
$(1/\epsilon)^{-L_0 -\tilde{L}_0}$ on $|\cA_k \rangle$. 
The end result is that the states differ by
 a scale factor: $|k;|z|=1\rangle=\epsilon^{  h_k + \tilde{h}_k}|\cA_k\rangle$.
The new state is an eigenstate of the dilatation operator.

In a similar way, we associate to 
the operator $\cA_i^{\dagger \prime}(z',\zb'=0)$ an ``out-state''
\footnote{The ``in'' and ``out-
states'' thus defined, are conjugates of each other: 
$\langle\cA|=|\cA\rangle^{\dagger}$ \cite{Ginsparg:1988ui}.}   
$\langle\cA_i|$
defined by $\lim_{z', \zb' \to 0}\langle 0|\cA_i^{\dagger \prime}(z',\zb')$. 
This state can be thought of as living
on a circle of radius $|z'|=\epsilon$ surrounding $z',\zb'=0$, 
in the limit $\epsilon \to 0$.
Now consider the path integral over the region $0\le|z'|< e^{-2\pi L}$.
This amounts to radially propagating a state at $|z'| = e^{-2\pi {L}}$ 
to the state at $z',\zb'=0$.
Notice that we have chosen radial evolution in the direction of
 increasing $|z|$ or equivalently in the direction of
decreasing $|z'|$. 
To find the finite radius state, we consider the
left action of the inverse radial evolution operator 
$(e^{-2\pi L}/\epsilon)^{  L_0 + \tilde{L}_0}$ on 
$\langle\cA_i|$. 
This operation gives
$\langle i; |z'|=e^{-2\pi L}|= \langle \cA_i|
 (e^{2\pi L}\epsilon)^{ -h_i - \tilde{h}_i}$. 

Now consider the $w$-correlator in eq. (\ref{finalfact1})
\begin{equation}
\corr{ \cA_k^{\dagger}(w,\wb=0) \cO_2(w,\wb=e^{2\pi s})
 \cA_i^{\prime}(w',\wb'=0)}_{S^2} \label{samplecorr2}
\end{equation}
where we choose to radially order the operators with respect to $|w'|$. 
The $w$-annulus is constructed in a similar way. The patch $0\le|w'|<e^{-2\pi L}$ is removed replacing the operator
at $w', \wb'=0$, $\cA_i^{\prime}(w',\wb'=0)$, by the state 
$|i; |w'|=e^{-2\pi L}\rangle=(e^{2\pi L}\epsilon)^{  h_i + \tilde{h}_i}|\cA_i\rangle$. 
Similarly the patch
$0\le |w|<1$ is removed replacing the operator at $w,\wb=0$, $\cA_k^{\dagger}(w,\wb=0)$, by the state  
$\langle k; |w|=1|=\langle \cA_k|\epsilon^{ -h_k - \tilde{h}_k}$.

In this way each correlator is replaced with a matrix element of 
a single operator. The $z$-correlator 
(\ref{samplecorr}) is replaced with
\begin{equation}
{(e^{2\pi L}\epsilon)^{h_i + \tilde{h}_i} \over \epsilon^{h_k  + \tilde{h}_k}}
\Big\langle i;|z'|=e^{-2\pi L}\Big|\cO_1(z,\zb=e^{2\pi s})\Big|k;|z|=1\Big\rangle
\end{equation}
while the $w$-correlator (\ref{samplecorr2}) is replaced with
\begin{equation}
{\epsilon^{h_k+\tilde{h}_k} \over (e^{2\pi L}\epsilon)^{h_i+\tilde{h}_i}}
\Big\langle k;|w|=1\Big|\cO_2(w,\wb =e^{2\pi s})\Big|i;|w'|=e^{-2\pi L}\Big\rangle
\end{equation}
Notice that the multiplicative scale factors cancel when we multiply the two expressions together.

The norms in the denominator of eq. (\ref{finalfact1}) can be also written in terms of the finite radius states.
We can think of the norms as normalized sphere amplitudes obtained by gluing each cut-off disk from the original 
z-sphere with the
corresponding cut-off disk from the w-sphere, as shown in Figure \ref{fig:detailtorus}.
From the definition of the metric, eq. (\ref{innerproductdef}), and the local gluing relation $zw=1$, we may write 
\begin{equation}
\langle k|k\rangle =\Big\langle \cA_k^{\dagger}(w,\wb=0)\cA_k(z,\zb=0)\Big\rangle=\Big\langle k; |w|=1\Big|k;|z|=1\Big\rangle
\end{equation}
Since the gluing relation of the prime coordinates is $w'z'=q$, we have that
\bea
&&\langle i|i\rangle =q^{h_i}{\qb}^{\tilde{h}_i}
\Big\langle \cA_i^{\dagger \prime}(z',\zb'=0)\cA_i^{\prime}(w',\wb'=0)\Big\rangle= \nn \\ 
&&  q^{h_i}{\qb}^{\tilde{h}_i}
\Big\langle i; |z'|=e^{-2\pi L}\Big|i;|w'|=e^{-2\pi L}\Big\rangle \label{radiusnorms}
\eea
To obtain the last equation, we rescale from the coordinate $z'$ to $\tilde {z}=z'/q$ so 
that $w'\tilde z=1$. The factors of $q^{h_i}{\qb}^{\tilde{h}_i}$ transform
the operator at $z',\zb'=0$ to the $\tilde z$-frame. 
We see that when the norm $\langle i|i\rangle$ is expressed as an inner product between finite radius states at
$|z'|=e^{-2\pi L}$ and $|w'|=e^{-2\pi L}$, the relative factor appearing 
cancels the factors of $q^{h_i}{\qb}^{\tilde{h}_i}$ in the
numerator of eq. (\ref{finalfact1}).

Therefore we can replace the RHS of eq. (\ref{finalfact1}) with
\bea
   && (q\qb)^{-c/24}\sum_{i} \sum_{k}   
  \frac{\Big\langle i;|z'|=e^{-2\pi L}\Big|\cO_1(z,\zb=e^{2\pi s})
\Big|k;|z|=1\Big\rangle_{annulus}  }
{\braket{i;|z'|=e^{-2\pi L}}{i;|w'|=e^{-2\pi L}}
\braket{k;|w|=1}{k;|z|=1}}\cr
 && \times\Big\langle k;|w|=1\Big|\cO_2(w,\wb =e^{2\pi s})
\Big|i;|w'|=e^{-2\pi L}\Big\rangle_{annulus}  \cr
&&=    \sum_{i} \sum_{k}  
  \frac{\langle i; x = iL |\cO_1^{[x]}( x = i s )
|k; x =0 \rangle_{cyl}\langle k; x =0 |\cO_2^{[x]}( x =  -is )
\Big|i; x =-iL  \rangle_{cyl}   }
{\braket{i;x=-iL}{i;x=-iL}
\braket{k;x=0}{k;x=0}}
 \label{finalfact2}
\eea 
In the second line, we express the equation in terms of cylinder amplitudes described by coordinates 
$x,\bar{x}$ ($z=e^{-2\pi i x}, w=e^{2\pi i x}$). 
The coordinate $x$ will be periodically identified, $x \sim x+2iL$, 
to be made compatible with
the gluing relations. 
Notice that the operators $\cO_1$ and $\cO_2$ must be transformed properly under the coordinate change.
The power of $ (q \bar q)^{-c/24} $ in the first line is absorbed in the change in the overall 
normalization of the partition function under the change of coordinates 
from annulus to cylinder, which follows from the constant shift in the 
Hamiltonian: $H_{cyl}=L_0 + \tilde{L}_0-(c+\tilde c)/24$.

The first gluing of the two annuli
 along circle $|z|=|w|=1$, using $zw=1$, gives a single annulus or equivalently a cylinder of length $2 L$,
 and is accompanied with a sum over a complete set of states $|k\rangle$ on the unit circle. 
Thus (\ref{finalfact2}) can now be written as
\bea
&&(q\qb)^{-c/24}\sum_{i}
  {\Big\langle i;|z'|=e^{-2\pi L}\Big| \cO_1(z,\zb=e^{2\pi s}) 
                                                   \cO_2(w,\wb =e^{2\pi s})
    \Big|i;|w'|=e^{-2\pi L}\Big\rangle_{annulus} 
 \over \Big\langle i;|z'|=e^{-2\pi L}\Big|i;|w'|=e^{-2\pi L} 
\Big\rangle_{sph} } \cr 
&& = \sum_{i}
  {\Big\langle i; x = i L \Big| \cO^{[x]}_1( x = i  s ) 
                            \cO^{[x]}_2( x = -is )
    \Big|i; x = -i L \Big\rangle_{cylinder} 
 \over \Big\langle i; x=-i L\Big|i;x=-i L 
\Big\rangle }
\label{annulusampl}
\eea
We emphasize that the numerator in the first line is an annulus transition amplitude and the 
denominator is a sphere amplitude. 
The final gluing identifies the inner and outer radii 
of the annulus,  at $|w'|=e^{-2\pi L}$ and $|z'|=e^{-2\pi L}$, through 
$z'w' = e^{-4\pi L}$, or equivalently the ends of the cylinder by $x \sim x+2iL$, to produce the torus with
$\tau=2iL$. 
Then the final sum over states
in (\ref{annulusampl}) allows us to express it as a trace, or 
equivalently as the torus two-point function $\corr{\cO_1(z,\zb=e^{2\pi s})\cO_2(w, \wb=e^{2\pi s})}_{T^2}$.

It is useful to rewrite eq. (\ref{finalfact2}) more geometrically
 ( see figure \ref{fig:detailtorus} ) 
 in order to exhibit 
its coordinate independence
\bea\label{geomfact} 
&&  \corr{\cO_1(p_1)\cO_2(p_2)}_{T^2}   \cr 
&& = \sum_{i,k} 
{   \langle  i ; C_2^L | \cO_1(p_1) | k ; C_1^L \rangle \langle k ; C_1^R |   \cO_2(p_2)  | i ; C_2^R \rangle \over 
   \langle i ; C_2^L | i ; C_2^R  \rangle  \langle k ; C_1^R | k ; C_1^L \rangle } \cr 
&& = 
\sum_{i,k} 
  { \langle \cO_1(p_1) \cA_i^{\dagger } ( C_2^L ) \cA_k ( C_1^L ) \rangle   
   \langle \cO_2 (p_2)  \cA_i ( C_2^R ) \cA_k^{\dagger}  ( C_1^R ) \rangle 
 \over {  \langle  \cA_{i}^{\dagger} (C_2^L)  \cA_i(C_2^R ) \rangle  
     \langle \cA_k^{\dagger}  ( C_1^L )  \cA_k ( C_2^R ) \rangle }}
\eea     
In the final line we have expressed the factorization in terms of 
operators which create states on finite size circles. 
The action of these operators on the vacuum is defined in terms of the radial evolution 
of states created by local operators. For example  $ \cA_k ( |z| =1 ) |0 \rangle   \equiv |k;|z|=1\rangle $, where the 
the operator $  \cA_k ( |z| =1 ) $ can be viewed as creating a state at finite radius. 
Macroscopic loop operators are discussed in CFT and 2D gravity in \cite{ginsmoore}. 
The final line of (\ref{geomfact}) is identical to the RHS of (\ref{gonefac}).

We have presented the factorization equation in terms of the gluing of two annuli.
It is instructive to view it conversely in terms of the cutting of the torus. 
 Start with a path integral on a torus,
expressed in terms of a generic set of fields $ \phi $  
\bea\label{pifact}  
\langle \cO_1 ( p_1 )  \cO_2  ( p_2) \rangle_{G=1}  
= \int [d\phi]    e^{ - S ( \phi )  }  \cO_1 ( p_1) \cO_2 ( p_2) 
\eea 
Now we cut along two circles denoted by $C_1$ and $C_2$ to get two cylinders. 
These cylinders can be conformally mapped to the annuli in Figure \ref{fig:2dgen1f1}. 
The fields on the left and right are denoted by $\phi_L$
and $\phi_R$. The boundary values on the circles are written as $\phi_{b_1} , \phi_{b_2}$.
Hence the correlator can be written as 
\bea 
&& \langle \cO_1 ( p_1 )  \cO_2  ( p_2) \rangle_{G=1}  \cr 
&& = \int [d \phi_{b_1}]  [d \phi_{b_2}]    \int [d \phi_L] |_{ \phi_{b_1}}^{ \phi_{b_2} }
  e^{- S ( \phi_{L}  )  } \cO_1 ( p_1 )  \int [d\phi_R] |_{\phi_{b_1}}^{ \phi_{b_2} } 
 e^{- S ( \phi_{R }  )  } \cO_2  ( p_2) 
\eea  
The fields $ \phi_L$ and $ \phi_R $ are integrated subject to boundary conditions 
$ \phi_{b_1} , \phi_{b_2}$ at the circles $ C_1, C_2$. Each of the left/right path integrals  
give rise to wavefunctionals of fields on these circles that are correlated by the insertions of the local operators. 
Using the correspondence between 
wavefunctionals and Hilbert space states, the integrals $ \int d \phi_{b_1}  \int d \phi_{b_2}  $ 
can be replaced by sums over states. These are the  states
 summed over in eqs. (\ref{geomfact}) (\ref{finalfact2}).  
These cutting and gluing relations appear in their simplest form in topological field theories, 
see for example \cite{wit2dgrav}\cite{dijkhou}.

\subsubsection{Reflection Positivity}
Consider again putting $\cO_2$ at $w,\wb=e^{2\pi s}$ in eq. (\ref{finalfact1}), 
which corresponds to $z, \zb=e^{-2\pi s}$ since $z$
and $w$ are glued with $zw = 1$, and choose now $\cO_1$ to be its
conjugate at $z, \zb=e^{2\pi s}$.  Then eq. (\ref{finalfact1}) becomes
\begin{align}
  &
  \corr{\cO_2^\dagger(z,\zb=e^{2\pi s})\cO_2(w,\wb=e^{2\pi s})}_{T^2}
  \nn \\
  = &    ( q \bar q )^{-c/24} 
 \sum_{i} \sum_{k} q^{h_i} \ov{q}^{\tilde{h}_i}
  \frac{1}{\braket{i}{i}\braket{k}{k}}\corr{\cO_2^\dagger(z,\zb=e^{2\pi s}) \cA_i^{\dagger \prime}(z',\zb'=0)    \cA_k(z,\zb=0)}_{S^2} \nn \\
  & \times
  \corr{\cA_k^{\dagger}(w,\wb=0)\cA_i^{ \prime} (w',\wb'=0)\cO_2(w,\wb=e^{2\pi s})}_{S^2} \nn \\
  = &  ( q \bar q )^{-c/24}  \sum_{i} \sum_{k} q^{h_i} \ov{q}^{\tilde{h}_i}
  \frac{\left|\corr{\cO_2^\dagger(z,\zb=e^{2\pi s}) \cA_i^{\dagger \prime}(z',\zb'=0)    
\cA_k(z,\zb=0)}_{S^2} \right|^2 }{\braket{i}{i}\braket{k}{k}}  \label{finalfact3}
\end{align}
Finally note that the set $\{\cA_i\}$ contains all local operators of the theory.  
If $\cA(z,\zb)$ is an operator in this set,
then so is $\cA^{\dagger}(z,\zb)$. The two have the same weights and norm with respect
to the metric $\cG_{ij}$ defined in (\ref{innerproductdef}). Thus we can also write the formula above as 
\begin{align}
  &
  \corr{\cO^\dagger(z,\zb=e^{2\pi s})\cO(w,\wb=e^{2\pi s})}_{T^2}
  \nn \\
  = &( q \bar q )^{-c/24} \sum_{i} \sum_{k} q^{h_i} \ov{q}^{\tilde{h}_i}
  \frac{\left|\corr{\cO^\dagger(z,\zb=e^{2\pi s}) \cA_i^{\prime}(z',\zb'=0)    
\cA_k(z,\zb=0)}_{S^2} \right|^2 }{\braket{i}{i}\braket{k}{k}}  \label{finalfact}
\end{align}

If the modular parameter $\tau$ is purely imaginary, so that $q$ is
real and positive, then each and every summand is real and positive.
This demonstrates reflection positivity for the torus.
Because every summand is real and positive we can discard some of the
intermediate states in the sum to get an inequality with the left-hand
side larger than the right-hand side. In the case of the matrix CFT,
we choose to keep only states that are totally holomorphic or totally
antiholomorphic.  Furthermore we throw away all states except those
with first derivatives $\d Z$ and $\d Z^\dagger$. We only keep
gauge-invariant polynomials in these fields, which can be written as
Schur polynomials.  These are diagonal
\begin{equation}
  \corr{\chi_R(\d Z^\dagger) \chi_S(\d Z)} \propto \delta_{RS}
\end{equation}

\subsection{Probabilities and Inequalities in 2D} 

We will now do some specific checks of the factorization equation
(\ref{finalfact}). We keep gauge-invariant products of the primary
field $\partial Z$ in the sum only. We choose to work in the Schur polynomial basis $\chi_R(\d Z(z))$ for which
the metric is diagonal. We will
obtain an inequality with the position dependences and torus moduli appearing
explicitly.

In the following we will write $R(z)$ for $\chi_R(\d Z(z))$ and 
$R^\dagger (z)$ for $\chi_R(\d Z^\dagger(z))$.
By the analysis above, we get an inequality for the torus
correlator of the form
\begin{align}
  \label{eq:torusequationfirst}
  &   ( q \bar q )^{c/24} 
\big\langle R^\dagger(z=e^{2\pi s}) R(w=e^{2\pi s})\big\rangle_{T^2, \tau} \nn \\
  &> \sum_{R_1, R_2} e^{-4\pi {L}\Delta_1} \frac{\big\langle R^\dagger (z=
      e^{2\pi s}) R_1^{\prime} (z'=0) R_2(z=0)\big\rangle\, \big\langle R_2^\dagger(w=0)
      R_1^{\dagger \prime} (w'=0)  R (w= e^{2\pi
        s})\big\rangle}{\braket{R_1}{R_1} \braket{R_2}{R_2}}
\end{align}
where $\Delta_1$ is the conformal dimension of the operator $R_1$, and $q = e^{2\pi i \tau} =
e^{-4\pi {L}}$ since $\tau = 2i{L}$. We denote the conformal dimension of $R_2$ by $\Delta_2$. In order for the pair of
operators $R_1$ and $R_2$ to contribute, $\Delta_1+\Delta_2=\Delta_R$ where $\Delta_R$ is
the conformal dimension of $R$.
To check this inequality explicitly, we must work out all the individual terms appearing in the inequality. Note that the left hand side
 is an unnormalized correlator, 
given by the insertion of operators in the path integral, without 
dividing by the torus partition function. 
The right hand side   is insensitive to the
 normalization of the sphere correlators,
 so we will set the sphere normalization factor 
to $1$ in the following.

\subsubsection{The metric on the Schur Polynomials}

We follow conventions so that for a
single real scalar field, the 2-point function of its holomorphic derivatives on the sphere is given by 
\begin{equation}
  \corr{\d X(z_1)\d X(z_2)}_{S^2} = - \frac{1}{(z_1 - z_2)^2}
\end{equation}
and similarly for a single complex scalar field $Z$:
\begin{equation}
  \corr{\d Z(z_1)\d Z(z_2)}_{S^2} = - \frac{1}{(z_1 - z_2)^2} \label{propagatorZ}
\end{equation}

Then the Schur polynomials satisfy
\begin{equation}
  \corr{\chi_{R}(\d Z^\dagger (z_1)) \chi_{S} (\d Z(z_2))}_{S^2} = 
   \delta_{RS}f_R \frac{(-1)^{\Delta_R}}{(z_1 - z_2)^{2\Delta_R}} \label{2ptschur}
\end{equation}
where $f_R$ is defined by
\begin{equation}
  f_R = \frac{\Dim_R \Delta_R!}{d_R}
\end{equation}
In this expression, $\Dim_R$ is the dimension of the $U(N)$ representation $R$ and $d_R$
is the dimension of the symmetric group $S_{\Delta_R}$ representation $R$. 
To derive this 2-point function, we repeat all the steps of the corresponding 
four dimensional computation of \cite{cjr}, but noting that now each field contraction will give a factor of the propagator
(\ref{propagatorZ}). The relevant color combinatorics are the same as in the case of \cite{cjr}.

Using eq. (\ref{2ptschur}), we can compute the diagonal elements of the metric given by
\begin{equation}
  \braket{R_i}{R_i} = \corr{R_i^{\dagger \prime}(z'=0)R_i(z=0)}_{S^2}
\end{equation}
Changing the coordinate of $R_i^{\dagger \prime}$ to $z$ using $z'z =
1$, and remembering that it is a primary field, we get
\begin{align}
  \braket{R_i}{R_i} & = \lim_{z_0 \to \infty} \corr{(-z_0^2)^{\Delta_i}
      R_i^{\dagger}(z= z_0)R_i(z=0)} \nn \\
  & = \lim_{z_0 \to \infty}\left[
    \frac{(-z_0^2)^{\Delta_i} (-1)^{\Delta_i}f_{R_i}}{(0-z_0)^{2\Delta_i}}
  \right] \nn \\
  & = f_{R_i}
\end{align}

\subsubsection{Three point function calculations}

We want to work out
\begin{equation}
  \corr{R^\dagger (z= e^{2\pi s}) R_1^{\prime}(z'=0)
    R_2(z=0)}_{S^2}
\end{equation}
which we will do by changing the coordinate of $R_1^{\dagger \prime}$
to $z$ via $zz'=1$, and using the general formula of \cite{cjr}
\begin{equation}
\corr{R^\dagger(z)R_1(z_1)R_2(z_2)}=g(R_1,R_2;R)f_R{(-1)^{\Delta_1+\Delta_2} \over (z_1-z)^{2\Delta_1}(z_2-z)^{2\Delta_2}}
\end{equation}
We get
\begin{align}
  & \corr{R^\dagger (z= e^{2\pi s}) R_1^{\prime}(z'=0)
    R_2(z=0)} \nn \\
 = & \lim_{z_0 \to \infty}\corr{R^\dagger (z= e^{2\pi s})
      (-z_0^2)^{\Delta_1} R_1(z = z_0)
      R_2(z=0)} \nn \\
  = &  \lim_{z_0 \to \infty}\left[ \frac{ (-z_0^2)^{\Delta_1} (-1)^{\Delta_1+ \Delta_2}g(R_1,R_2;R) f_R
      }{(e^{2\pi s}-z_0)^{2\Delta_1}(e^{2\pi
        s}-0)^{2\Delta_2}} \right] \nn \\
  = & g(R_1,R_2;R) f_R(-1)^{\Delta_2} e^{-4\pi s\Delta_2}
\end{align}
Here, $g(R_1,R_2;R)$ is the group theoretic LR coefficient associated with the three representations $R_1$, $R_2$ and $R$
of $U(N)$.
For the other 3-point correlator we obtain the same result:
\begin{align}
  \corr{R_2^\dagger(w=0) R_1^{\dagger\prime}(w'=0)  R (w=
      e^{2\pi s})}_{S^2} = g(R_1,R_2;R) f_R(-1)^{\Delta_2} e^{-4\pi s\Delta_2}
\end{align}

\subsubsection{The torus two point function}

The torus Green's function in complex $x$ coordinates, such that $x
\sim x+1 \sim x+\tau$, is given by
\begin{equation}
  G'(x,\ov x;y, \ov y) =-\log  \left|\theta_1\left(x-y ; \tau \right) \right|^2 + \frac{2\pi}{\tau_2}[\text{Im}(x-y)]^2
\end{equation}
where $\theta_1$ is a theta function\footnote{$\a'$ has been set equal to $2$ in the 
corresponding formula of \cite{Polchinski}.}.  For a single complex field, this implies
\begin{align}
  Z_{T^2}^{-1}\corr{\d_x Z^\dagger(x) \d_{y} Z(y)}_{T^2} & = -
  Z_{T^2}^{-1}\d_x^2\corr{Z^\dagger(x) Z(y)}_{T^2} \nn \\
  & =  \d_x^2 \left(\log
    \vartheta_{11}\left(x-y;\tau\right)\right) -
  \frac{2\pi}{\tau_2} \nn \\
  & = -  \wp\left(x-y;\tau\right)
\label{weierstrasspcorr}
\end{align}
where $\wp$ is the Weierstrass elliptic function. The factor of $Z_{T^2}$, the torus partition function, 
appears because the Weierstrass function  is 
the {\it normalized} correlator. Notice that factorization produces the un-normalized torus path integrals.   
Transforming to $z=e^{-2\pi i x}$ and $w=e^{2\pi i y}$ coordinates, 
we also have that 
\begin{equation}
  \corr{\d_z Z^\dagger(z) \d_w Z(w)}_{T^2}  = \frac{1}{(-2\pi i z)}\frac{1}{(2\pi i w)}\corr{\d_x Z^\dagger(x) \d_{y} Z(y)}_{T^2}
\end{equation}

\begin{figure}[t]
\begin{center}
\includegraphics{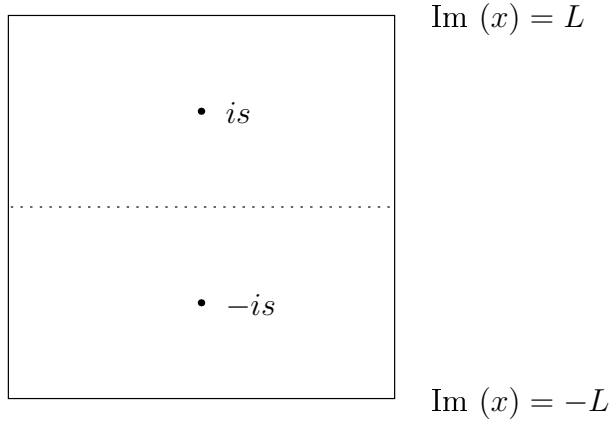}
\caption{The torus correlator obtained after gluing } \label{fig:2dgen1f3}
\end{center}
\end{figure}

We are interested in the two point function for which the operators are inserted at
$x=is$ ($z = e^{2\pi s}$) and $y = -is$ ($w = e^{2\pi s}$), with
$\tau = 2i{L}$--see Figure \ref{fig:2dgen1f3}.
So we obtain that
\bea
  && Z_{T^2}^{-1}  \corr{\d_z Z^\dagger(z=e^{2\pi s}) \d_w Z(w=e^{2\pi
      s})}_{T^2}   = \frac{e^{-4\pi s}}{(2\pi)^{2}}Z_{T^2}^{-1}
{ \corr{\d_x Z^\dagger(x=is) \d_{y} Z(y=-is)}_{T^2}  } \cr 
 && =- \frac{1}{(2\pi)^{2}}e^{-4\pi s}
  \wp\left(2is;2i{L} \right)   \equiv \Gamma( is , -is ) 
\label{propagatortorusZ}
\eea
where we introduced the notation $\Gamma(is,-is)$ for brevity.
The function $-\wp\left(2is;2i{L} \right)$ is positive
for all real values of $s$, as expected from the property of
 reflection positivity.

We can check that (\ref{propagatortorusZ}) leads to the correct pole structure in the limit of small $s$.
Using the expansion of the Weierstrass elliptic function
(\ref{weierstrasspcorr}) for small $x-y=2is$ and $\tau =
2i{L}$, we obtain
\begin{align}
   \Gamma(is,-is)    & =  - \frac{1}{(2\pi)^{2}}e^{-4\pi s}
  \wp\left(2is;2i{L} \right) \nn \\
  & = - \frac{1}{(2\pi)^{2}}e^{-4\pi s}\left[\frac{1}{-4s^2}   +
    \sum_{m,n \in \Z : m,n \neq 0} \left\{ \frac{1}{(2is+n+2mi{L})^2}-
      \frac{1}{(n+2mi{L})^2} \right\} \right] \nn \\
  & \simeq   \frac{1}{16\pi^2s^2}
  \label{inequalityapproximation}
\end{align}
The same result can be obtained from the sphere 2-point function, eq. (\ref{propagatorZ}), for $z_1=e^{-2\pi i s}$, 
$z_2=e^{2\pi i s}$ in the limit $s \to 0$. The pole structure is dictated by the operator product expansion.

Now we can compute the torus 2-point function of the Schur polynomial $R$. The color combinatorics are the same as in the
sphere-case. Each field contraction gives a factor of $\Gamma(is,-is)$. The number of such contractions is set by the (integer)
conformal dimension of $R$. So we obtain
\begin{equation}
  \corr{R^\dagger(z=e^{2\pi s}) R(w=e^{2\pi s})}_{T^2}  = Z_{T^2}\Gamma(is,-is)^{\Delta_R} f_R
\end{equation}

\subsubsection{The inequality}

If we insert into  (\ref{eq:torusequationfirst})
 all the elements, the inequality becomes
\begin{equation}
Z_{T^2} \Gamma(is,-is)^{\Delta_R} f_R >   ( q\bar q )^{-c/24} \sum_{R_1, R_2}
 e^{-4\pi {L}\Delta_1} \frac{e^{-8\pi s\Delta_2 }g(R_1,R_2;R)^2
  f_R^2}{f_{R_1}f_{R_2}}
\end{equation}
In terms of the Weierstrass elliptic function, we
obtain
\begin{equation}
  \left(- \frac{1}{(2\pi)^{2}} \wp\left(2is;2i{L} \right)
  \right)^{\Delta_R} >   \frac{( q\bar q )^{-c/24}}{Z_{T^2}}  
\sum_{R_1,R_2} e^{-4\pi{L}\Delta_1+ 4\pi s (\Delta_1-\Delta_2) } 
  \frac{g(R_1,R_2;R)^2
    f_R}{f_{R_1}f_{R_2}} \label{eq:toreqsec}
\end{equation}
In the strict large $L$ limit, the factor  
$\frac{( q\bar q )^{-c/24}}{Z_{T^2}} $ is equal to one because only 
the state corresponding to the unit operator contributes to the partition sum \footnote{We can gap non-zero momentum and winding states
by considering a compact version of the CFT.}. 
At finite $L$ it is smaller than one and makes the 
inequality easier to satisfy. 
 If we embed the Matrix CFT in a supersymmetric 
theory and use periodic boundary conditions for the fermions in the gluing process, this
 extra factor is exactly one (assuming that the ground state is unique). Hence in general we expect the stronger inequality
  \bea 
 \left(- \frac{1}{(2\pi)^{2}} \wp\left(2is;2i{L} \right)
  \right)^{\Delta_R} >  
\sum_{R_1,R_2} e^{-4\pi{L}\Delta_1+ 4\pi s (\Delta_1-\Delta_2) } 
  \frac{g(R_1,R_2;R)^2
    f_R}{f_{R_1}f_{R_2}}\label{eq:torusequationsecond}
\eea
to be valid, although our main interest is at large $L$.

\begin{figure}[t]
\begin{center}
\includegraphics{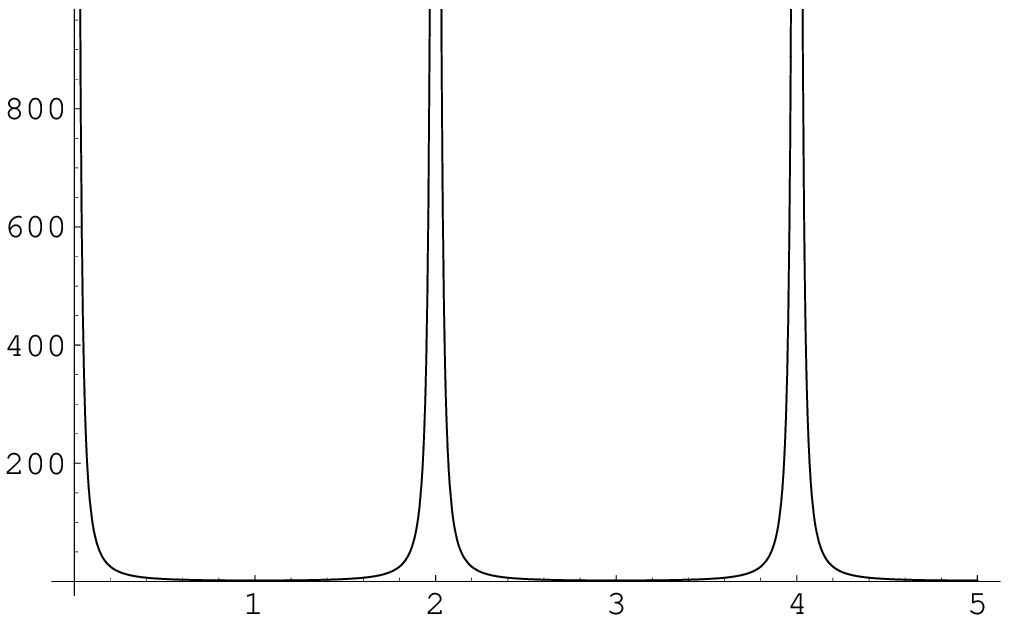}
\caption{A plot of $- \wp\left(2is;2i{L} \right)$ against $s$ with $L =
1$} \label{fig:weierstrass}
\end{center}
\end{figure}

We can now do various checks of the inequality
(\ref{eq:torusequationsecond}). If we further
 restrict to intermediate operators
for which $\Delta_1 =\Delta_2 = \Delta$ ($\Delta=\Delta_R/2$), we get
\begin{equation}
\left(- \frac{1}{(2\pi)^{2}} \wp\left(2is;2i{L} \right)
\right)^{2\Delta} >      
     e^{-4\pi \Delta {L}} \sum_{R_1, R_2}
\frac{g(R_1,R_2;R)^2
  f_R}{f_{R_1}f_{R_2}}
\end{equation}
where the RHS is now a constant as a function of $s$.  
The function $-\wp\left(2is;2i{L} \right)$ is a real, positive function of $s$ which is periodic. 
The period is $L$ with respect to $s$ (or
$2L$ with respect to the separation of the two operator insertions, $2s$).
In each period
this function reaches a
minimum at the middle of its period.  Thus the LHS of the inequality reaches a
minimum at $s={L/2}$ (see Figure \ref{fig:weierstrass}).  At this point
we have
\begin{align}
  -\wp(i{L};2i{L})& =\pi^2\left(4 \sum_{n> 0}\left\{ \textrm{coth}(2n\pi {L})\textrm{cosech}(2n\pi
   {L}) \right\} +\frac{1}{3}\right)
\end{align}
For this and other limits of the Weierstrass elliptic function see
Appendix \ref{appweier}.

This means that the inequality will be true for all $s$ provided that
\begin{equation}
\label{gibbongibbon}
\left( \sum_{n> 0}\left\{ \textrm{coth}(2n\pi {L})\textrm{cosech}(2n\pi
   {L}) \right\}  +\frac{1}{12}
\right)^{2\Delta} >e^{-4\pi \Delta {L}} \sum_{R_1, R_2}
\frac{g(R_1,R_2;R)^2
  f_R}{f_{R_1}f_{R_2}} 
\end{equation}
We can see how this is satisfied for all values of $L$.  For large ${L}$, the LHS tends to
$(1/12)^{2\Delta}$ which will be much bigger than $e^{-4\pi \Delta
  {L}}$.  For small ${L}$ the hyperbolic functions $\textrm{coth}(2n\pi {L})$ and
$\textrm{cosech}(2n\pi {L})$ both blow up.

\begin{figure}[t]
\begin{center}
\includegraphics{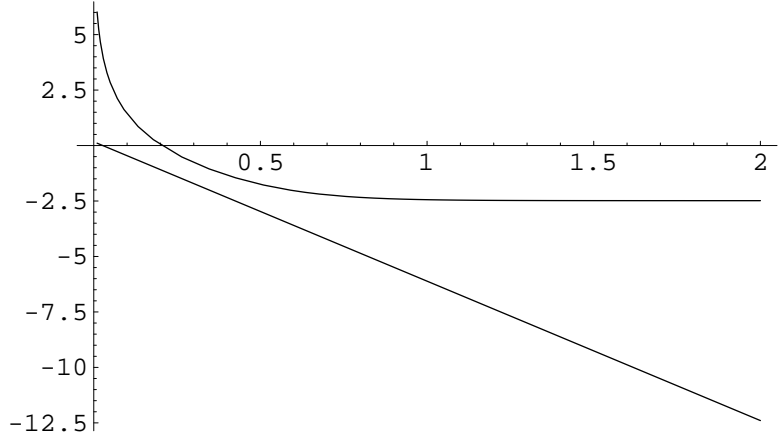}
\caption{A plot of of the logarithms of the LHS of
  (\ref{gibbongibbon}) (top) against (\ref{goatgaot2})
  (bottom) against $L$ for our chosen representations.  We have in
  fact taken the $N$th root of each side.  We can ignore the
  $3/\sqrt{8}$ factor in (\ref{goatgaot2}) because it adds a small constant to
  the lower graph which does not affect the inequality for any value
  of $N$.}
\label{fig:2dinequality}
\end{center}
\end{figure}

We will now check our result for the transition from a size $N$ $AdS$
giant, a single-row representation written $R=[N]$, to two smaller
$AdS$ giants, $R_1, R_2 = [N/2]$, so that $\Delta=N/2$.  Then
the RHS of (\ref{gibbongibbon}) is given by
\begin{align}
  \frac{f_{[N]}}{f_{[N/2]}^2} e^{-2\pi N {L}} & =
  \frac{(2N-1)!(N-1)!}{\left((3N/2-1)!\right)^2} e^{-2\pi N{L}} 
   \nn \\
  & \sim  \frac{3}{\sqrt{8}} \left(\frac{32}{27}\right)^N e^{-2\pi N{L}} 
 ( ~  1 + \cO ( 1/ N ) ~ )  
 \label{goatgaot2}
\end{align}
In Figure \ref{fig:2dinequality}, the LHS of (\ref{gibbongibbon}) is
plotted against (\ref{goatgaot2}), for the specific Schur
polynomials chosen, as a function of ${L}$ to verify that the inequality
holds for all $L$ and $N$.

For $\Delta_1 \ne \Delta_2 $ the $s$-dependence of the RHS of
(\ref{eq:torusequationsecond}) is no longer trivial.  Some numerical
checks of the inequality have been made for this situation.

\subsubsection{Probability interpretation in the large ${L}$, $N$ limits}
We can now obtain a well defined probability for a transition to occur,
 from a state of charge $N$,
created by the operator $\chi_{[N]}(\partial Z)$, to two 
states of charge $N/2$
each created by the operator  $\chi_{[N/2]}(\partial Z)$
 in the large $L$ and $N$ limits.

The large $L$ limit is the appropriate limit.
 The discussion in Section \ref{torusgluing}, of gluing at two punctures, should be related to 
   the limit in moduli space where $ \tau = 2i {L} $ and $ {L} \rightarrow
   \infty $. This can be seen as follows.  Consider a sphere 
  with disks of radii $ e^{-2 \pi r_1 } , e^{ - 2 \pi r_2 }$ removed 
  near the N and S poles. 
 The region from the equator to the first disk 
  is mapped to a cylinder of radius one and length $  r_1   $ using the 
  exponential map $ z = e^{ - 2i  \pi x } $.
  The region from the equator to the second disk is mapped to a cylinder of 
  radius $1$ and length $ r_2  $ by the map $ z' = e^{ - i 2 \pi y } $. 
 Hence we have a cylinder with radius $1$ and length $ r_1 + r_2  $. 
 When this is glued to another similar 
   cylinder of unit radius and size $ r_1 + r_2  $, we get a torus 
  with $ \tau $ parameter $ 2 i ( r_1 + r_2  ) $.  As the disks reach zero
  size in the limit $ r_1 , r_2 \rightarrow \infty $, 
  we get a torus with $ \tau \rightarrow i \infty $. 

Because the LHS of (\ref{eq:torusequationsecond}) approaches a constant for large ${L}$,
the probability
is given by
\begin{equation}\label{examprob2d} 
  P([N] \to [N/2],[N/2]) \sim \frac{3}{\sqrt{8}}\exp \left\{(-2\pi {L} +
    \log(32/27)+\log 12)N\right\}
\end{equation}
which is less than $1$ because $L$ is taken large.
 We have used (\ref{goatgaot2}),
 which computes the contribution to the RHS for this
 particular process. Notice that the factor $L$ governs the spatial separation of the states on the cylinder.

\subsection{Miscellaneous comments}

\subsubsection{Zero coupling gauge theory v/s unconstrained free fields}\label{zeroversusfree}

The two dimensional Matrix CFT we are considering is invariant under global $U(N)$ transformations. 
The symmetry imposes restrictions on the type of internal states that contribute in factorization relations.

Suppose that the external states in a factorization equation 
are invariant under global $U(N)$ transformations:
\begin{equation}
Q_{a}|A\rangle = 0, \qquad Q_{a}|B\rangle = 0
\end{equation}
where the operators $Q_{a}$ denote the generators of $U(N)$ transformations. These satisfy $[Q_a, Q_b]=if_{abc}Q_c$,
with $f_{abc}$ the $U(N)$ Lie algebra structure constants. 
Consider the overlap
\begin{equation}
\langle B|U(t_1,t_2)|A\rangle 
\end{equation}
where $U(t_1,t_2)$ is the time (radial) evolution operator. The
Hamiltonian (dilatation operator) is invariant under global $U(N)$
transformations so that $[ Q_{a},U]=0$.  Now we insert a complete set
of orthonormal states at time $t'$ (or radius $r'$), which can be
taken to be eigenstates of the mutually commuting Cartan generators,
to obtain a factorization relation
\begin{equation}
\langle B|U(t_1,t_2)|A\rangle = \sum_C \langle B|U(t_1,t')|C\rangle\langle C|U(t',t_2)|A\rangle 
\end{equation}
Then the intermediate states $|C\rangle$ must also be invariant under
global $U(N)$ transformations.  We can argue for this as follows:
\begin{eqnarray}
& \langle B|U(t_1,t')|C\rangle 
= \langle B|\left(e^{i\alpha_{a}Q_{a}}U(t_1,t')e^{-i\alpha_{a}Q_{a}}\right) |C\rangle \nn \\
& = \left( \langle B|e^{i\alpha_{a}Q_{a}}\right) U(t_1,t')\left( e^{-i\alpha_{a}Q_{a}}|C\rangle\right)
= \langle B|U(t_1,t')|C\rangle e^{-i\alpha_{a}q_{a}}
\end{eqnarray}
where the vector $q_{a}$ denotes the charges of the state $|C\rangle$.
Since the above must hold for arbitrary $\alpha_{a}$, it is clear that
only intermediate states with $q_{a}=0$ contribute. Therefore, the
operator corresponding to $|C\rangle$ must also be invariant. So only $U(N)$ 
 invariant operators contribute in the genus zero factorization. 
If we consider a factorization equation of a genus-1 correlator such
as eq. (\ref{gonefac}), then the symmetry implies that the net charge
of the internal operators $B_1$ and $B_2$ contributing must add
to zero, assuming that the external operators are invariant.

If a theory has a local gauge symmetry, such as the four dimensional
$\cN=4$ Super Yang Mills theory we are interested in, then there are
further constraints on the types of internal operators contributing in
factorization equations.  Consider for example correlators of local
gauge invariant operators of the $\cN=4$ theory on the $S^3\times S^1$
manifold.  
Suppose we factorize such higher genus correlators in terms of correlators of local operators on
$S^4$.  Invariance of
the theory under local gauge transformations implies that in order for
the internal local operators and to contribute, each must
be a local gauge invariant operator.

The considerations above address the following puzzle. Consider the
$\cN=4$ Super Yang Mills theory on $S^3\times S^1$ in the limit of
vanishing coupling constant. Even in this limit, the zero mode of
$A_0$ on the sphere does not decouple from matter fields, and because
of the non-trivial topology of the manifold, it cannot be gauged away. 
The relevant gauge invariant quantity is 
the Wilson line of $A_0$ around the $S^1$ circle.
As a consequence, thermal two point functions of local gauge invariant
operators composed of adjoint scalars are different from the two point
functions of the same operators in a theory with unconstrained, free
scalar fields only--see \cite{Brigante:2005bq}. This difference is
also reflected in factorization equations in terms of local operators:
in the zero coupling gauge theory, each of the internal local
operators must be a local gauge invariant operator to contribute,
while in the unconstrained free scalar theory only the net charge of
the operators has to vanish. 
We expect that correlators of local gauge invariant operators 
in the two theories 
should agree in the limit of large radius for the 
$S^1$ circle.  When the circle becomes uncompact, $A_0$ can be gauged away.
This was discussed, in the case of large $N$ in \cite{Brigante:2005bq}. 
In addition, it was argued in \cite{Brigante:2005bq} that to leading order
 in $1/N$, non-renormalization theorems 
protecting the two point and three
point functions of $1/2$ BPS operators against 't Hooft coupling 
corrections survive in the low temperature phase of the theory.
The large $N$ and large radius limits are the relevant limits for our 
computations in section \ref{spherecircle}.

\subsubsection{Windings from torus factorization sums}

If our intermediate states $ \cA_{i}'$ and $\cA_j$ appearing in the
factorization equation differ in their holomorphicity, then we can
interpret some of the summands in the torus factorization as paths
winding around a non-trivial cycle in the torus.  For example, when
$\cA_i' (z',\bar{z}'=0)= \partial Z^{\dagger\prime}(z',\bar{z}'=0)$
and $\cA_k(z,\bar{z}=0)=: \partial Z(z,\bar{z}=0)\partial
Z(z,\bar{z}=0):$, we obtain a path with winding number 1.  See the
Appendix section \ref{sec:windings} for a detailed discussion.

\section{Factorization in the 4D CFT}\label{sec:fact4d}

\subsection{Introduction}

The factorization arguments described in the
 previous section extend naturally to conformal field theories in
four dimensions.
To obtain sphere factorization identities, we glue together
 two $S^4$s around one puncture
to produce a single $S^4$.  To obtain genus-1 factorization identities,
 we glue together
two $S^4$s at two punctures to get a genus-1 surface which is
conformally  equivalent to the $S^1 \times S^3$ manifold.
The argument for the factorization of the correlation functions in the
$3+1$-dimensional CFT follows from the path integral discussion 
in Section \ref{opstatecorr}.  In the sum over intermediate states
 we keep only the Schur polynomials in a single complex scalar
$\Phi$.

\subsection{Metric}\label{sec:metric}

In order to define a positive metric on the space of operators, we choose the
scalar 2-point function on $\R^4$ to satisfy the convention
\begin{equation}
  \Delta_x G(x-y) = - \delta^4(x-y) \label{s4correlator}
\end{equation}
This gives 
\begin{equation}
  G(x-y) = \frac{1}{4\pi^2 |x-y|^2}
\end{equation}

The metric on the space of Schur polynomials is given by
\begin{equation}
  \corr{R^{\dagger \prime}(r'=0) R(r=0)}
\end{equation}
where $r' = 1/r$.  To compute the correlator, we map $R^{\dagger \prime}$ back to the $r$-coordinate frame. 
Under the coordinate transformation $r' \to r
= 1/r'$, the metric changes as follows
\begin{equation}
  dr^{\prime 2} + r^{\prime 2} d\O^2 \to \frac{1}{r^{4}}(dr^{2} + r^{ 2} d\O^2)
\end{equation}
and so the primary fields transform as 
\begin{equation}
  \Phi^{\prime}(x') \to \O(x)^{-\Delta/2}\Phi(x) =   r^{2\Delta}\Phi(x)
\end{equation}
where $\O(x)=1/r^4$ is the conformal factor \cite{Ginsparg:1988ui}. Thus for the metric element we obtain
\begin{align}
  \corr{R^{\dagger \prime}(r'=0) R(r=0)} & = \lim_{r_0\to \infty}
  \corr{r_0^{2\Delta}R^\dagger(r=r_0) R(r=0)} \nn \\
  & = \left(\frac{1}{4\pi^2}\right)^\Delta f_R
\end{align}

\subsection{The genus zero factorization in four dimensions}\label{sphere4}

Following the two dimensional example, we start with two 4-spheres, one with
coordinates $(r,\O_i)$ and the other with coordinates $(s,\O_i')$. Next we cut out a 4-ball
of unit radius around the origin in each, and glue them together using
$rs=1$.  The factorization identity implies an inequality given by
\begin{align}
  & \big\langle R_1^\dagger(s=e^{x_1}) \cdots R_k^\dagger(s=e^{x_k}) R_k(r=e^{x_k})\cdots  R_1(r=e^{x_1})  \Big\rangle \nn \\
  & > \sum_{R} \frac{\big\langle R_1^\dagger(s=e^{x_1} ) \cdots R_k^\dagger(s=e^{x_k}) R(r=0) \big\rangle \big\langle  R^\dagger(s=0) R_k(r=e^{x_k})\cdots  R_1(r=e^{x_1})   \big\rangle}{\big\langle
    R^\dagger R\big\rangle} \label{spherefactin4d}
\end{align}
where we set $x_j > 0$ for $j=1, \dots, k$ so that the operator insertions are outside
the cut-off region.  We have suppressed the angular coordinates of the
operators $R_j$ in \eqref{spherefactin4d}, but these can be arbitrary in general.

In general the correlator on the LHS of the inequality is not extremal, so it may
have a non-trivial dependence on the 't Hooft coupling constant. To avoid this
complication, we do our calculations in the limit that the correlator becomes
extremal, i.e. when $x_1 = x_2 = \cdots = x_k$.

In the large separations limit, $x_j \to \infty$, we recover the combinatorial
factorization identities discussed in \cite{cr}.

\subsection{The genus one  factorization in four dimensions} 

We parameterize four dimensional flat space $\R^4$ with spherical coordinates so that the metric is given by
\bea 
ds^2 = dr^2 + r^2 d\Omega_3^2 
\eea 
This metric is conformal to the standard metric on $S^3 \times \R $ under the coordinate transformation $ r = e^{ \tau } $:
\bea 
ds^2  = e^{ 2 \tau } ( d \tau^2 + d \Omega_3^2 )
\eea 

In the two dimensional example, we started with two copies of $ S^1 \times I  $ 
described by coordinates $ 1 \le  |z| \le e^{ 2 \pi L } $ and
 $ 1 \le  |w| \le e^{ 2 \pi L } $.  In the four dimensional case, we start with 
two cylinders $S^3 \times I $ described by coordinates $ ( r , \Omega_i )$ 
and $  ( s , \Omega^{\prime}_i ) $ with the radial variables in the range 
\bea 
&&   1 \le r \le e^T  \cr 
&& 1 \le s \le e^T 
\eea 
In most of the following expressions, we suppress the angular dependence since the angles, 
 in all of the gluings, are identified trivially. 

Introduce also the coordinates $r' = 1/r$ and $s'=1/s$.  We now glue the two
cylinders $S^3 \times I $ at the inner ends $r=1$, $s=1$ with $rs=1$.
We then glue the outer ends at $r=e^T$, $s=e^T$ with $r's' = e^{-2T}$
(i.e. $rs=e^{2T}$). The gluing produces an $S^3 \times S^1$ manifold with $\tau \sim \tau + 2T$ \footnote{  
In our notation, $2T$ stands for the inverse temperature with regards to the thermal theory on $S^3\times S^1$. We 
hope that the notation does not cause confusion to the reader.}.

\subsection{The genus one factorization and inequality }

The derivation of factorization of correlators 
on genus-$1$ surfaces in two dimensions uses basic features of CFT, such as 
the operator-state correspondence and properties of the path integral representation of 
correlators. The same steps can be run through in four dimensions. 
Now we are looking at correlators on $\Sigma_4 ( G=1) $, which is 
obtained by gluing two copies of $ S^3 \times I$, each obtained by cutting out 
the neighborhoods of two points in an $S^4$ manifold. 
We obtain 
\bea\label{geomfac4}  
&& \langle R^{\dagger}  ( P_1) R (  P_2   ) \rangle_{G=1} \cr
&&= 
 \sum_{ i , j  } {  \langle R^{\dagger} ( P_1 ) \cA_i^{\dagger}  ( C_2^L  )  \cA_{k} ( C_1^L )  \rangle 
                 \langle  \cA_k^{\dagger} ( C_1^R )   \cA_i ( C_2^R )  R ( P_2)  \rangle 
\over \langle \cA_{i}^{\dagger}   ( C_2^L  )  \cA_{i} ( C_2^R ) \rangle \langle  \cA_{k}^{\dagger}   ( C_1^L  )  \cA_{k} ( C_1^R )  \rangle
} \nnm \\
\eea 
The surfaces $ C_i^L $ and $C_i^{R}$ are now 3-spheres. 
Eq. (\ref{geomfac4}) is the 4d analog of eq. (\ref{geomfact}).   
By scaling, we can express the RHS in terms of correlators of local operators 
on $\R^4$ 
\bea\label{faclocal4}  
&&  \langle R^{\dagger} ( r = e^x ,  \Omega_i ) R ( s = e^x  , \Omega_i ) \rangle \cr 
&& = 
Z_0 \sum_{i , j }   e^{- 2 T  \Delta_i   } 
{  \langle R^{\dagger} ( r = e^x , \Omega_i ) \cA_i^{\dagger \prime } ( r' =  0 )    \cA_{k}   ( r  = 0 ) \rangle 
 \langle   \cA_k^{\dagger  } ( s = 0 ) \cA_i ^\prime ( s' = 0 )  R ( s = e^x , \Omega_i )   \rangle    
\over \langle i | i \rangle \langle k | k \rangle  } \nnm \\ 
\eea 
This is the 4d analog of eq. (\ref{finalfact0}). 
$Z_0$ is the large $T$ limit of the Euclidean partition function
on $S^3 \times S^1$, analogous to the $ ( q\bar q )^{-c/24} $ term in
two dimensions. It depends only on the Casimir energy of the ground state. We will not need it explicitly. 
In going from a path integral expression to an 
operator expression, we must specify a   
time-ordering. We specialize to the case where $ P_2 $
 and $P_1$ are related by Euclidean 
time reversal so that we can expect positivity of the RHS of the equations above. 
We will further restrict the sum to the case where  
 $ \cA_{i}^{\dagger } $ and $ \cA_k $ are given
respectively by the Schur Polynomials $ \chi_{R_1} ( \Phi )  $ and $ \chi_{R_2} ( \Phi ) $. 
 By checking 
the resulting inequality, we will obtain well-behaved probabilities.

We want to demonstrate the inequality
\begin{align}
   & \langle R^{\dagger}  (s=e^x, \Omega_i )
 R (  r=e^x  , \Omega_i  )  \rangle_{G=1}\nn \\
 & > Z_0 \sum_{R_1,R_2}e^{- 2 T  \Delta_1   } 
 \frac{\langle R^{\dagger} ( r = e^x , \Omega_i ) R_1^{\prime} ( r'  = 0 )
   R_2  ( r =0 ) \rangle \,
\langle  R_2^{ \dagger } ( s =0 ) R_1^{\dagger \prime } ( s' = 0 )
R ( s = e^{x} , \Omega_i   ) \rangle}{\langle R_1^\dagger R_1\rangle \,\langle R_2^\dagger R_2\rangle} \label{eq:4dinequality45}
\end{align}

We work out the first three-point function to get 
\begin{align}
& \langle R^{\dagger} ( r = e^x , \Omega_i ) R_1^\prime ( r'  = 0 )
   R_2  ( r =0 ) \rangle \nn \\
& = \lim_{r_0 \to \infty} \langle R^{\dagger  } ( r = e^x , \Omega_i ) r_0^{2\Delta_1} R_1 ( r = r_0 )
   R_2  ( r =0 ) \rangle \nn \\
& = (4\pi^2)^{-\Delta_1 - \Delta_2} e^{-2x\Delta_2}g(R_1,R_2;R) f_R
\end{align}
Similarly for the second correlator we get
\begin{align}
\langle  R_2^{ \dagger } ( s =0 ) R_1^{\dagger \prime } ( s' = 0 )
R ( s = e^{x} , \Omega_i   ) \rangle =  (4\pi^2)^{-\Delta_1 - \Delta_2} e^{-2x\Delta_2}g(R_1,R_2;R) f_R
\end{align}
Hence the right-hand side of the inequality \eqref{eq:4dinequality45} becomes
\bea\label{prodthree}  
\sum_{R_1,R_2}  (4\pi^2)^{-\Delta_1 - \Delta_2} { g ( R_1 , R_2 ; R )^2 f^2_R 
\over f_{R_1} f_{R_2}} e^{ - 2 T \Delta_1 } e^{ - 4 x \Delta_2 }
\eea 
Because of charge conservation, the only terms contributing to the RHS are those for which $\Delta_1+\Delta_2=\Delta_R$,
where $\Delta_R$ is the conformal dimension of the Schur operator $R$.

\subsection{The correlator on $S^3 \times S^1$}\label{spherecircle}

Let the metric on $S^3 \times S^1$ be given by 
\begin{equation}
  ds^2 = d\tau^2 + d\chi^2 + \sin^2 \chi(d\theta^2 + \sin^2 \theta d\phi^2)
\end{equation}
where $\tau \in [0,2T]$, $\chi,\theta \in [0,\pi]$ and $\phi \in
[0,2\pi]$.

If the differential operator $K$ admits a complete set of eigenvectors
$\Psi_n(x)$ with $K\Psi_n = \l_n \Psi_n$, then the corresponding Green's
function is given by
\begin{equation}
  G(x,y) = \sum_{n| \l_n \neq 0}\frac{\Psi_n^*(x)\Psi_n(y)}{\l_n}
\end{equation}
and it satisfies
\begin{align}
  K G(x,y) & = \sum_{n| \l_n \neq 0}\Psi_n^*(x)\Psi_n(y)\nn \\
  & =  \delta(x-y) - \sum_{n| \l_n = 0}\Psi_n^*(x)\Psi_n(y) \label{greensat}
\end{align}

For a conformally coupled scalar field in four dimensions, the differential operator $K$
is given by 
\begin{equation}
  K = \Delta - \frac{1}{6}R \label{conformalcouple}
\end{equation}
where $\Delta$ is the Euclidean Laplacian and the second term is the
coupling to the 4-dimensional curvature \cite{Birrell:1982ix}.  It is like a mass term and has the same
sign as a positive mass term in a Euclidean theory.  For $S^1
\times S^3$ with unit radii, only the curvature of
$S^3$ contributes, giving for the Ricci scalar curvature $R=6$.
Thus $K = \Delta - 1$.

On $S^3$ the spherical harmonics are given by \cite{Birrell:1982ix}
\begin{equation}
  \cY_{\mathbf{k}}(\O_i) = \Pi_{kJ}(\chi) Y^M_J(\theta,\phi)
\end{equation}
where $\mathbf{k} = (k,J,M)$, $Y^M_J$ are spherical harmonics on
$S^2$ and $\Pi_{kJ}$ is given by
\begin{equation}
  \Pi_{kJ} = \left[\ha \pi k^2 (k^2-1) \cdots  (k^2-J^2)
  \right]^{-1/2} \sin^J\chi\left( \frac{d}{d\cos\chi}\right)^{1+J} \cos
  k\chi
  \label{PIfunction}
\end{equation}
The quantum numbers $k$, $J$ and $M$ lie in the following ranges
\begin{align}
  k & = 1,2, \dots, \nn \\
  J & = 0,1, \dots, k-1 \nn \\
  M & = -J, -J +1, \dots, J
  \label{kJMrange}
\end{align}

The harmonics $\cY_{\mathbf{k}}(\O_i)$ satisfy 
\begin{equation}
  \Delta_{S^3} \cY_{\mathbf{k}}(\O_i) = -(k^2-1)\cY_{\mathbf{k}}(\O_i)
\end{equation}
and they are orthonormal.
Spherical harmonics on $S^1$ are given by
\begin{equation}
  h_m(\tau) = \cN e^{im\pi\tau/T}
\end{equation}
where $\cN = (2T)^{-\ha}$ is the normalization factor.  They satisfy
\begin{equation}
  \Delta_{S^1} h_m = -\left(\frac{m\pi}{T}\right)^2h_m
\end{equation}
Thus if 
\begin{equation}
  \Psi_n = h_m(\tau)\cY_{\mathbf{k}}(\O_i)
\end{equation}
where $n=(m,\mathbf{k})$, then
\begin{equation}
  \Delta_{S^3 \times S^1}\Psi_n = (\Delta_{S^3} + \Delta_{S^1})\Psi_n
  = \left[-(k^2-1)  -\left(\frac{m\pi}{T}\right)^2 \right]\Psi_n
\end{equation}

If we add the conformal coupling term as in (\ref{conformalcouple}), we
get
\begin{equation}
  K\Psi_n = \left(\Delta_{S^3 \times S^1}-1\right)\Psi_n = \left[-k^2 -\left(\frac{m\pi}{T}\right)^2 \right]\Psi_n
\end{equation}
This eigenvalue problem has no zero-mode solution.  In accordance 
with the $\R^4$ correlator
(\ref{s4correlator}), we actually choose the Green's function to satisfy
\begin{equation}
  K G(x,y) = - \delta^4(x-y)
\end{equation}
so that we get a positive metric on the space of operators. So the desired Green's
function is given by
\begin{align}
  G(x,y) & = - \sum_{n}\frac{\Psi_n(x)^* \Psi_n (y)}{\l_n} \nn \\
  & = \sum_{m,k,J,M}
  \frac{h_m(\tau)^*  \cY_{\mathbf{k}}^* (\O_i) h_m (\tau')\cY_{\mathbf{k}}
 (\O_i')}{k^2 + \left(\frac{m\pi}{T}\right)^2}
  \label{s1s3res}
\end{align}
where $k,J$ and $M$ are in the ranges set out in (\ref{kJMrange}) and
$m$ is an integer.

We want to work out
\begin{equation}
  \corr{R^\dagger(s=e^x) R (r=e^x)}_{G=1}
\end{equation}
where the angular coordinates are fixed to coincide.

If we change coordinates to $s = e^{-\tau}$, $r = e^{\tau}$, we get 
\begin{align}
  \corr{\Phi^\dagger(s=e^x) \Phi (r=e^x)}_{G=1} & = \frac{1}{rs}\corr{\Phi^\dagger(\tau=-x) \Phi (\tau=x)}_{G=1} \nn \\
  & = e^{-2x}\corr{\Phi^\dagger(\tau=-x) \Phi (\tau=x)}_{G=1}
\end{align}
Now insert the Green's function (\ref{s1s3res}) to get
\begin{equation}
Z_{G=1}^{-1} \corr{\Phi^\dagger(\tau=-x) \Phi (\tau=x)}_{G=1}  = 
 \sum_{m,k,J,M }
\frac{h_m(0)^* \cY_{\mathbf{k}}^* (\O_i) h_m(2x)\cY_{\mathbf{k}}(\O_i)}{k^2
  +\left(\frac{m\pi}{T}\right)^2}
\end{equation}
where we put each $S^3$ spherical harmonic at the same point on the
$S^3$ and $Z_{G=1}$ is the thermal partition function.   A clever choice of the angular point simplifies the sum. 
  Let that point be where $\chi = 0$ so that $\Pi_{kJ}$ is
zero for $J > 0$, since the term $\sin^J \chi$ at the front of the expression is
zero ($\cos k\chi$ is a polynomial in $\cos \chi$ so for $\chi=0$
the derivatives of $\cos k\chi$ give a constant).  Then the only terms
that contribute are those with $J=M=0$.  We get
\begin{align}
 \Pi_{k0} & = \left[\ha \pi k^2\right]^{-1/2} \frac{d}{d\cos\chi} \cos
  k\chi \nn \Big|_{\chi=0} \nn\\
  & = 2^{1/2}\pi^{-1/2} k
\end{align}
Then noting that $Y^0_0(\theta,\phi) = 2^{-1} (\pi)^{-1/2}$, we get
\begin{align}
\Gamma ( -x, x ) & \equiv { \corr{\Phi^\dagger(\tau=-x) \Phi (\tau=x)}_{G=1} \over Z_{G=1} } \cr 
 & =  \sum_{m\in \Z,k\geq 1}
\frac{\cN^2 e^{im2\pi x/T}2^{-1} \pi^{-2} k^2   }{k^2
  +\left(\frac{m\pi}{T}\right)^2} \nn \\
& =   \frac{1}{4\pi^2T}\sum_{m\in \Z,k\geq 1}
\frac{ k^2 e^{im2\pi x/T}   }{k^2 + \left(\frac{m\pi}{T}\right)^2} \nn \\
&  =  \frac{1}{4\pi^2T}\left[ 2 \sum_{m > 0,k\geq 1}
\frac{k^2 \cos(m2\pi x/T)}{k^2
  + \left(\frac{m\pi}{T}\right)^2} + \sum_{k\geq 1}
\frac{k^2}{k^2}\right]
\end{align}
where the second term in the last expression is the $m=0$ term.  When
plotted the truncated sums converge everywhere, except when $x$ is an
integer multiple of $T$.

\subsubsection{The Inequality}\label{sec:theinequality243}
The computations above lead to the spacetime inequality
\begin{align}
 & e^{-2x ( \Delta_1 + \Delta_2 ) } ( \Gamma(-x, x )) ^{\Delta_1 + \Delta_2}
f_R Z_{G=1}  \nn \\
& >  Z_0 
\left(\frac{1}{4\pi^2}\right)^{\Delta_1 + \Delta_2} \sum_{R_1, R_2} { g ( R_1 , R_2 ; R )^2 f_R^2
\over f_{R_1} f_{R_2} } e^{ - 2 T \Delta_1 } e^{ - 4 x \Delta_2 }
\end{align}
or
\begin{align}
  & \left( \frac{1}{4\pi^2T}\left[ 2\sum_{m > 0,k\geq 1}
\frac{k^2 \cos(m2\pi x/T)}{k^2+ \left(\frac{m\pi}{T}\right)^2} + 
\sum_{k\geq  1}  1 \right]\right)^{\Delta_1 + \Delta_2}   \nn \\
& >  { Z_0 \over Z_{G=1} } 
\left(\frac{1}{4\pi^2}\right)^{\Delta_1 + \Delta_2} \sum_{R_1, R_2} { g ( R_1 , R_2 ; R )^2 f_R 
\over f_{R_1} f_{R_2} } e^{ - 2 T \Delta_1+2x(\Delta_1 - \Delta_2)
}
\label{yakyak}
\end{align}
After canceling the $4\pi^2$ constants, we obtain
\begin{align}
   \left( \frac{1}{T}\left[ 2\sum_{m > 0,k\geq 1}
\frac{k^2 \cos(m2\pi x/T)}{k^2+ \left(\frac{m\pi}{T}\right)^2} + 
\sum_{k\geq  1 } 1 \right]\right)^{\Delta_1 + \Delta_2}  
 > { Z_0 \over Z_{G=1} }   \sum_{R_1, R_2} { g ( R_1 , R_2 ; R )^2 f_R 
\over f_{R_1} f_{R_2} } e^{ - 2 T \Delta_1 + 2 x(\Delta_1- \Delta_2)
}
\label{yy2}
\end{align}
This expression is very similar to the $S^1\times S^1$ inequality.  Note
however that the LHS does not have the analog of the $(1/12)$ term of 
eq. (\ref{gibbongibbon}), since the zero mode has been lifted by the conformal 
mass term. 

As in the 2d case, in the large $T$ limit, the factor 
$ { Z_{0} \over Z_{G=1} } $ tends to $1$. For the case of the 
thermal partition function, we have  $ { Z_{0} \over Z_{G=1} }  < 1  $
in general \footnote{For a comprehensive discussion of 
the thermal partition
function of the ${\cal{N}}=4$ Super Yang Mills theory on $S^3$ see \cite{amp}. For supersymmetric partition sums involving
BPS states see \cite{kmm}.}. 
If we perform the gluing with 
periodic boundary conditions for the fermions this factor 
will be $1$ , also just like the 2d case. 
Hence we expect the stronger inequality 
\begin{align}
   \left( \frac{1}{T}\left[ 2\sum_{m > 0,k\geq 1}
\frac{k^2 \cos(m2\pi x/T)}{k^2+ \left(\frac{m\pi}{T}\right)^2} + 
\sum_{k\geq  1 } 1 \right]\right)^{\Delta_1 + \Delta_2}  
 >    \sum_{R_1, R_2} { g ( R_1 , R_2 ; R )^2 f_R 
\over f_{R_1} f_{R_2} } e^{ - 2 T \Delta_1 + 2 x(\Delta_1- \Delta_2)
}
\label{yakyak2}
\end{align}
to hold, although again our main interest in this paper is at large $T$. 

For $\Delta_1 = \Delta_2 = \Delta$ the $x$ dependence of the RHS
vanishes, so it is sufficient to check the inequality at the minimum
of the LHS.  This minimum occurs at $x = \ha T$, i.e. where the points
are at maximum separation on the $S_1$.  At this point, we have
\begin{align}
 \frac{1}{T}\left[ 2\sum_{m > 0,k\geq 1}
\frac{k^2 \cos(m\pi)}{k^2+ \left(\frac{m\pi}{T}\right)^2} + 
\sum_{k\geq 1 } 1 \right]= 
 &\frac{1}{T}\left[2 \sum_{m > 0,k\geq 1} \frac{k^2(-1)^m }{k^2+
    \left(\frac{m\pi}{T}\right)^2} +  \sum_{k\geq 1} 1 
  \right] \nn \\
  = &\frac{1}{T}  \sum_{k\geq 1} \left[\left(-1 + kT
      \textrm{cosech}(kT)\right) + 1 \right]\nn\\
  = & \sum_{k\geq 1} k \textrm{cosech}(kT)
  \label{yummysummy}
\end{align}
The sum above is   convergent.  Thus the inequality becomes
\begin{align}
   \left(\sum_{k\geq 1} k \textrm{cosech}(kT)\right)^{2\Delta}  
 > \sum_{R_1, R_2} { g ( R_1 , R_2 ; R )^2 f_R 
\over f_{R_1} f_{R_2} } e^{ - 2 T \Delta}
\label{yakyak5}
\end{align}

For small $T$ the inequality holds because the RHS is
constant and the sum in the LHS blows up.  For large $T$ we can
approximate the sum (\ref{yummysummy}) by only taking the first term
in the sum and noticing that in this limit
\begin{equation}
  \textrm{cosech}(T) \to 2e^{-T}\label{yummyapprox}
\end{equation}

For $R=[N]$, $\Delta_1 = \Delta_2 = N/2$, $R_1, R_2 = [N/2]$, the RHS of
(\ref{yakyak5}) is given by
\begin{align}
  \frac{f_{[N]}}{f_{[N/2]}^2} e^{-T N} & =
  \frac{(2N-1)!(N-1)!}{\left((3N/2-1)!\right)^2} e^{-T N}
  \nn \\
  & \sim  \frac{3}{\sqrt{8}} \left(\frac{32}{27}\right)^N e^{-T N} \label{goatgaot}
\end{align}
For large $T$ and our choice of $R$ the inequality becomes
\begin{equation}
  2^{N}e^{-NT} > \frac{3}{\sqrt{8}} \left(\frac{32}{27}\right)^N e^{-T N}
\end{equation}
which is satisfied.

\begin{figure}[t]
\begin{center}
\includegraphics{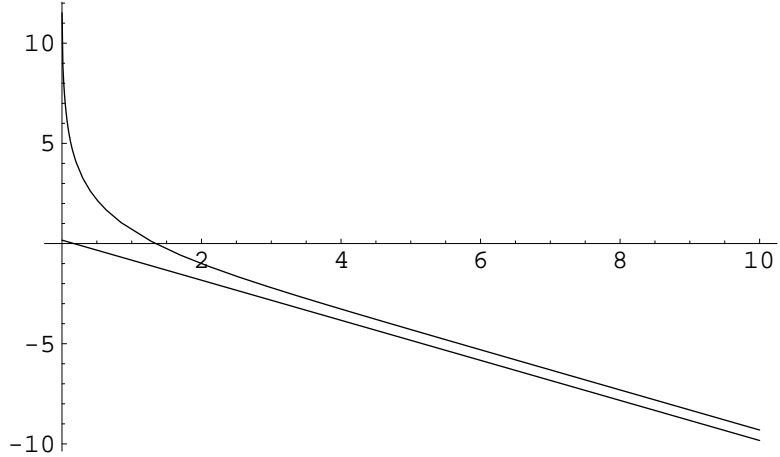}
\caption{A plot of of the logarithms of the LHS of (\ref{yakyak5})
  (top) against the RHS of (\ref{yakyak5}) (bottom) against $T$ for
  our chosen representations.  We have in fact taken the $N$th root of
  each side.  We can ignore the $3/\sqrt{8}$ factor on the RHS because it
  adds a small constant to the lower graph which does not affect the
  inequality for any value of $N$.} \label{fig:4dinequality}
\end{center}
\end{figure}

In Figure \ref{fig:4dinequality}, the LHS of (\ref{yakyak5}) is plotted against the RHS of (\ref{yakyak5}),
for our choice of Schur polynomials, as a function of ${T}$, to verify that the
inequality holds for all $T$.  For large $T$, as expected the graphs
are separated by a constant value $\log(27/16)$.

\subsection{Probability interpretation in the large $T$ limit}\label{sec:largeT}

We can now obtain a well-defined probability for a transition.  We
take the limit $T \to \infty$ and fix $x= \ha T$ so that the operators
are as far apart from each other as they can be.

In this limit we find for general $R$, $R_1$ and $R_2$
\begin{align}
P(R \to R_1, R_2) & =  \frac{1}{ (2e^{-T})^{\Delta_1 +\Delta_2}} \frac{ g ( R_1 , R_2 ; R )^2 f_R 
}{ f_{R_1} f_{R_2} } e^{ - T(\Delta_1 +\Delta_2)}  \nn \\
& =  \frac{1}{2^{\Delta_1 +\Delta_2}} \frac{ g( R_1 , R_2 ; R )^2 f_R 
}{ f_{R_1}f_{R_2}}
\end{align}
where we have used the approximation \eqref{yummyapprox} for the large
$T$ limit of the genus-1 correlator.  This probability is independent
both of the spacetime positions of the operators and of $T$.

\section{Results for probabilities }\label{sec:correctresults}

The calculations done here are given in the Appendix 
\ref{correctnorm}.

\subsection{$G =0$ factorization}

For the amplitude of several operators combining into a bigger
operator we use genus zero factorization.  The correlators are computed on 
 $\R^4$ and the results for probabilities  are invariant 
under the conformal transformation to $S^4$. 
In a large distance limit, 
the resulting normalization prescription is equivalent to the
overlap of states normalization we na\"ively used before. 
These sphere factorization relations are equivalent
to the factorization equations derived in \cite{cr}.
The gluing procedure is as in section \ref{sphere4}.
For example, the probability for two ``in'' states to evolve to a 
single ``out'' state is given by
\begin{align}
  & P\left(R_1(r=e^x, \O_i), R_2(r=e^y,\O_i) \to R(r=0)\right)\nn \\
  & = \frac{\left|\big\langle R_1^\dagger(r=e^x, \O_i)R_2^\dagger(r=e^y, \O_i)
      R(r=0) \big\rangle \right|^2}{\big\langle
    R_2^\dagger(s=e^y, \O_i)R_1^\dagger(s=e^x, \O_i)R_1(r=e^x, \O_i)
R_2(r=e^y, \O_i)\big\rangle\big\langle
    R^\dagger R\big\rangle} \label{skoda}
\end{align}
In our calculations we put $R_1$ and $R_2$ at the same position
$x=y$ so that the normalization factor in the denominator is an
extremal correlator. The results will then be valid beyond 
the zero  coupling limit $ g_{YM}^2 =0 $, 
where the actual computations are done. 
If we separate them in spacetime, then we have a non-extremal correlator 
in the denominator which can be computed at zero coupling, but 
which will  receive non-trivial corrections at finite coupling. 
We further take the $x,y \to \infty$ limit.  This maximizes the distance of
the operators $R_1$ and $R_2$ from $R$ and gives a probability
independent of the spacetime positions of the operators.

For two giants combining into another giant we get
\begin{align}
  P(\textrm{2 size } N/2 \textrm{ S giants} \to \textrm{1
    size } N \textrm{ S giant}) & = \frac{f_{[1^N]}}{\sum_S g\left([1^{N/2}],
  [1^{N/2}];S\right)^2 f_S   }<1 \nn \\
  P(\textrm{2 size } N/2 \textrm{ AdS giants} \to \textrm{1
    size } N \textrm{ AdS giant}) & = \frac{f_{[N]}}{\sum_S g\left([N/2],
  [N/2];S\right)^2 f_S   } <1 \label{smalltobig}
\end{align}

For the transition of Kaluza Klein gravitons to a giant we get
\begin{align}
  P(N \textrm{ size } 1 \textrm{ KK gravitons} \to \textrm{one
    size } N \textrm{ S giant}) & \sim \frac{1}{N^N} \nn \\
  P(N \textrm{ size } 1 \textrm{ KK gravitons} \to \textrm{one
    size } N \textrm{ AdS giant}) & \sim \left(2^{2N-1}\frac{1}{\sqrt{\pi
        N}}\right)\frac{1}{N^N} 
\end{align}
\begin{align}
  P(N/2 \textrm{ size } 2 \textrm{ KK gravitons} \to \textrm{one
    size } N \textrm{ S giant}) & \sim \sqrt{\frac{2}{e}}\frac{1}{(eN)^{N/2}} \nn \\
  P(N/2 \textrm{ size } 2 \textrm{ KK gravitons} \to \textrm{one
    size } N \textrm{ AdS giant}) & \sim \left(2^{2N-1}\frac{1}{\sqrt{\pi
        N}}\right) \sqrt{\frac{2}{e}}\frac{1}{(eN)^{N/2}} 
\end{align}
 We see that larger KK gravitons are more likely to evolve into a 
 giant graviton than several smaller ones. It would 
 be interesting to give a proof that this trend continues to hold 
 when  KK states of  more general small angular momenta are considered. 
 For the case of $N/k$   angular momenta equal to $k$, the obvious 
 guess extrapolating   the leading behavior of the 
above results is  $ N^{-N/k} $.   
 The results of Appendix \ref{21mixed} will be useful for the case 
 where   only angular momentum $1$ and $2$ are involved. 
 More generally we will need to establish some general  properties 
 of the relevant symmetric group quantities.  The information 
 theoretic  ideas on overlaps from \cite{balbabel1} may be explored
 as a tool.

Strictly traces can only be interpreted as Kaluza-Klein states
when the individual traces involved are small as above. 
It is of interest, nevertheless,  to compute probabilities for 
extrapolated KK-states where large powers  are involved. 
We find 
\begin{align}
  P(1 \textrm{ size } N \textrm{ KK graviton} \to \textrm{one
    size } N \textrm{ S giant}) & \sim \sqrt{\pi N} \frac{1}{2^{2N}} \nn \\
  P(1 \textrm{ size } N \textrm{ KK graviton} \to \textrm{one
    size } N \textrm{ AdS giant}) & \sim \left(2^{2N-1}\frac{1}{\sqrt{\pi
        N}}\right) \sqrt{\pi N} \frac{1}{2^{2N}} = \frac{1}{2}
\label{asymm2}
\end{align}
For transitions to outgoing KK gravitons  we must use the
 basis dual to the trace basis. For the case of a single 
trace, and an initial giant, 
we find the same probability whether we have a sphere giant or an AdS 
giant
\begin{equation}
\label{asymm1}
  P( \textrm{one
    size } N \textrm{ giant} \to  \textrm{one size }
 N \textrm{ KK graviton})  =  \frac{1}{N}
\end{equation}
These transitions do not  decay exponentially as $N$ becomes large.
Note also the asymmetry between \eqref{asymm1} and \eqref{asymm2}, which 
is another illustration of the probabilities on the choice of measurement.

\subsection{$G=1$ factorization}

For the amplitude of 1 giant graviton into 2 smaller giants we must use 
genus-1 factorization.  We take two 4-spheres, one with coordinates
$(r,\O_i)$, the other with $(s,\O_i')$, cut out two 4-balls at radii 1
and $e^T$ from the origin in each, and glue the spheres together so that
$rs=1$ near the first gluing and $rs=e^{2T}$ near the second.  Also
introduce a primed coordinate $r'$ on the first sphere with $rr'=1$
and $s'$ on the second with $ss'=1$.

The probability is then given by
\begin{align}
 & P\left(R(r=e^x, \O_i) \to R_1^\prime(r'=0) R_2(r=0)\right)\nn \\
 & = Z_0e^{-2 T\Delta_1}\frac{\left|\big\langle
       R^\dagger(r=e^x, \O_i) R_1^\prime(r'=0)R_2(r=0) \big\rangle
    \right|^2}{\big\langle
    R^\dagger (s=e^x, \O_i) R(r=e^x, \O_i)\big\rangle_{G=1}  \big\langle
    R_1^{\dagger}R_1\big\rangle\big\langle R_2^\dagger R_2\big\rangle} \label{skoda2}
\end{align}
where $x \in [0,T]$ so that the operator is outside the cut-off area.  We
take the limit $T \to \infty$, where the factor $ Z_0 e^{-2T\Delta_1 }$ 
goes to $1$  (see discussion in   Section \ref{sec:theinequality243}). In addition we 
 fix $x= \ha T$ so that the operators
are far apart from each other, maximizing the distance of the
insertion of $R$ from the two boundaries of the cut $S^4$.  This procedure will
give a probability independent of the spacetime dependencies of the
operators, as discussed in Section \ref{sec:largeT}.  In this limit we
find
\begin{align}
P(R \to R_1, R_2)  =  \frac{1}{2^{\Delta_1 +\Delta_2}} \frac{ g ( R_1 , R_2 ; R )^2 f_R 
}{ f_{R_1} f_{R_2}} \label{eq:comparison1}
\end{align}

For the transition of a giant into two smaller giants
\begin{align}
  P(\textrm{1
    size } N \textrm{ S giant} \to \textrm{two size } N/2 \textrm{ S giants})
    & \sim \sqrt{\frac{\pi N}{2}} \left(\frac{1}{2}\right)^{2N} \nn \\
  P(\textrm{1
    size } N \textrm{ AdS giant} \to \textrm{two size } N/2 \textrm{ AdS giants})
    & \sim \frac{3}{\sqrt{8}}  \left(\frac{16}{27}\right)^{N} \label{bigtosmall}
\end{align}
These are well-normalized probabilities and demonstrate that
(\ref{skoda2}) with a higher genus correlator in the denominator gives
the proper implementation of the multi-particle normalization.  In the
old multi-particle normalization prescription, we got a divergent result for this
transition of $AdS$ giants
\begin{equation}
 \frac{\left|\big\langle
       \chi_{[N]}(\Phi^\dagger) \chi_{[\frac{N}{2}]}(\Phi)\chi_{[\frac{N}{2}]}(\Phi) \big\rangle
    \right|^2}{\big\langle
   \chi_{[N]}(\Phi^\dagger)\chi_{[N]}(\Phi)\big\rangle  \big\langle
    \chi_{[\frac{N}{2}]}(\Phi^\dagger)\chi_{[\frac{N}{2}]}(\Phi)\big\rangle\big\langle
    \chi_{[\frac{N}{2}]}(\Phi^\dagger) \chi_{[\frac{N}{2}]}(\Phi)\big\rangle}  \sim \frac{3}{\sqrt{8}}  
\left(\frac{32}{27}\right)^{N}\label{eq:comparison3} 
\end{equation}  
The factor of $2^{-N}$ from equation \eqref{eq:comparison1} provides
the correction to \eqref{eq:comparison3} to give the correctly
normalized result \eqref{bigtosmall}.

We can also compute the transition of a giant to two Kaluza-Klein gravitons
giving
\begin{align}
  P(\textrm{1
    size } N \textrm{ S giant} \to \textrm{two size } N/2 \textrm{ KK gravitons})
    & \sim  \left(\frac{2}{N}\right)^2\sqrt{\frac{\pi N}{2}}
    \left(\frac{1}{2}\right)^{2N} \nn \\
  P(\textrm{1
    size } N \textrm{ AdS giant} \to \textrm{two size } N/2 \textrm{ KK gravitons})
    & \sim \left(\frac{2}{N}\right)^2\frac{3}{\sqrt{8}}  \left(\frac{16}{27}\right)^{N}
\end{align}
These are well-normalized probabilities.  In the old multi-particle
normalization scheme, we had a diverging result for this transition 
\begin{equation}
 \frac{\left|\big\langle
       \chi_{[N]}(\Phi^\dagger)  \tr(\Phi^{\frac{N}{2}})\tr(\Phi^{\frac{N}{2}})\big\rangle
    \right|^2}{\langle\chi_{[N]} ( \Phi^\dagger )\chi_{[N]}
    ( \Phi )\rangle  \big\langle
    \tr(\Phi^{\dagger\frac{N}{2}})\tr(\Phi^{\frac{N}{2}})\big\rangle\big\langle
    \tr(\Phi^{\dagger\frac{N}{2}})\tr(\Phi^{\frac{N}{2}})\big\rangle}  \sim \frac{1}{6\sqrt{2}}  \left(\frac{32}{27}\right)^{N}
\end{equation}
 
An interesting question is whether a Schur polynomial operator can
only evolve into other Schur polynomials.  We might ask whether in
the large $T$ limit
\begin{align}
\sum_{R_1, R_2} P(R \to R_1, R_2)
\end{align}
adds up to 1.  We can calculate this sum when $R$ is a sphere (or $AdS$)
giant because, by the Littlewood Richardson rules, it can only split
into other sphere (or $AdS$) giants.  We find that this guess does not work
\begin{align}
\sum_{k} P([1^N] \to [1^{k}],[1^{N-k}]) < 1
\end{align}
which means that the infinite sums over additional outgoing states 
do contribute a finite amount. 

\subsection{Higher genus factorization}
For higher genus $G= n-1$ factorization, a natural guess for the analogous
equation to \eqref{eq:comparison1} is
\begin{align}
P(R \to R_1, R_2, \dots ,R_n)  =  \frac{1}{k_n^{\Delta_1 +\Delta_2 +\cdots + \Delta_n}} \frac{ g ( R_1 , R_2, \dots, R_n ; R )^2 f_R 
}{ f_{R_1} f_{R_2}\cdots  f_{R_n} }
\end{align}
where $k_n$ is a constant.  We know $k_1 = 1$ and $k_2 =2$.  We
assume that this equation holds in a long-distance limit, when the
operators are in a symmetric configuration far apart from each other.

We can work out limits on $k_n$ by considering the transition of an
$AdS$ giant into $n$ smaller $AdS$ giants
\begin{align}
P([N] \to n \times [N/n])  & =  \frac{1}{k_n^N} \frac{ f_{[N]}
}{ f_{[N/n]}^n }\nn \\
 & \sim  \frac{1}{\sqrt{2}}  \left[\frac{(n+1)}{n}  \right]^{\frac{n}{2}} \left[\frac{4n^{n+1}}{k_n (n+1)^{n+1}}  \right]^{N}  
\end{align}
in the large $N$ limit.  Given that $4n^{n+1}(n+1)^{-n-1}$ tends up to
$4/e$, $k_n > 4/e$ would certainly ensure that the probability is not
larger than 1, although this condition is clearly too strong for
$n=1$.  $k_n = n$ would satisfy this condition and works for $n=1,2$
but this is no more than a guess.

For the transition of an $AdS$ giant of $R$-charge $\Delta_R$ to
KK gravitons we find
\begin{align}
P([\Delta_R] \to \tr(\Phi^{\Delta_1}), \dots \tr(\Phi^{\Delta_n}))  & = \frac{1}{k_n^{\Delta_R}}  \frac{1}{\Delta_1 \cdots \Delta_n} \frac{f_{[\Delta_R]}}{f_{[\Delta_1]} \cdots f_{[\Delta_n]} }
\end{align}
and for a sphere giant
\begin{align}
P([1^{\Delta_R}] \to \tr(\Phi^{\Delta_1}), \dots \tr(\Phi^{\Delta_n}))  & = \frac{1}{k_n^{\Delta_R}}  \frac{1}{\Delta_1 \cdots \Delta_n} \frac{f_{[1^{\Delta_R}]}}{f_{[1^{\Delta_1}]} \cdots f_{[1^{\Delta_n}]} }
\end{align}

For genus $G=2$ we have for the transition of an $AdS$ giant into KK
gravitons
\begin{align}
P(\textrm{1 size } N \textrm{ AdS giant} \to \textrm{three size } N/3 \textrm{ KK gravitons}) & = \sqrt{\frac{2}{3}} \frac{36}{N^3}  \left(\frac{81}{64k_3}  \right)^N \nn \\
P(\textrm{1 size } N \textrm{ AdS giant} \to \textrm{one size } N-2 \textrm{ and 2 size 1 KKs}) & =\frac{(2N-1)(2N-2)}{(N-2)N^2}\frac{1}{k_3^N}
\end{align}
which makes it more likely for a giant to evolve into 3 medium-sized
KK gravitons than into one large one and two tiny ones.

\section{Bulk interpretation of the gluing properties of
correlators}\label{sec:bulkinterpreta}

 The factorization properties of the CFT correlators  
 allow the construction of correlators 
 on a 4-manifold of more complicated topology in terms of correlators
 on manifolds of simpler topology.  For example the theory on $S^3 \times S^1$ 
can be reconstructed 
 by starting from correlators on $S^4$. As we have emphasized above, 
 these relations imply that to get properly
 normalized probabilities from correlators 
 on $S^4$ (or the conformally equivalent $\R^4$) we need, in general, 
 correlators on more complicated topologies. 

 In the CFT the correlators of local operators can be interpreted in
 terms of transition amplitudes between states. These states
 can be identified as wavefunctionals of the fields on $S^3$
 boundaries of four dimensional balls, $B^4$,
 cut out around the local
 operators. Hence the amplitudes are given by path integrals
 with  boundary conditions on the CFT fields, specified at the $S^3$ 
 boundaries. Using this CFT interpretation of correlators as
 transition amplitudes, and the bulk-boundary correspondence of
 AdS/CFT, it is natural to interpret the correlators as gravitational
 transition amplitudes, obtained by Euclidean bulk path integrals,
 subject to boundary conditions for bulk fields that are specified in the
 neighborhood of the local operator insertions in the boundary CFT.
 This is indeed compatible with perturbative computations
 \cite{gkp,witten,fmmr,lmrs} for operators of small $R$-charge. The
 work of LLM \cite{llm} relating local operators to bulk geometries suggests
 that we can interpret correlators of operators with large $R$ charge in
 terms of bulk transition amplitudes between geometries (LLM-like in
 the case of half-BPS operator insertions) defined in the neighborhood
 of the boundary insertions. Note that although the 
 bulk path integral is over Euclidean metrics, the asymptotic geometries are 
 AdS-like, and so they admit a Lorentzian continuation. For a recent discussion of Euclidean 
 quantum gravity in an M-Theory context see \cite{gibbhawkfest}.
 The above bulk spacetime picture of correlators implies, for example, that a 
 three point function of gauge theory operators can be viewed as a 
 transition from a disjoint union of LLM geometries to a single LLM geometry. 
 This is a topology changing process. We will note, in Section \ref{sec:holtopchange}, that 
 this holographic setup for topology change implies constraints on the 
 interpolating topologies.

In this section we will investigate some of the implications of this
picture.  Some of our discussion will be in terms of the
five-dimensional bulk, where the sphere part of $ AdS_5 \times S^5 $
is captured through dimensional reduction to gravitational fields on
$AdS_5$ and higher KK modes coming from the five sphere.

One strength of the interpretation of correlators as transition
amplitudes computed via bulk Euclidean path integrals is immediately
apparent. Since the factorization properties of correlators on the CFT
side follow from the path integral implementation of geometrical
gluing relations, it is reasonable to expect that a simple
bulk-gravitational explanation of these relations among correlators
might follow from the postulate that the correlators can also be
interpreted as gravitational transition amplitudes defined in terms of
path integrals with asymptotic geometries (LLM-like geometries in the
case of half-BPS operators of large $R$ charge). Gluing on the CFT side
is then lifted to gluing on the gravity side.  In CFT, an important
ingredient in relating path integral gluing to relations among
correlators of operators is the correspondence between operators and
states, viewed as wavefunctionals.  Such a connection in gravity is
not directly understood, but we will be lead to some discussion of it
based on AdS/CFT considerations in Section \ref{sec:opwavecorr}.

In addition to SYM correlators on $S^4$ we will be interested in
correlators on manifolds which can be obtained by simple cutting and
pasting procedures of copies of $S^4$. We can cut out the open four-ball 
neighborhoods  $ B^4_{\circ} $  of $n$ points  of $S^4$ and to 
get a manifold denoted by $ S^4 \setminus
\sqcup_{\alpha=1}^n ( B^4_{\circ}  )_{\alpha } $. 
This can also be written as $ \overline { S^4 \setminus
\sqcup_{\alpha=1}^n ( B^4  )_{\alpha } }  $, 
indicating that we can remove closed balls, and then take the
 closure 
\footnote{ $B^k$ will denote  closed  balls and $B^k_{\circ} $ open ones. }. 
Take two copies of $ S^4 \setminus
\sqcup_{\alpha=1}^n ( B^4_{\circ}  )_{\alpha } $ 
and glue  along the $S^3$ boundaries. The analogous
construction in two dimensions gives the genus $n-1$ surface. We will
denote the corresponding manifold in 4D as $\Sigma_4 ( n-1) $ and
refer to it as having genus $n-1$ by analogy to the 2D case. The
subscript denotes the dimension, and the argument denotes the
genus. These manifolds can also be obtained as the boundary in $\R^5$
of the neighborhood of a graph with $n-1$ loops. In the following we
will also find it useful to consider neighborhoods of graphs in $B^5$,
with endpoints of the graph lying on the $S^4$ boundary of the
$B^5$. These graphs, denoted as Witten graphs, appear in the
perturbative computation of correlators in AdS.
  They will play a role in understanding how to lift gluings of 
$ S^4 \setminus
\sqcup_{\alpha=1}^n ( B^4_{\circ}  )_{\alpha } $ to the bulk.

\subsection{Bulk geometries for $S^3 \times S^1$ boundary from Witten graphs }\label{sec:bulkgluinggenus1}

Consider  the case of $S^3 \times S^1$.  Start from 2-point
functions on $S^4$. Cut out two disjoint copies of $B^4_{\circ} $
 around the insertion points,  
 obtaining a manifold with topology $ S^3 \times I $. 
Using the scaling symmetry on $S^4$, we can
obtain states at the boundaries of $S^3 \times I $.  Two copies of
$S^3 \times I $ can be glued to get $ S^3 \times S^1 $.  The $S^4$ is
the boundary of Euclidean $AdS_5$, which has topology $B^5$.  We would
like to understand how the gluing lifts to the bulk.  It is well known
that the supergravity partition function for the $S^3 \times S^1$ manifold receives
contributions from two different bulk topologies, namely $ B^4 \times
S^1 $ and $ S^3 \times B^2 $ \cite{witten}\cite{wittenpt}.  Hence the procedure for
lifting the gluings from boundary to bulk should account for both
these possibilities. We will demonstrate that this is accomplished 
 simply by using Witten graphs.

\begin{figure}[t]
\begin{center}
\resizebox{!}{5cm}{\includegraphics{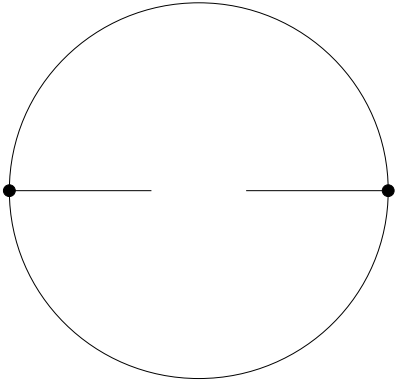}}
\caption{Disconnected graph $G_1$ in $B^5$ associated with two
  insertions on the boundary $S^4$} \label{fig:dgraph1}
\end{center}
\end{figure}

\begin{figure}[t]
\begin{center}
\resizebox{!}{5cm}{\includegraphics{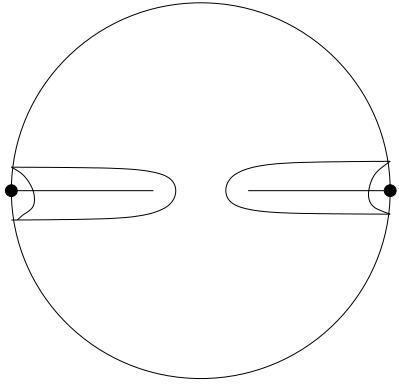}}
\caption{Neighborhood of the graph $G_2$}
\label{fig:dgraph2}
\end{center}
\end{figure}

Given two points on $S^4$ bounding a $B^5$, a very simple graph to
consider is the disconnected one consisting of two lines, joining
points in the bulk to the points on the boundary (see Figure
\ref{fig:dgraph1}).  We will denote this disconnected graph $G_1$.
The neighborhood of each line is a $B^4$ fibered over an interval and
collapsing to zero size at one end. This is homeomorphic to $B^5$.
Hence the neighborhood of the graph is a disjoint union of two small $
B^5$'s. Now consider the original $B^5$ with this neighborhood
removed, i.e the complement in $B^5$ of the neighborhood of the graph.  Take the closure.  
Let us call this $\overline{ B^5 \setminus N ( G_1,B^5 )} $ where $ N (
G_1,B^5 ) $ indicates a neighborhood\footnote{More exactly
  we write $N ( G,B^5 )= \{x \in B^5 : || G -x || \leq
  \epsilon \}$ where we are using the metric inherited from the
  trivial embedding of $B^5$ in $\R^5$.  We do not use the metric of
Euclidean AdS in
  this definition.} in the $B^5$ of the graph fixed by a small number
$\epsilon$.  The original $S^4$ boundary  now has two $B^4_{\circ} $ 
 removed. 
 It has two    $S^3$ boundaries 
 (see Figure \ref{fig:dgraph2}), exactly
the geometry we would consider purely from the point of view of CFT on
$S^4$. After excising these graph neighborhoods from $B^5$ 
( and taking the closure ), the original $S^4$
boundary has become $ S^3 \times I $. The remaining 5D manifold still
has topology $B^5$, and  its $S^4$  boundary can be  described as
$$ B^4 \cup ( S^3 \times I ) \cup B^4  $$ 
The two $B^4$'s are joined to $S^3 \times I $ at the two ends of $I$
on $S^3$'s.

\begin{figure}[t]
\begin{center}
\resizebox{!}{5cm}{\includegraphics{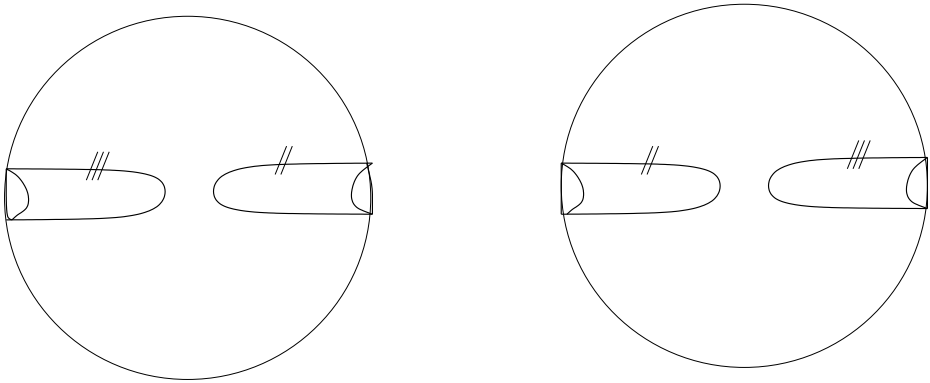}}
\caption{Gluing two copies of the $B^5$ with graph neighborhood removed}\label{fig:dgraph3} 
\end{center}
\end{figure}

Take two copies of this $\overline{B^5 \setminus N ( G_1, B^5 )}$ which is 
topologically the same as $  B^4 \times B^1  \cong B^5 $,
and do two gluings (see Figure
\ref{fig:dgraph3}). The outcome is $ B^4 \times S^1$ with boundary
$S^3 \times S^1$.  Thus we have obtained one of the bulk geometries
holographically dual to $S^3 \times S^1$ by lifting to the bulk the
CFT gluing of two copies of $S^3 \times I$.

\begin{figure}[t]
\begin{center}
\resizebox{!}{5cm}{\includegraphics{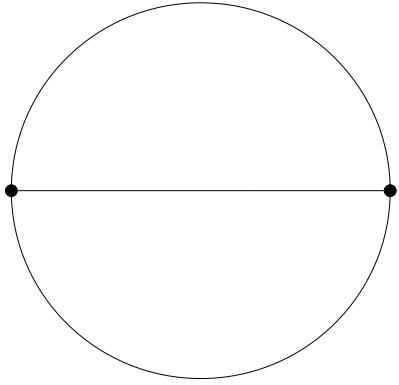}}
\caption{Connected graph $G_2$ in $B^5$ associated with two insertions
  on $S^4$}\label{fig:cgraph1}
\end{center}
\end{figure}

\begin{figure}[t]
\begin{center}
\resizebox{!}{5cm}{\includegraphics{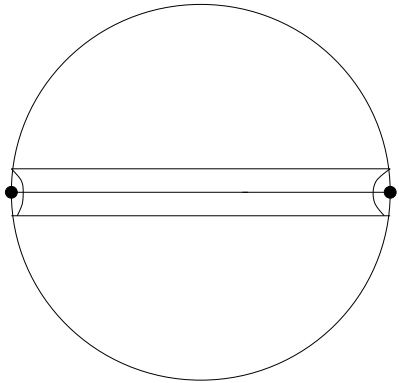}}
\caption{Neighborhood of the connected graph $G_2$ of topology $B^4
  \times I$}\label{fig:cgraph2}
\end{center}
\end{figure}

\begin{figure}[t]
\begin{center}
\resizebox{!}{5cm}{\includegraphics{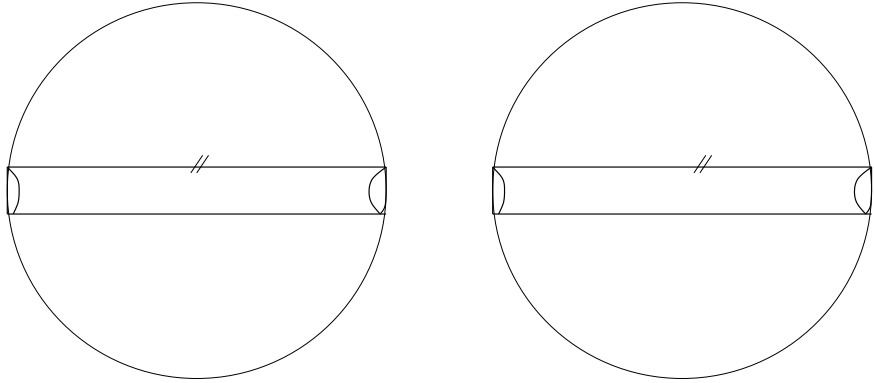}}
\caption{Gluing two copies of the $B^5$ with graph neighborhood removed } \label{fig:cgraph3}
\end{center}
\end{figure}

Now we want to understand, through the bulk lifting of boundary
gluings, the bulk geometry $ S^3 \times B^2$ which also has boundary
$S^3 \times S^1$.  Again we start with two points in the $S^4$
boundary of $B^5$. Now draw the graph which joins the two points and
extends through the bulk (see Figure \ref{fig:cgraph1}). We will call
this graph $G_2$.  The neighborhood of the graph is $B^4 \times I $.  Excise this neighborhood from the
$B^5$.  The manifold   $\overline {B^5
\setminus N ( G_2 , B^5 )}$ (see Figure \ref{fig:cgraph2}),  has
topology $ S^3 \times B^2 $, which has boundary $ S^3 \times S^1$.
The $S^1 $ consists of the interval $I$ which bounds the excised
region, joined to a semicircular interval on the original $S^4$
boundary.  Now take two of these $\overline{ B^5 \setminus N ( G_2 ,
B^5 ) }$.  Glue along the interior $S^3 \times I $ as indicated in
Figure \ref{fig:cgraph3}.  Since $B^2 $ joined to another $ B^2 $
along an interval is $B^2$, the outcome of this gluing of $ S^3 \times
B^2 $ to $ S^3 \times B^2$ along $ S^3 \times I $ is $ S^3 \times
B^2$. This is the second topology with boundary $S^3 \times S^1$ which
appears in \cite{witten}.

\subsubsection{Further topologies with $ S^3 \times S^1$ boundary}

If we use more complicated Witten graphs, with loops inside the bulk
$B^5$, we get more complicated bulk manifolds with boundary topology
$S^3 \times S^1$.  We do not know if they support metrics which are
extrema of the supergravity action. But they will certainly contribute
in the bulk path integral corresponding to the partition function on
$S^3 \times S^1$.  A natural question is whether, using the most
general Witten graphs and the most general gluing maps, we can produce
the most general bulk topology with the specified boundary topology.

\subsection{Gluing to higher genus 4-manifolds and 
corresponding bulk topologies    }

In this section we will show how to build  bulk topologies 
 corresponding to the higher genus four-manifolds $ \Sigma_4 ( n-1 )
$. As mentioned earlier we can obtain $ \Sigma_4 ( n-1 ) $ by starting
with two copies of $S^4$ with $n$ punctures, excising $n$ copies of
$B^4_{\circ}$  around the punctures  and gluing along the $S^3$
boundaries. Following the lead from the discussion of $ \Sigma_4 ( 1 )
$, we will consider tree-level Witten graphs with $n$ boundary
points. For our topological considerations, graphs related by merging
two internal  vertices by shrinking a connecting  edge will be
equivalent. Distinct graphs will correspond to different ways of
separating the $n$ points into subsets. This is the same as the number
of ways of partitioning $n$, usually denoted by $p(n)$.
 All the points in one subset 
will be joined up by one vertex in the bulk. When all the $n$ points are
connected by one vertex, we have a connected graph $V_n$. When 
they are separated into different subsets, we have disconnected graphs. 
In the description of the bulk topologies corresponding to 
disconnected graphs, it will be useful to use the concept of handle 
attachment, which appears in the theory of handlebody decompositions
\cite{gs}.  We will start with a brief review of handlebody decompositions. 
For a physics discussion of these see \cite{dowgar,hartn}.

\subsubsection{Handlebody decompositions}

To give a handlebody decomposition of a manifold $M$, 
we start with a $d$-dimensional ball $B^d$ (a 0-handle) and then add
handles to it until we obtain a manifold homeomorphic to $M$.
A $k$-\emph{handle} is a manifold $B^k \times B^{d-k}$ which we glue
onto $M$ along the boundary $ \partial B^k \times B^{d-k} = 
S^{k-1} \times B^{d-k}$.

\begin{figure}[t]
\begin{center}
\includegraphics{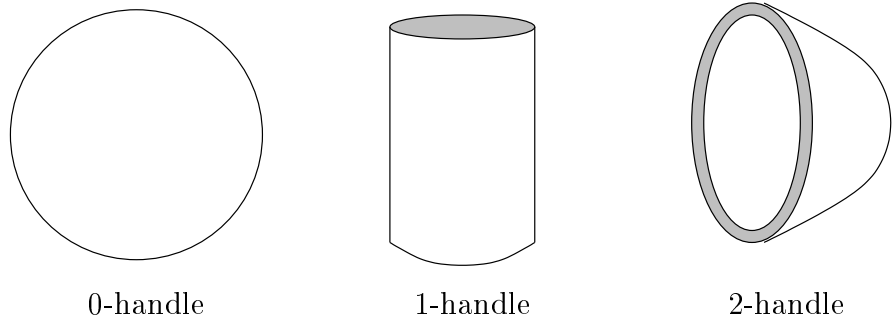}
\caption{The different handles in 3 dimensions } \label{fig:handles}
\end{center}
\end{figure}

For the different handles in three dimensions, $d=3$, see Figure
\ref{fig:handles}.  A 0-handle $B^3$ is a filled ball.  A 1-handle
$B^1 \times B^2$ is a filled cylinder which we can bend to attach it
to the manifold at the two ends of the cylinder ($S^0 \times B^2$, two
disconnected filled circles). A 2-handle $B^2 \times B^1$ can be
thought of as a thickened hemisphere (like a squash ball cut down the
middle) which we glue along the base $S^1 \times B^1$.  The $B^1$
interval provides the thickening.  A 3-handle $B^3$ is a filled ball
which we glue along its surface $S^2$. 
In general handlebody decompositions are not unique.

\subsubsection{Gluing for the complements of connected Witten graphs}

\begin{figure}[t]
\begin{center}
\includegraphics{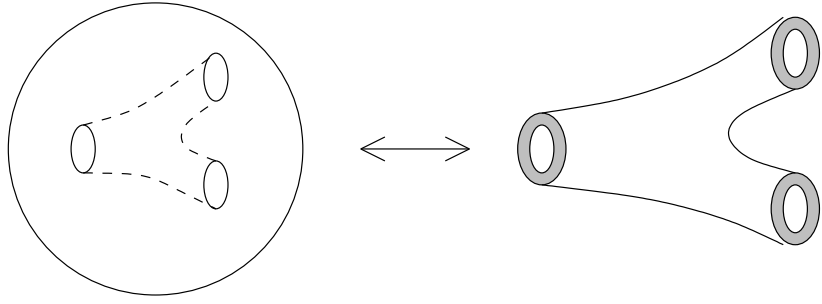}
\caption{$\overline{B^3 \setminus N(V_3,  B^3)}$ is homeomorphic to
  $\d^{(i)} N(V_3,  B^3) \times B^1$: thickened pants}
\label{fig:trousers}
\end{center}
\end{figure}

Now we want to understand how to glue the five-manifolds 
related to  connected Witten graphs.  We
have two copies of $\overline{B^5 \setminus N(V_n,  B^5)}$. Each is obtained 
by removing from $B^5$ the neighborhood $ N ( V_n , B^5 ) $
of the Witten graph $V_n$, and taking the closure of the resulting manifold.  
 This procedure restricts, on the  
boundary of the $B^5$, to the excision of $n$ copies of $B^4_{\circ}$ 
around $n$ points on the $S^4$. It thus provides a bulk lifting of 
the usual CFT construction of removing open neighborhoods 
of operator insertions. The neighborhood $N(V_n,  B^5)$ has topology 
$B^5$ and boundary $S^4$. The interior boundary of this neighborhood 
will be defined as the intersection of the boundary of $N(V_n,  B^5)$ with 
 $\overline{B^5 \setminus N(V_n,  B^5)}$. This interior boundary\footnote{More
  formally $\d^{(i)} N ( G ,  B^5 )= \{x \in B^5 : || G -x || =
  \epsilon \}$.  $\d^{(i)} N ( G ,  B^5 )$ differs from $\d N ( G ,  B^5 )$ because $\d N ( G ,  B^5 )$ includes the four-balls around the insertion points of  the Witten graphs.} 
 will be denoted by $\d^{(i)} N(V_n,  B^5)$. For concreteness 
see the left picture  in Figure \ref{fig:trousers}  
for  $\overline{ B^3 \setminus N ( V_3 , B^3 )} $. It is clear that
 $\d^{(i)} N(V_3 ,  B^3 )$ is the usual two dimensional pants diagram. 
We will be gluing two copies of $\overline{B^5 \setminus N ( B^5 , V_n )}$ 
along   $\d^{(i)} N(V_n,  B^5)$. 

\begin{figure}[t]
\begin{center}
\includegraphics{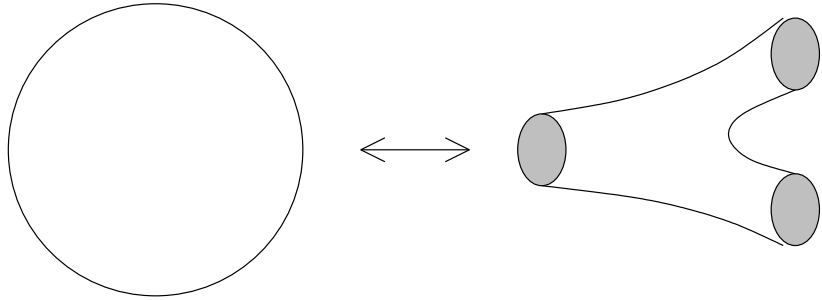}
\caption{$B^3$ is homeomorphic to
  $N(V_3,  B^3)$}
\label{fig:prepants}
\end{center}
\end{figure}

The crucial observation  is that $\overline{B^5 \setminus N(V_n, 
B^5)}$ is homeomorphic to the thickening, \break 
 $\d^{(i)} N(V_n,  B^5) \times B^1$,  of the internal surface
$\d^{(i)} N(V_n,  B^5)$.  We see this by first
noting that $B^5$ is homeomorphic to $N(V_n,  B^5)$ (see
Figure \ref{fig:prepants} for the case $d=3$, $n=3$).
Then if we remove $N(V_n,  B^5)$ from the $B^5$, and take the closure, 
 we just get a
thickening $\d^{(i)} N(V_n,  B^5)
\times B^1$ of the internal surface  (see Figure \ref{fig:trousers}
 for the case of $d=3$,
$n=3$).  Since $\overline{B^5 \setminus N(V_n, 
B^5)}$ is homeomorphic to $\d^{(i)} N(V_n,  B^5) \times B^1$, $\overline{B^5 \setminus N(V_n, 
B^5)}$ is homotopic to $\d^{(i)} N(V_n,  B^5)$ by the trivial homotopy retract that shrinks the $B^1$ to a point.

\begin{figure}[t]
\begin{center}
\includegraphics{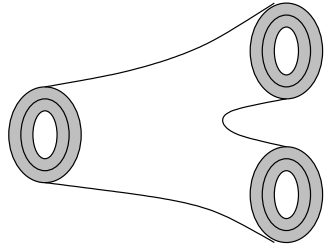}
\caption{$\d^{(i)}N(V_3,  B^3) \times B^1$ glued to $\d^{(i)}N(V_3,
   B^3) \times B^1$ along the internal surface}
\label{fig:invert}
\end{center}
\end{figure}

This means that we can `invert' the second copy of $\overline{B^5 \setminus
N(V_n,  B^5)}$ and glue it inside the internal surface
$\d^{(i)}N(V_n,  B^5)$ of the first copy of $\overline{B^5 \setminus N(V_n,
 B^5)}$ (see Figure \ref{fig:invert} for $d=3, n=3$).  To
invert the second copy we take the manifold $\d^{(i)}N(V_n,  B^5)
\times B^1$ and invert the direction of the $B^1$ coordinate.

The resulting manifold is the same as $\d^{(i)}N(V_n,  B^5) \times
B^1$, but now double the thickness.  Thus it has the same topology as
the original manifold $\overline{B^5 \setminus N(V_n,  B^5)}$. 
It is interesting to note that 
$ \partial \overline{B^5 \setminus N(V_n,  B^5)} $ is made of two 
copies of $ S^4 \setminus \sqcup_{\alpha =1 }^n B^4_{\circ} $ joined 
at the $S^3$'s, hence it is $ \Sigma_{4}( n-1) $. This is consistent 
with the result that the gluing has not changed the topology of the 
bulk manifold or its boundary. 

\subsubsection{Gluing for the complements of disconnected graphs}

Suppose a Witten graph $G$ is composed of $m$ disconnected
components $G = V_{n_1} \sqcup V_{n_2} \sqcup \cdots \sqcup V_{n_m}$ where
$\sqcup$ means disjoint union.

If we now glue $\overline{B^5 \setminus N(G,  B^5)}$ to a copy of itself
along the internal surface $\d^{(i)}N(G, B^5)$ the resulting
manifold is the same one $\overline{B^5 \setminus N(G,  B^5)}$ with
$(m-1)$ 1-handles attached.

\begin{figure}[h]
\begin{center}
\includegraphics{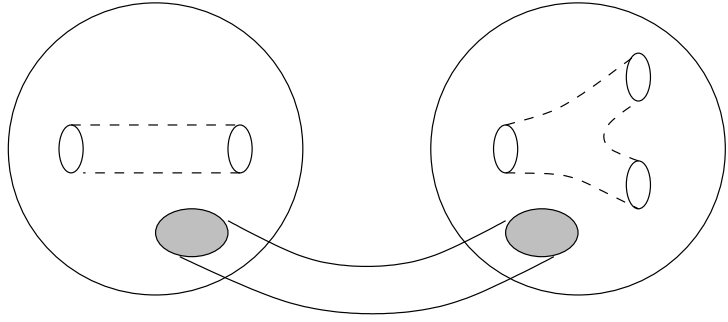}
\caption{$\overline{B^3 \setminus N(V_3 \sqcup V_2,  B^3)}$; a 1-handle
  links the connected parts}
\label{fig:distort}
\end{center}
\end{figure}

We can see this if we deform $\overline{B^5 \setminus N(G, B^5)}$ into
$\overline{B^5 \setminus N(V_{n_1}, B^5)}$, $\overline{B^5 \setminus
N(V_{n_2}, B^5)}$, \dots and $\overline{B^5 \setminus N(V_{n_m},
B^5)}$ linked in a line by 1-handles (see Figure \ref{fig:distort}).
Locally each $\overline{B^5 \setminus N(V_{n_i}, B^5)}$ glues to its
copy as above for the connected Witten graphs.  Now there are two
1-handles in each link between the connected parts, generating $(m-1)$
non-trivial 1-cycles.  We can also obtain this manifold by starting
with $\overline{B^5 \setminus N(G, B^5)}$ and attaching $(m-1)$
1-handle loops.

The figure also makes it clear that the boundary of $\overline{B^5
\setminus N(G, B^5)}$ is a connected sum of $ \Sigma ( n_1 -1 ) ,
\Sigma ( n_2 -1 ) \cdots \Sigma ( n_m -1 ) $.  This boundary is
topologically $ \Sigma ( n -m )$ where $n = \sum_{i=1}^m n_i $.  After
the gluing we have $\overline{B^5 \setminus N(G, B^5)}$ with $(m-1)$
1-handles attached.  Each 1-handle increases the genus by one, so the
glued manifold has boundary $ \Sigma ( n -1 )$, as expected.

\subsubsection{Homology groups  } 

For the complement of the connected  Witten graph, $\overline{ B^5 \setminus N ( V_n , B^5 )}$,  
the  homology groups,  derived in Appendix Section \ref{sec:celldecompositioncomplementconnected}, are
\begin{itemize}
\item $H_0(\overline{ B^5 \setminus N ( V_n , B^5 )}) = \Z$ 
\item $H_1(\overline{ B^5 \setminus N ( V_n , B^5 )}) = \{0\}$ 
\item $H_2(\overline{ B^5 \setminus N ( V_n , B^5 )}) = \{0\}$ 
\item $H_3(\overline{ B^5 \setminus N ( V_n , B^5 )}) = \Z^{n-1}$
\item $H_4(\overline{ B^5 \setminus N ( V_n , B^5 )}) = \{0\}$
\item $H_5(\overline{ B^5 \setminus N ( V_n , B^5 )}) = \{0\}$
\end{itemize}
The Euler character follows 
\begin{equation}
  \chi (\overline{ B^5 \setminus N ( V_n , B^5 )}) = \sum_j (-1)^j b_j = 1
  +(-1)^3(n-1) = 2-n
\end{equation}

For the homology of $\overline{B^5 \setminus N(V_{n_1} \sqcup V_{n_2}
\sqcup \cdots \sqcup V_{n_m}, B^5)}$, the complement of a disconnected
Witten graph, see Appendix Section
\ref{sec:celldecompositioncomplementdisconnected}.

For the genus $n-1$ 4-manifold  $ \Sigma_4 ( n-1 ) $  the homology groups, derived in Appendix Section \ref{sec:cellboundary},
are 
\begin{itemize}
\item $H_0(\S_4(n-1)) = \Z$ 
\item $H_1(\S_4(n-1)) = \Z^{n-1}$ 
\item $H_2(\S_4(n-1)) = \{0\}$
\item $H_{3}(\S_4(n-1)) = \Z^{n-1}$ 
\item $H_{4}(\S_4(n-1))= \Z$ 
\end{itemize}
For the case of a 
 2-dimensional boundary the same methods give the 
 standard homology of a Riemann surface with genus $g = n-1$
\begin{itemize}
\item $H_0(\S_2(n-1)) = \Z$
\item $H_1(\S_2(n-1)) = \Z^{2(n-1)}$
\item $H_{2}(\S_2(n-1))= \Z$
\end{itemize}

A simple check on these results is provided by 
 the Mayer-Vietoris sequence, which relates
homology groups associated with $ X \cup Y $ , $ X \times Y $ and $ X
\cap Y $. In our case $X = \overline{B^5  \setminus N( V_n , B^5 )}$.  $Y$ is $ N ( V_n , B^5 ) $,
 the closed neighborhood of the graph, which has the 
topology of a ball.  $ X \cup Y $ is $B^5$. $X \cap Y$ is the interior
 part of the boundary of $N ( V_n , B^5 )$, called $\d^{(i)} N(V_n,
B^5)$.  
The  Mayer-Vietoris sequence  implies that 
\bea\label{mvchi} \chi ( X ) + \chi ( Y )
= \chi ( X \cap Y ) + \chi ( X \cup Y ) \eea 
In this case
\begin{align}
   \chi ( Y ) &= \chi ( B^5 ) = 1 \nn \\
   \chi ( X \cap Y ) &=   2 - n     \nn \\
   \chi ( X \cup Y ) &= \chi ( B^5 ) = 1 
\end{align}
Hence we deduce $ \chi ( X ) =  2 -  n  $.
In the above we have used the fact that the Euler character 
of   $ X \cap Y = \d^{(i)} N ( V_n , B^5  )$ is  $ 2- n $. In the case of 
3D bulk and 2D boundary this is the familiar Euler character 
of $S^2$ with $n$ disks removed.  In the case of 5D bulk and 3D 
boundary this is derived in the Appendices. One way is to 
use an explicit  cell decomposition  (see  Appendix Section \ref{sec:celldecompositioncomplementconnected}).  
The other way is to use  the fact that
 $S^4 \setminus \sqcup_{\alpha }  ( B^4_{\circ} )_{\alpha }  $
 is a quotient of $ B^4 \setminus \sqcup_{\alpha }  ( B^4_{\circ} )_{\alpha }  $, 
which retracts to an n-wedge of spheres (see Appendix Section \ref{sec:shortexact}).  

Note that we expected $\chi(\overline{B^5 \setminus N(V_n, B^5)}) = \chi(\d^{(i)}
N(V_n, B^5))$ since $\overline{B^5 \setminus N(V_n, B^5)}$ is homotopic to
$\d^{(i)} N(V_n, B^5)$ and the Euler characteristic is homotopy
invariant.

Now we take the gluing of $ X $ with another copy of 
$X$ along $ \d^{(i)} N( V_n , B^5  )$. Let us call the resulting space $ Z $. 
Then
\bea 
&& X \cup X = Z \cr 
&&  X \cap X = \d^{(i)} N ( V_n , B^5 ) 
\eea 
 Use \ref{mvchi} again to find 
\be
 \chi ( X  ) + \chi ( X )  = \chi ( Z ) + (2-n)   
\ee 
which gives $ \chi ( Z  ) = 2 - n  $.
We have explained that $ X \cong Z $ so we have here 
checked  that $\chi ( Z ) = \chi ( X ) $, i.e. 
the Euler characters before and after gluing are the same.

\subsection{Holographic topology change}\label{sec:holtopchange}

  A clear prescription for the computation of correlators in Euclidean 
 AdS space is given in \cite{witten,gkp} in the context of 
 perturbation theory, using the correspondence between single trace 
 CFT  operators  and Kaluza-Klein states in the bulk. 
 As emphasized in \cite{bbns} the perturbative set-up does  not work 
 for operators of large $R$ charge. A clear extension of the perturbative
 prescription  for the computation of Euclidean correlators is not  
 available. The work of LLM suggests that for the Schur polynomial
 operators, in the regime of sufficiently large $R$ charge, 
 the right way to find the bulk computation of the correlators is 
 to use the LLM geometries. For a Euclidean setting it is natural to
 do a Wick rotation of the LLM geometry and cutoff the region where the 
 $S^3$ of AdS is larger than a fixed size.  

\begin{figure}[h]
\begin{center}
\resizebox{!}{5cm}{\includegraphics{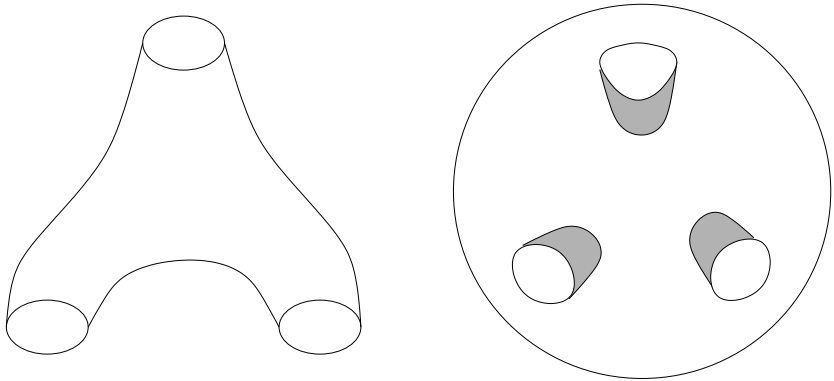}}
\caption{On the left is a cobordism between two $S^1$'s (or $S^3$'s) and
  one $S^1$ (or $S^3$); on the right is the analogous interpolation
  for the bulk}
\label{fig:boundaryinterpolate}
\end{center}
\end{figure}

 For concreteness consider a three-point function. 
 For each operator insertion we have a cut-off LLM-like
 geometry, which determines boundary conditions for the 
 bulk path integral over metrics ( and other fields ).  
 The three-point correlator can be viewed as a transition amplitude between a
disconnected pair of LLM-like geometries and a single LLM-like
geometry.  This is reminiscent of cobordisms, where a manifold
interpolates between several disconnected boundary components. These
appear in formal discussions of 2D CFT and topological field theory
\cite{dijkhou}.  In low dimensional Matrix models, this is
discussed as macroscopic loop amplitudes \cite{ginsmoore}.  A
2-dimensional example is the pants diagram which describes a
transition from a disjoint union of two circles to a single circle
(see Figure \ref{fig:boundaryinterpolate}).

In the CFT we can think of our transition amplitudes in terms of 
 cobordisms. For a three-point function for example,
 we take $S^4$ and cut out 3 balls around
the operator insertions. We then map our operators to states in
Hilbert spaces associated to the three $S^3$ boundaries.  
The $S^4$ with three balls removed is an interpolation ( cobordism )
 between the disjoint union of two $S^3$'s and a single $S^3$.  
The correlator is computed as a CFT path integral on this cobordism. 
To lift this picture to the bulk let us assume for simplicity that we
can discuss this in the purely five dimensional perspective. We take
our Euclidean AdS, which we think of as a  5-dimensional
 ball, and remove $B^5$'s at the boundary around the operator
insertions (see Figure \ref{fig:boundaryinterpolate}).  We can then
 insert  cut-off LLM-like geometries 
associated with the operators in the balls. 
We integrate over all metrics ( and other fields ) in the remaining 
bulk. At the interface of the bulk with the LLM geometries
 (shaded gray in Figure \ref{fig:boundaryinterpolate}) the 
 boundary conditions are specified by the cut-off LLM. 
 The bulk must also be
asymptotically Euclidean AdS in the remaining boundary regions.
We have a transition from a pair of disjoint geometries to  a single one.
  This is not a cobordism because the initial and final conditions do not 
 correspond to distinct boundaries of the bulk, but rather 
 to different  marked regions of the single boundary, with boundary conditions 
 determined by LLM geometries. 

\begin{figure}[h]
\begin{center}
\includegraphics{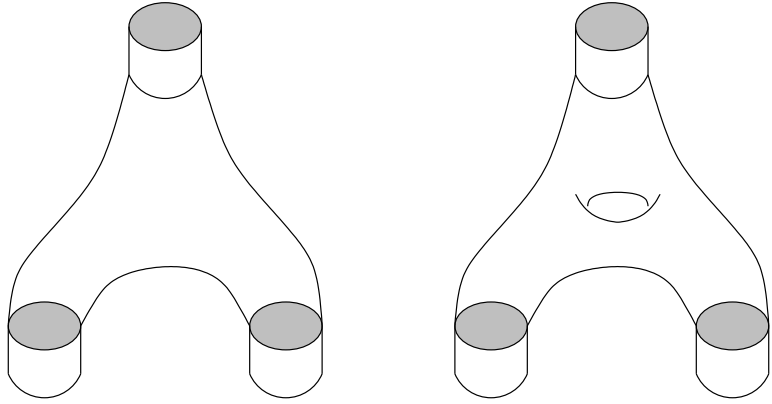}
\caption{Two different interpolating bulk geometries}
\label{fig:interpolate}
\end{center}
\end{figure}

It is worth emphasizing an important difference between this
picture of topology change with the one natural from a  
traditional  gravitational path integral perspective, which uses cobordisms 
and does not implement holography. From the traditional perspective,
a five dimensional gravity theory   would sum over all possible topologies consistent with
 the initial and final topologies living at 4D boundaries, 
 using some appropriate weights \cite{horo,dowgar}.  But from  the holographic 
picture, we have 5D geometries which provide 4D boundary conditions on separate regions 
of the 4D boundary. The topology of the complete 4D boundary is fixed, since this is where the 
dual non-gravitational field theory lives.  Fixing the boundary
topology constrains the bulk topology. This is easiest to see in the
even simpler case of 2D boundary theory and 3D bulk. There are many
ways to interpolate between two copies of $B^2 \times I$ and a single
copy, involving boundary topology of arbitrary genus (with three
boundary circles).  Figure \ref{fig:interpolate} gives the genus zero
and genus one cases.  But the boundary CFT gives a well-defined
amplitude for each fixed genus.  The CFT does not give a prescription
for summing over these different topologies.  The story is the same in
the case of 4D CFT/5D bulk. We are given, from super-Yang Mills
theory, a three point function for a fixed boundary topology.  We can
construct bulk topologies interpolating between two balls $ B^4 \times
I$ and a single one by using neighborhoods of graphs with three
vertices, with any number of loops (note that in previous discussions
we considered graphs with no loops).  SYM does not give a way of
summing over these different 4D topologies.  Hence holography acts 
as a constraint on interpolating topologies.

Given that the  bulk-boundary correspondence in ADS/CFT \cite{maldainflat} 
can be interpreted in terms of a Hartle-Hawking
 wavefunction \cite{harthawk}, it is interesting to note that we
 here have a situation where 
different regions of the Hartle-Hawking boundary are associated 
with different geometries ( LLM-like for the case of half-BPS 
operators ). The search for a framework that can 
handle probabilities  for multiple geometries ( universes )
  appearing in a multiverse 
 in the context of eternal inflation scenarios
  has been an active topic of discussion
\cite{fs,bousso,page,vilenk}. It will be interesting to 
 apply  the lessons on the correct normalization  of
 probabilities in quantum gravity, given by AdS/CFT,  to the spacetimes of 
 interest in eternal inflation.  One qualitative lesson we may extract 
 from sections 3-6 in this paper, is that properly normalized
 answers to  questions regarding physics on one spacetime,
 require knowledge of correlators on more complicated topologies.  
 A systematic framework for exploring the relevance of these ideas
 to eternal inflation could perhaps be found along the lines \cite{fssc}. 

 Finally we note that a different perspective on 
 topology change in the context
 of LLM has been discussed in \cite{horava}. In the latter discussion, the 
 topology changing process is described entirely in terms of the Fermi sea. 
 In the present picture the fermions are only relevant as a 
 description of the half BPS states in the asymptotic regions, while the bulk 
 involves fluctuating geometries and in general goes beyond the half BPS 
 sector.

\subsection{Towards holographic topological gravity theory} 

The observations in this section can be viewed as hints
towards the  definition of axioms for a   holographic 
topological gravity theory. Such a theory in $D$ dimensions
is related to a conformal or topological field theory in $D-1$ 
dimensions. The partition function of the holographic topological theory, 
obtained by a sum over topologies with fixed boundary,  is 
equal to that of the boundary  field  theory. 
 Operator insertions at a point 
in the $(D-1)$ dimensional theory 
are associated to states in a Hilbert space, living on 
the $D-2$ dimensional boundary of a neighborhood of the point. 
The usual gluing relations of the boundary theory are lifted 
to the bulk via the Witten graph construction we have described. 
The above remarks on holographic topology change should also 
have a natural role in an axiomatic holographic topological gravity theory.

\subsection{Operator-Wavefunctional correspondence in quantum gravity}\label{sec:opwavecorr}

In Section \ref{sec:bulkgluinggenus1} we posed a geometrical question on how to lift boundary
gluings, associated to a choice of punctures for insertions of local
CFT operators, to bulk gluings.  The solution we described made use of
Witten graphs which have the punctures as end-points and join them up
through vertices in the bulk.  The same gluing on the base space of
the CFT was lifted to different gluings of the bulk manifolds, along
the interior boundary of the neighborhoods of the Witten graphs. From the CFT
perspective, geometrical gluing relations translate into relations
between amplitudes, after we use the correspondence between local
operators and wavefunctionals of fields on a sphere surrounding the
local operator.  Interpreting the bulk gluings in an analogous manner
in terms of wavefunctionals of gravity ( and other fields ) in AdS ,
we are lead to conclude that the insertion of a physical  observable (
corresponding to a  CFT operator ) on the boundary
 of AdS leads to wavefunctionals on the
interior boundaries of the neighborhoods of all the possible Witten
graphs. This is to be contrasted with the much simpler
operator-wavefunctional correspondence in CFT. Admittedly we have only
given indirect evidence for this more complicated
operator-wavefunctional correspondence in gravity, and it would be
interesting to derive it more directly.  It appears superficially to
be a consequence of the greater non-locality we expect in a quantum
theory of gravity \cite{lpstu,gmh}. A more direct derivation of this
multiplicity of wavefunctionals related to a set of operator
insertions on the boundary should also clarify the relation between
the topological use of Witten graphs here and their perturbative use.
The work of \cite{freidel} has some of the elements needed to make this 
connection. In that work, Feynman integrals $I_{\Gamma}$ are related 
to expectation values of observables  ( in a spin foam model ),  
i.e $ I_{ \Gamma } = \langle 0 |  \cO  | 0 \rangle $. 
Since our gluing story suggests the consideration of wavefunctionals 
associated to Feynman graphs, it is natural to explore if they are related 
to   states    $ \cO  | 0 \rangle $ appearing in  \cite{freidel}. 
It is an interesting future direction to explore the extension of this
kind of connection between observables and wavefunctionals in quantum
gravity to more general spacetimes.

\section{Summary and Outlook } 

We started with a puzzle regarding the unexpected growth of 
 normalized correlators of gauge theory operators
on $S^4$ ( or $\R^4$ )   corresponding to AdS giants. 
We have found a resolution of the paradox by observing that the 
 proper normalization which leads to a probabilistic interpretation 
involves the division by correlators on 4-manifolds of more 
complicated topology, which we called higher genus manifolds by analogy 
to the two dimensional case.  The appropriate behavior of the probabilities, 
 that they are less than one and add up to one when all outcomes 
 are taken into account, follows from factorization equations of 
 4D CFT which relate correlators on higher genus to those on lower genus. 
 These points were illustrated in two dimensions before moving on to 
 the 4D case. 

 These factorization properties follow by implementing
 geometrical gluing relations at the level of the path integral of 
 the CFT. In AdS/CFT the CFT can be viewed as living  on a 4D boundary of a 5D 
 bulk,  where the extra five dimensions are reduced away \`a la Kaluza Klein.  
 As a first step towards a bulk understanding of these 
 properties, we considered how to lift the gluings of  the 4D 
 boundaries to the 5D bulk. Witten graphs played a central role 
 in this story.

 There are several avenues for future research suggested by this work. 

\begin{itemize}

\item 
We have observed a trend that products of traces are more likely
to overlap with  Schur operators $ \chi_R ( \Phi ) $, 
 if they involve traces of higher powers. It will be instructive 
  to see how general this is. If these results are extended to 
 the case of decay of brane-antibrane systems in AdS, they could  
 be related to the fact that brane decay is more likely to produce 
 longer strings. 

\item 
 We have explored the idea that LLM geometries determine boundary conditions
 for  the bulk  path integrals. The results  of section \ref{sec:bulkinterpreta} 
 can be viewed as indirect supporting evidence for such a point of view. A 
 more direct approach is desirable.  A
 satisfactory formulation should allow an 
 extension of  the Euclidean gravitational path integral 
 prescription for computation of perturbative correlators 
 of single trace operators to the case of operators of very large charge
 corresponding to Young diagrams (or fermion excitations). 

\item  
 We have outlined some aspects of a holographic topological gravity theory
 in Section \ref{sec:bulkinterpreta}.
 It is an open problem to give  complete  definitions and exhibit 
 non-trivial examples.

\item 
While our discussion has focused on  local operators, 
there is also a substantial literature on Wilson loops 
in $\cN=4$ SYM, including connections to free fermions, see for example
 \cite{maldaloop,drukgro,okusem}.
 We may expect that while summing over local operators 
 leads to gluing along $S^3$, summing over Wilson loops will 
 be related to gluing along $ S^1 \times S^2$. One does have to deal with the 
 additional subtlety that, in the case of a general  Wilson loop, conformal 
 transformations ( which were used in the operator-state map )   will also 
 transform the loop itself. We expect that many aspects of our discussion 
 of the lifts from boundary gluings to bulk gluings will carry over. 
 The topological role of neighborhoods of Witten graphs 
 would now be extended to  neighborhoods of worldsheets of strings,  
 bounded by the Wilson loops, and  extending into the bulk. It will be 
 interesting to calculate normalized probabilities
 in the larger context involving 
 both Wilson loops and local operators. 

\item 
 The lessons we have learned on the correct normalization of probabilities 
 should be applied more generally in quantum gravity, in particular 
 to the problem of probabilities in the multiverse.
 We have made some preliminary remarks in 
 this direction in Section \ref{sec:holtopchange}.

\end{itemize}

\vskip.5in 

{ \bf Acknowledgements } We thank Andreas Brandhuber, Nick Dorey, Patrick 
Dorey,  Paul Heslop, Simon McNamara, Costis Papageorgakis, Andrew Sellers,
Rodolfo Russo,  Gabriele Travaglini and  Jan Troost for discussions. SR is 
supported by a  PPARC Advanced Fellowship  and in part  by the
 EC Marie Curie Research Training Network MRTN-CT-2004-512194.

\newpage
\begin{appendix}

\section{Appendix   }\label{appendixcalc}

\subsection{Multiparticle-normalized transitions of S and AdS-giants } 

We want to work out the normalized correlators for transitions from $AdS$
and $S$ giant graviton states into multiple KK gravitons.  We will use two
normalizations: the multi-particle normalization and the
overlap-of-states normalization.

For example the multi-particle-normalized $S$ transition amplitude is given by
\begin{equation}
  \label{eq:Sdecaymulti}
  \frac{\langle \chi_{[1^L]} (\Phi^{\dagger}) (\tr (\Phi^J) )^{L/J}
    \rangle}{ || \chi_{[1^L]} ( \Phi ) ||\,\,  ||\tr ( \Phi^J )||^{L/J}}
\end{equation}
and the overlap-of-states-normalized $AdS$ transition is given by
\begin{equation}
  \label{eq:AdSdecayoverlap}
  \frac{\langle \chi_{[L]} (\Phi^{\dagger}) (\tr (\Phi^J) )^{L/J}
    \rangle}{ || \chi_{[L]} ( \Phi ) ||\,\,  ||(\tr ( \Phi^J ))^{L/J}||}
\end{equation}
where we do not insist that $L \sim N$ so that we can be as general as possible.

The norms of the the $S$ and $AdS$ giants are given respectively by
\begin{align}
  || \chi_{[1^L]} ( \Phi ) ||^2 & = f_{[1^L]} = \frac{N!}{(N-L)!} \\
  || \chi_{[L]} ( \Phi ) ||^2 & = f_{[L]} = \frac{(N+L-1)!}{(N-1)!}
\end{align}
We can compute the norms involving traces in certain limits. These tractable cases are: 
\begin{itemize}
\item $L << N$ and any $J \leq L$ for the overlap normalization;
\item $J << N$ and any $L$ for the multi-particle normalization;
\item $J=1,2$ and any $L$ for the overlap normalization (see Sections
  \ref{J1} and \ref{J2});
\item $J=L,L/2$ for both normalizations (see Section \ref{JL}).
\end{itemize}

For $L << N$ for the overlap-of-states normalization and $J << N$ for the multi-particle normalization, we can use the
result proved below that for large $N$ and $JM < < N$
\begin{equation}
  \label{eq:crucialequation}
    ||(\tr ( \Phi^J ))^{M}||^2 = \langle(\tr(\Phi^J))^{M}
    (\tr(\Phi^{\dagger J}))^{M}  \rangle  \sim M!  J^{M} N^{JM}
\end{equation}
from which we see that the multi-particle normalization factor for $J
<< N$ is
\begin{equation}
  \label{eq:multinormalisation}
    ||(\tr ( \Phi^J ))||^{L/J}  \sim J^{L/2J}N^{L/2}
\end{equation}
If we fix $L$ and vary $J$, then we find that $J^{L/2J}N^{L/2}$
increases to a peak of $e^{L/2e}N^{L/2}$ at $J = e$ and then decreases
sharply approaching zero.
The overlap-of-states normalization factor for $L << N$ is given by
\begin{equation}
  \label{eq:overlapnormalisation}
  ||(\tr ( \Phi^J ))^{L/J}|| \sim \sqrt{(L/J)!}J^{L/2J}N^{L/2} 
\end{equation}
which decreases even faster as a function of $J$.  It still peaks
around $J=1,2$.
The fact that both of these normalizations are decreasing functions of
$J$ in these bounds means that giant gravitons are always more likely
to undergo transitions into larger KK modes than smaller ones.

Now we can proceed
\begin{eqnarray}
  \label{eq:Sdecay}
  \langle \chi_{[1^L]} (\Phi^{\dagger}) (\tr (\Phi^J) )^{L/J}
    \rangle & = & \sum_{R_1 \cdots R_{L/J}} g(R_1, \dots, R_{L/J};[1^L])
    \chi_{R_1}\left(J\right) \cdots
    \chi_{R_{L/J}}\left(J\right)f_{[1^L]} \ret
& =&     \left(\chi_{[1^J]}\left(J\right)\right)^{L/J}
f_{[1^L]}\ret
& =&     (-1)^{(J-1)L/J}|| \chi_{[1^L]} ( \Phi ) ||^2
\end{eqnarray}
We obtain the first line by writing each trace $\tr (\Phi^J)$ as a sum of Schur polynomials\footnote{For details for this and other similar 
identities see Appendix \ref{sec:idnotcon}.} over representations of the symmetric group $S_J$.  
Each trace sum includes representations $R_i$ corresponding to Young diagrams with $J$ boxes only. 
$\chi_{R_i}(J)$ is the character of 
a cycle of length $J$, 
e.g. $(12 \dots J)$.  In the
second line we have noted that we can only build $[1^L]$ in tensor
products of representations which are also single columns. Thus the LR coefficient  $g(R_1, \dots, R_{L/J};[1^L])$ is non-zero 
only when each $R_i=[1^J]$.

Similarly
\begin{eqnarray}
  \label{eq:Sdecay2}
  \langle \chi_{[L]} (\Phi^{\dagger}) (\tr (\Phi^J) )^{L/J}
    \rangle & = & \sum_{R_1 \cdots R_{L/J}} g(R_1, \dots, R_{L/J};[L])
    \chi_{R_1}\left(J\right) \cdots
    \chi_{R_{L/J}}\left(J\right)f_{[1^L]} \ret
& =&     \left(\chi_{[J]}\left(J\right)\right)^{L/J}
f_{[1^L]}\ret
& =&  || \chi_{[L]} ( \Phi ) ||^2
\end{eqnarray}

The multi-particle-normalized $S$ transition for $J << N$ is given by
\begin{eqnarray}
  \label{eq:Sdecaymulti2}
  \frac{\langle \chi_{[1^L]} (\Phi^{\dagger}) (\tr (\Phi^J) )^{L/J}
    \rangle}{ || \chi_{[1^L]} ( \Phi ) ||\,\,  ||\tr ( \Phi^J
    )||^{L/J}} & =& \frac{(-1)^{(J-1)L/J}|| \chi_{[1^L]} ( \Phi ) ||}{||\tr ( \Phi^J
    )||^{L/J} }\ret
  & \sim & (-1)^{(J-1)L/J}  J^{-L/2J} N^{-L/2} \sqrt{\frac{N!}{(N-L)!}}
\end{eqnarray}
and to get the overlap-normalized version for $L << N$ just divide by
$\sqrt{(L/J)!}$.

For $L=N$ and $J << N$ we get for the multi-particle normalization
\begin{eqnarray}
  \label{eq:SdecaymultiN}
  \frac{\langle \chi_{[1^N]} (\Phi^{\dagger}) (\tr (\Phi^J) )^{N/J}
    \rangle}{ || \chi_{[1^N]} ( \Phi ) ||\,\,  ||\tr ( \Phi^J
    )||^{N/J}} & \sim & (-1)^{(J-1)N/J}  J^{-N/2J} N^{-N/2}
  N^{N/2}e^{-N/2} (2\pi N)^{\frac{1}{4}}\ret
& = & (-1)^{(J-1)N/J} (2\pi)^{\frac{1}{4}} e^{-N/2 + \frac{1}{4}
  \log(N) -(N/2J)\log(J) }
\end{eqnarray}
which is exponentially decreasing for all $J$.

The multi-particle-normalized $AdS$ transition for $J << N$ is given by
\begin{eqnarray}
  \label{eq:Sdecaymulti3}
  \frac{\langle \chi_{[L]} (\Phi^{\dagger}) (\tr (\Phi^J) )^{L/J}
    \rangle}{ || \chi_{[L]} ( \Phi ) ||\,\,  ||\tr ( \Phi^J
    )||^{L/J}} & =& \frac{|| \chi_{[L]} ( \Phi ) ||}{||\tr ( \Phi^J
    )||^{L/J} }\ret
  & \sim &   J^{-L/2J} N^{-L/2} \sqrt{\frac{(N+L-1)!}{(N-1)!}}
\end{eqnarray}
and to get the overlap-normalized version for $L << N$ just divide by
$\sqrt{(L/J)!}$.

For $L=N$ and $J << N$ we get for the multi-particle normalization
\begin{eqnarray}
  \label{eq:SdecaymultiN2}
  \frac{\langle \chi_{[N]} (\Phi^{\dagger}) (\tr (\Phi^J) )^{N/J}
    \rangle}{ || \chi_{[N]} ( \Phi ) ||\,\,  ||\tr ( \Phi^J
    )||^{N/J}} & \sim &  J^{-N/2J} N^{-N/2}
  2^N N^{N/2}e^{-N/2} 2^{\frac{1}{4}}\ret
& = &  2^{-\frac{1}{4}} e^{-N/2 + N\log(2) -(N/2J)\log(J)}
\end{eqnarray}
The factor on the $N$ in the exponential is $-1/2 + \log(2) -
(1/2J)\log(J)$, which is positive for all $J$.  Thus this
exponentially increases for all $J$.  This shows that the multi-particle 
normalization does not give well-defined probabilities.

\subsection{Overlap normalizations : general formulas } 
Consider the correlator
\begin{equation}
  \label{eq:prodtracecorr}
  \langle  ~ ( \tr \Phi^J)^M  ( \tr \Phi^{\dagger J})^M  \rangle
\end{equation}
which appears in overlap normalizations. By using the diagrammatic method 
of \cite{cr} we can get 
\begin{equation}
  \label{eq:dsfdsfdsf5}
  \langle  ~ ( \tr \Phi^J)^M  ( \tr \Phi^{\dagger J})^M  \rangle =
  J^M M! \sum_{\s_2 \in
    [J^{M}]} N^{C\left(  \s_1  \s_2 \right)}
\end{equation}
where $\s_1$ is a fixed permutation in the symmetric group 
conjugacy class $ [J^M]\subset S_{JM}$,  characterized by $M$ cycles of length $J$, 
and $\s_2$ runs over all the elements in this conjugacy class.   $C(\s_1  \s_2)$ is the number of cycles in the permutation $\s_1  \s_2$. 
By converting to the Schur basis, we can also get the equivalent form
\begin{equation}
  \label{eq:prodtracecorr2}
  \langle\tr(\Phi^J)^M \tr(\Phi^{\dagger J})^M  \rangle = \sum_{R}
  \chi_R \left( \sigma_1  \right) \chi_R
  \left( \sigma_1  \right) f_R
\end{equation}
 where $\sigma_1 $ is again a fixed  permutation in the conjugacy class 
$ [J^M]$.

Getting explicit formulas for the sum in (\ref{eq:dsfdsfdsf5})
requires additional work.  For large $N$ and $JM << N$ the leading
terms will be
\begin{equation}
  \label{eq:dsfdsfdsf6}
  \langle  ( \tr \Phi^J)^M  ( \tr \Phi^{\dagger J})^M  \rangle =
  J^M M! \left( N^{JM} + O(N^{JM-2}) \right)
\end{equation}
The first term comes from (\ref{eq:dsfdsfdsf5}) when $\s_1 \s_2 =
1^{JM}$.  There is no $N^{JM-1}$ term because that would require $\s_1
\s_2 \in [1^{JM-2}2]$, a permutation with odd sign.  This is
impossible because $\s_1 \s_2$ must have even sign since $\s_1$ and
$\s_2$ are in the same conjugacy class.  This is an important fact
because if we now raise (\ref{eq:dsfdsfdsf6}) to a multiple of $N$, as
we do for the multi-particle normalization, we see that we only need
the first term for the large $N$ approximation.

For $J=1,2$ we can work out more explicit formulas for the sum in
(\ref{eq:dsfdsfdsf5}) (see Sections \ref{J1} and \ref{J2}); also any
$J$ for $M=1$ (see Section \ref{JL}).

\subsection{Overlap normalization for $ (\tr \Phi )^M $   
}\label{J1}

We know that 
\begin{equation}
  \label{eq:1cycles}
  \langle ~ ( \tr  \Phi)^M (\tr \Phi^{\dagger} )^M  \rangle = M! N^M
\end{equation}
For $M=N$ we find
\begin{equation}
  \label{eq:1cyclesN}
  \langle ~ ( \tr \Phi )^N (\tr \Phi^{\dagger} )^N  \rangle \sim
  N^{2N}e^{-N}\sqrt{2\pi N}
\end{equation}
which will make both the $S$ and the $AdS$ correlator very small
indeed.

For the transition from $L$  KK modes,  of angular momentum $1$ each,
to an $S$-giant,   we get
\bea 
{ \langle  \chi_{[1^L]} ( \Phi ) ( \tr   \Phi^{\dagger}    )^L  \rangle   \over 
  || (\tr  \Phi  )^L ||~  || \chi_{1^L} ( \Phi ) || }
=   \sqrt { { 1 \over N^L} { N ! \over L! (N-L)! } }
\eea 

For $ L=N$
\begin{align}
  \label{eq:Sdecayoverlaptoad}
  \frac{\langle \chi_{[1^N]} (\Phi^{\dagger}) (\tr  \Phi  )^{N}
    \rangle}{ || \chi_{[1^N]} ( \Phi ) ||\,\,  ||(\tr  \Phi)^{N}||} = &
  \frac{|| \chi_{[1^N]} ( \Phi ) ||}{||(\tr  \Phi )^{N}||} \ret
  = &   \sqrt{\frac{N!}{N!N^N}}\ret
  =  & N^{-N/2}
\end{align}

For the transition from $L$ KK modes of angular momentum $1$ each to 
an $AdS$-giant, we get
\bea 
{ \langle \chi_{[L]} ( \Phi ) ( \tr  \Phi^{\dagger}   )^L \rangle  \over 
  || (\tr  \Phi  )^L ||~  || \chi_{[L]} ( \Phi ) || } 
= \sqrt { { (N+L-1)! \over (N-1)!   L! N^L } } 
\eea 
When $ L=N $
\begin{align}
  \label{eq:AdSdecayoverlaptoad}
  \frac{\langle \chi_{[N]} (\Phi^{\dagger}) (\tr \Phi )^{N}
    \rangle}{ || \chi_{[N]} ( \Phi ) ||\,\,  ||(\tr  \Phi )^{N}||} =
  & \frac{|| \chi_{[N]} ( \Phi ) ||}{||(\tr \Phi )^{N}||} \ret
  = &  \sqrt{\frac{N(2N)!}{2N N!}\frac{1}{N!N^N}}\ret
  \sim & 2^{N-1/2}(\pi N)^{-1/4}N^{-N/2}
\end{align}
which are both very small.

\subsection{Overlap normalization for $ ( \tr ( \Phi^2 ) ) ^M $ }\label{J2}

We can show (see below)  that
\begin{align}
  \label{eq:2cyclesstart}
  \langle ~ ( \tr \Phi^2)^M  ( \tr \Phi^{\dagger 2})^M  \rangle =  & 2^{2M}M!
  \l(\l+1)(\l+2) \cdots (\l+M-1) \ret
   = & 2^{2M}M! \frac{(\l+M-1)!}{(\l-1)!}
\end{align}
where $\l = N^2/2$.  For $M =N/2$ we find that
\begin{align}
  \label{eq:2cyclesstartlargeN}
  \langle  ~ ( \tr \Phi^2)^{N/2}  ( \tr \Phi^{\dagger 2})^{N/2} 
 \rangle \sim  & 2^{N}(N/2)!
  \frac{(N^2/2+N/2)!}{(N^2/2)!} \ret
  \sim & 2^N \sqrt{\pi N}e^{-N} (N/2)^{N/2} (N^2/2 + N/2)^{N^2/2+N/2}
  (N^2/2)^{-N^2/2} \ret
  = & 2^N \sqrt{\pi N} e^{-N} (N/2)^{N/2}(N^2/2)^{N/2}(1 +
  1/N)^{N^2/2+N/2}\ret
  = & 2^N \sqrt{e \pi N} e^{-N/2}(N/2)^{N/2}(N^2/2)^{N/2}\ret
  = & \sqrt{e \pi N} e^{-N/2}N^{3N/2}
\end{align}
where we have used $(1+1/N)^N \sim e$ and  $(1+1/N)^{N^2} \sim e^N$.

For the transition from $N$ KK modes, with angular momentum $2$ each, to 
an S-giant, we get 
\begin{align}
  \label{eq:Sdecayoverlaptoad2}
  \frac{\langle \chi_{[1^N]} (\Phi^{\dagger}) (\tr  \Phi^2  )^{N/2}
    \rangle}{ || \chi_{[1^N]} ( \Phi ) ||\,\,  ||(\tr  \Phi^2 )^{N/2}||}
  \sim (2/e)^{1/4} N^{-N/4} e^{-N/4}
\end{align}
For the transition to the $AdS$ giant  we get
\begin{align}
  \label{eq:AdSdecayoverlaptoad2}
  \frac{\langle \chi_{[N]} (\Phi^{\dagger}) (\tr \Phi^2 )^{N/2}
    \rangle}{ || \chi_{[N]} ( \Phi ) ||\,\,  ||(\tr  \Phi^2  )^{N/2}||}
 \sim 2^{N-1/4}(e\pi)^{-1/4}  N^{-N/4-1/4} e^{-N/4}
\end{align}
which are both very small, but larger than the $J=1$ results.

\subsubsection{Overlap normalization for $ ( \tr \Phi^2 ) ^M $ from
  Casimir diagrammatics}\label{2mcalculation}

 There is a nice formula, which can be derived using a 
 diagrammatic method for the class-algebra of symmetric groups.

 The coefficient of $N^{2K} $ is obtained by summing over all possible 
 ways of writing $ K  = \sum_{i=1 } k_i $ where $k_1, k_2 , k_3 ...  $
 are   non-negative integers which also obey $ \sum_{i} i  k_i = M$  
\bea\label{norm2m} 
&&   \langle  (\tr  \Phi^2  )^M  ( \tr    (\Phi^{\dagger} )^  2  )^M \rangle  \nnm \\
&& = \sum_{K} M!^2  N^{ 2K } \sum_{\{k_i\} }  
     \prod_i F_i  \nnm \\ 
\eea 

The $F_i$ are  given by
\begin{align}\label{Fi}  
 F_i &= {  N_i^{k_i} \over k_i! (i!)^{2k_i} } \nn \\
  N_i & = (2i) ( (2i-2)!! )^2  = 2i (2i-2)^2(2i-4)^2 \cdots 2^2  = (2i)^{-1}2^{2i} (i!)^2
\end{align}
from which we have
\begin{equation}
 F_i  = { 2^{(2i-1)k_i} \over i^{k_i} k_i! } 
\end{equation}

A simple manipulation can be used to rewrite (\ref{norm2m}) as
\bea\label{norm2mf1}  
&&   \langle  (\tr  \Phi^2  )^M  ( \tr  ( \Phi^{\dagger})^  2  )^M \rangle  \nnm \\
&& = 2^{2M} 
  (M!)^2\sum_K  2^{-K} N^{2K} \sum_{\{k_i\} }\prod_i \frac{ 1}{i^{k_i} k_i!} 
\eea 
 We can associate the set $k_i$ with a conjugacy class in 
$S_M$, described by $k_i$ cycles of length $i$, so that $K$ 
is the total number of cycles. 
Rewriting (\ref{norm2mf1}) 
in terms of conjugacy classes  $[S_M]$, and then as a sum over $S_M$,  we get 
\begin{align}
  \label{eq:partofMcong}
  \langle ( \tr \Phi^2)^M  ( \tr (\Phi^{\dagger })^2 ) ^M  \rangle 
=  & 2^{2M}M!
  \sum_{ [ \s ] \in [ S_M ] }
  \frac{M!}{|\Sym([\s])|} \left(\frac{N^2}{2}\right)^{ C([\s])}
  \ret
= & 2^{2M}M!
  \sum_{ [\s]  \in [S_M ] }  | [\s] |
  \left(\frac{N^2}{2}\right)^{C ([\s])}\ret
= & 2^{2M} M!
  \sum_{\s \in S_M}  \left(\frac{N^2}{2}\right)^{C(\s)}
\end{align}
where $  | [\s] | $ is the number of symmetric group elements in the 
conjugacy class $[\s]$.   $C([\s])$ is the number of cycles in $[\s]$.  
$|\text{Sym}([\s])|$ is the size of the symmetry group of the permutation $\s$.  
The above  happens to be the formula for the 
 dimension of the totally symmetric representation
$[M]$ of $U(N^2/2)$.  If $\l = N^2/2$ then we get
\begin{align}
  \label{eq:partofMcong2}
  \langle\tr(\Phi^2)^M \tr(\Phi^{\dagger 2})^M  \rangle =  & 2^{2M}M!
  \l(\l+1)(\l+2) \cdots (\l+M-1)
\end{align}

\begin{figure}[h]
\begin{center}
\resizebox{!}{3.5cm}{\includegraphics{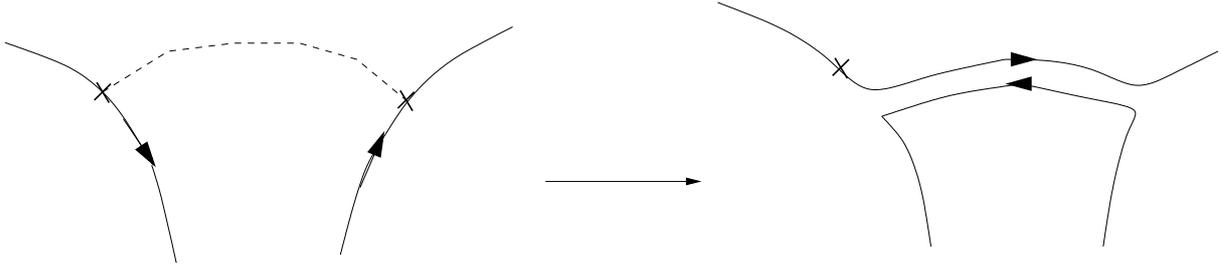}}
\caption{Merging two crosses}
\label{fig:mergecrossmove}
\end{center}
\end{figure}

\noindent
{ \bf   Proof :  }  
 The derivation of (\ref{norm2mf1}) can be related to 
 the class algebra multiplication of $ [2^M] . [2^M ] $ in $S_{2M}$.
  A useful technique for the calculation is based on the realization 
 of these operators in $ V^{\otimes 2M  }$ in terms of $ U(N ) $  Casimir 
 operators. For example 
\bea 
T_{[2]} = \Sigma_{\sigma \in [2] }  ~ \sigma =
 { 1\over 2  } \sum_{a_1 \ne a_2 } \rho_{a_1} ( E_{i_1 i_2} ) 
\rho_{a_2} ( E_{i_2 i_1} ) \equiv  { 1 \over 2 } E_{ [ 2 ] }  
\eea   
where $  E_{i_1 i_2} $ is the matrix with $1$ in
 the $(i_1,i_2) $ entry and zero elsewhere. 
The sum of elements in a conjugacy class can in general 
be related to such Casimirs ( called cycle operators )
 by equations of the form 
 \bea 
  T_{\vec l  } \equiv   { 1 \over | \Sym ( [ \vec l  ])  |  }  E_{ \vec l } 
\eea
 This is described in detail in \cite{chen, partensky} and can be used 
to give a diagrammatic algorithm for computation of products 
in the class algebra of symmetric groups \cite{sramthesis}.  
 The $E_{l_1, l_2 ,.. } $ operators are associated with 
circles having crosses marked on them, with  the number of crosses 
being $l_1, l_2 , ... $. When we are multiplying two of these operators 
we sum over ways of  joining the crosses from the two sets,  
with lines. These lines are then simplified with the move in 
Figure \ref{fig:mergecrossmove}. This move is a diagrammatic 
representation of the effect of multiplying the $U(N)$ generators.  
    
\begin{figure}[h]
\begin{center}
\resizebox{!}{4cm}{
\includegraphics{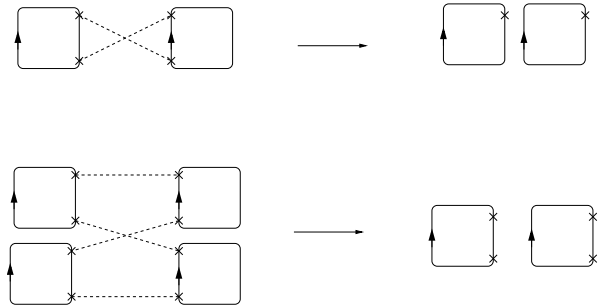} } 
\caption{First two types of diagrams for multiplication $[2^M].[2^M]$ }
\label{fig:diagsfor2m}
\end{center}
\end{figure}

The $E_{[2^M]}$ operator can be represented diagrammatically by 
$M$ oriented circles with two crosses each, which we will describe in words as 
$M$ copies of  $ C_2 $. When multiplying, we can draw the circles 
from the first  $E_{[2^M]}$ on the left and those from the second on the
 right. 
 The multiplication  involves a few basic 
 products of the form
\bea 
 C_{2} . C_{2}  && \sim  C_{1}.C_{1} \nnm \\
 ( C_{2})^2 . (C_{2})^2 && \sim  C_{2}.C_{2 } \nnm \\ 
 ( C_{2})^3 .  ( C_{2} )^3  && \sim C_{3}.C_{3 } \nnm \\ 
 && \vdots 
\eea 
In the first type of multiplication, there are lines joining 
crosses from one circle on the left and one on the right. 
In the second type of multiplication, there are lines joining 
crosses from two circles on the left and two circles on the right. 
These two types of multiplication are shown in Figure \ref{fig:diagsfor2m}. 
Let  $k_1$ be the number of $ (C_1)^2 $ coming from the multiplications 
 of the first type. The $k_2$ is the number 
 of $ ( C_2)^2 $ coming from multiplications
 of the second type etc.
The resulting diagram corresponds to the 
 cycle operator $ E_{1^{2 k_1} , 2^{2 k_2} , .. } $, related to the 
conjugacy class $ [ 1^{2 k_1} , 2^{2 k_2} , .. ]$, 
which is weighted, according to  (\ref{eq:dsfdsfdsf5}),    by $N^{2K } $.

The $T$ operators  we are multiplying 
are related to $E$ operators by the factor 
\bea 
c_1 = \left( { 1 \over |  \Sym ( T_{ 2^M }  )  | } \right)^2
\eea
 
To get a multiplication labeled by $k_1, k_2 ..$ 
we need to choose $k_1$ operators $C_{2}$  from the $M$ in $ E_{[2^M]} $, 
$2k_2$ operators $C_2 $ from the $M$ etc. This gives a 
factor 
\bea 
{ M! \over \prod_i ( ik_i)!  } 
\eea 
We  square because the same factor occurs in each of the 
$T_{2^M} $ we are multiplying, to get 
\bea 
c_2 =  \left(  { M! \over \prod_i ( ik_i)! }   \right)^2 
\eea 
Given the $ik_i$ copies of $C_{2} $ we group them in $k_i$ sets 
of $i$ in the following number of ways 
\bea 
\prod_{i} { ( ik_i)!  \over  (i!)^{k_i}  k_i! }
\eea  
We again square since this arises from each factor in the 
product 
\bea 
c_3 = \prod_{i} \left({ ( ik_i)!  \over  (i!)^{k_i}  k_i! }\right)^2
\eea 
There are $k_i!$ ways of connecting  the $k_i$ copies of 
  circles  with two crosses, from the two factors giving 
\bea 
c_4 = k_i! 
\eea 
Having fixed a set of $i$ $C_2$ to be connected to 
another $i$ $C_2$ there is a factor of 
\bea 
&&  ( 2i ) ( 2i -2)^2 (2i-4)^2 ... 2^2 \nnm \\ 
&& = (2i) ( (2i-2)!! )^2 \nnm \\ 
&&  = N_i 
\eea
Since this occurs $k_i$ times we have 
\bea 
c_5 = N_i^{k_i} 
\eea  
The resulting $ E_{[i^{2 k_i} ] } $-operator must be converted to 
$T_{ [ i^{2k_i } ]}   $ by a factor 
\bea 
c_6 =|  \Sym ( [ i^{2k_i } ]  ) | 
\eea 
Finally there is a factor of 
\bea 
 c_7 =    { |   \Sym  [  2^M ] |       |^2 \over |  \Sym  [ i^{2k_i } ]   | }
\eea 
which arises in converting the trace normalization problem to a 
class algebra problem .

Collecting the factors $ c_1 ..c_7 $ we get 
$F_i $ given in (\ref{norm2m}).

\subsubsection{The recursion method for $ || ( \tr ( \Phi^2 ) )^M ||$  }

A neat derivation of (\ref{eq:1cycles}) and (\ref{eq:2cyclesstart})
can be obtained by recursion.

Let $\cA^1_{M}$ be defined by
\begin{equation}
  \label{eq:1cyclesasasdasdads}
  \cA^1_{M} \equiv \langle(\tr(\Phi))^{M} (\tr(\Phi^{\dagger}))^{M}  \rangle
\end{equation}

Now choose a single $\Phi$ and Wick contract it with a single
$\Phi^\dagger$ (of which there are $M$ choices).  This Wick
contraction gives us a factor of $N$ and leaves us with $M-1$
uncontracted $\Phi$s and $M-1$ uncontracted $\Phi^\dagger$.  This
gives us a recursion relation
\begin{align}
  \cA^1_{M} = M N \cA^1_{M-1}
\end{align}
Applying this $M$ times and noting that $\cA^1_0 = 1$ we get
\begin{align}
  \cA^1_{M} = M! N^{M}
\end{align}
as expected.

\begin{figure}[h]
\begin{center}
\includegraphics{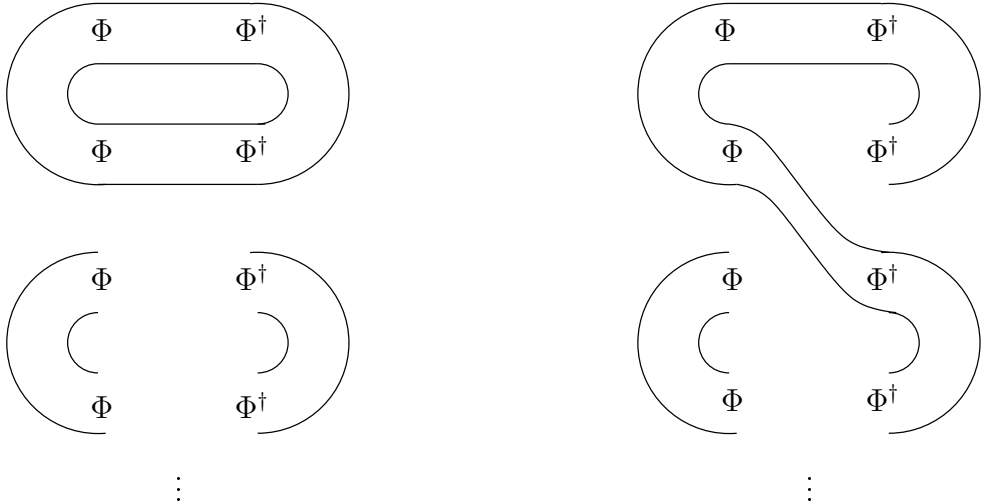}
\caption{For $\cA^2_{M}$ there are two different ways of contracting
  a $\tr(\Phi^{\dagger 2})$ with two $\Phi^\dagger$s.}
\label{fig:recursion3}
\end{center}
\end{figure}

Let $\cA^2_{M}$ be defined by
\begin{equation}
  \label{eq:1cyclesadasd}
  \cA^2_{M} \equiv \langle(\tr(\Phi^2))^{M} (\tr(\Phi^{\dagger 2}))^{M}  \rangle
\end{equation}
Now choose a $\tr(\Phi^2)$ and Wick contract the two $\Phi$s with two
$\Phi^\dagger$s.  There are two different ways of doing this.  The
first way, on the left of Figure \ref{fig:recursion3}, is to contract
them with a $\tr(\Phi^{\dagger 2})$ giving a factor of $N^2$.  There
are $M$ $\tr(\Phi^{\dagger 2})$s and two ways of pairing up the
$\Phi$s and $\Phi^\dagger$s.  This leaves us with $M-1$ uncontracted
$\tr(\Phi^2)$s and $M-1$ uncontracted $\tr(\Phi^{\dagger 2})$s.
The other way, on the right of Figure \ref{fig:recursion3}, is to pair
up the $\Phi$s with $\Phi^\dagger$s from different $\tr(\Phi^{\dagger
  2})$s.  There are $2M$ choices for the first $\Phi^\dagger$ and
$2M-2$ for the second.  Again this leaves us with $M-1$ uncontracted
$\tr(\Phi^2)$s and $M-1$ uncontracted $\tr(\Phi^{\dagger 2})$s.
Altogether we have
\begin{align}
  \cA^2_{M} = & 2M(N^2 +2M-2)A^2_{M-1}\ret
  = & 2^2M(\l + M-1)A^2_{M-1}
\end{align}
where $\l = N^2/2$.  This becomes
\begin{align}
  \cA^2_{M} =2^{2M}M! \frac{(\l+M-1)!}{(\l-1)!}
\end{align}
as expected.

\subsection{$J=L,L/2$ for both normalizations}\label{JL}

We know from \cite{cr} that
\begin{equation}
  \label{eq:Jcycles}
  \langle\tr(\Phi^L) \tr(\Phi^{\dagger L}) \rangle =
  \frac{1}{L+1}\left(\frac{(N+L)!}{(N-1)!} - \frac{N!}{(N-L-1)!}  \right)
\end{equation}
by considering the equation
\begin{equation}
  \label{eq:sdfswbjq}
   \langle\tr(\Phi^L) \tr(\Phi^{\dagger L}) \rangle = \sum_R \chi_R \left({L}\right) \chi_R
  \left({L} \right) f_R
\end{equation}
and noting that $\chi_R \left({L}\right)$ is only non-zero for
hooks.

If $L=N$ we find
\begin{equation}
  \label{eq:Ncycles}
  \langle\tr(\Phi^N) \tr(\Phi^{\dagger N}) \rangle \sim 
  \frac{(2N)!}{N!}
\end{equation}
For the transition of a sphere giant we get
\begin{align}
    \frac{\langle \chi_{[1^N]} (\Phi^{\dagger}) (\tr (\Phi^N) )
    \rangle^2}{ || \chi_{[1^N]} ( \Phi ) ||^2\,\,  ||\tr ( \Phi^N
    )||^2}  \sim & \frac{(N!)^2}{(2N)!} \nn \\
 \sim &(\pi N)^{\frac{1}{2}}2^{-2N}
\end{align}
which is very small.  For the $AdS$ transition we get
\begin{align}
    \frac{\langle \chi_{[N]} (\Phi^{\dagger}) (\tr (\Phi^N) )
    \rangle^2}{ || \chi_{[N]} ( \Phi ) ||^2\,\,  ||\tr ( \Phi^N
    )||^2}  \sim &  \frac{(2N-1)!}{(N-1)!}
\frac{N!}{(2N)!}\ret
  & = \frac{1}{2}
\end{align}
which is a large probability.

We can also write down a formula for $J = L/2$
\begin{align}
  \label{eq:Jcycleswerw}
  \langle(\tr(\Phi^{L/2}))^2 (\tr(\Phi^{\dagger L/2}))^2 \rangle = &
  \sum_R \sum_{R_1,R_2,S_1,S_2} g(R_1,R_2;R)g(S_1,S_2;R) \ret
&\chi_{R_1}
  \left({L/2}\right)\chi_{R_2}
  \left({L/2}\right)\chi_{S_1}
  \left({L/2}\right)\chi_{S_2} \left({L/2}\right) f_R
\end{align}
given that we know $\chi_{R_1} \left({L/2}\right)$ will
only be non-zero for hooks $\chi_{[(L/2 - r),1^r]}
\left({L/2}\right) = (-1)^r$.  This gives us
\begin{align}
  \label{eq:Jcyclessdfs}
  \langle(\tr(\Phi^{L/2}))^2 (\tr(\Phi^{\dagger L/2}))^2 \rangle = &
  \sum_R \sum_{r_1,r_2,s_1,s_2} (-1)^{r_1+r_2+s_1+s_2}g([(L/2 -
  r_1),1^{r_1}],[(L/2 - r_2),1^{r_2}];R) \ret
  &g([(L/2 - s_1),1^{s_1}],[(L/2 - s_2),1^{s_2}];R) f_R
\end{align}
where $r_i$ and $s_i$ are integers characterizing the hooks.

Another approach to finding (\ref{eq:Jcycleswerw}) is to use the
Murnaghan-Nakayama rule on
\begin{align}
  \label{eq:Jcycleswerw2}
  \langle(\tr(\Phi^{L/2}))^2 (\tr(\Phi^{\dagger L/2}))^2 \rangle = &
  \sum_R \left(\chi_R\left({L/2} \circ {L/2}\right)\right)^2 f_R
\end{align}

\subsection{General formula for 
$|| (\tr \Phi )^{M_1} (\tr \Phi^2 )^{M_2} ||^2$ }\label{21mixed}

\begin{figure}[h]
\begin{center}
\resizebox{!}{3.8cm}{
\includegraphics{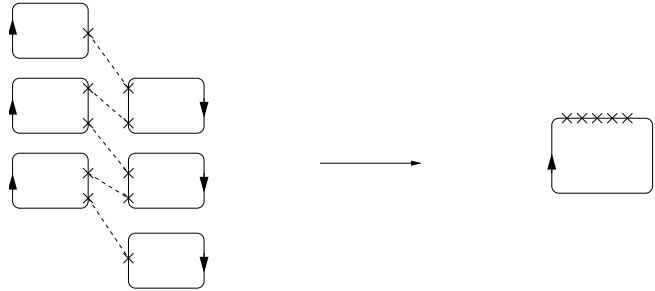} } 
\caption{Diagram for $p_2$  }
\label{fig:1m12m2figpi}
\end{center}
\end{figure}

\begin{figure}[t]
\begin{center}
\resizebox{!}{3.8cm}{
\includegraphics{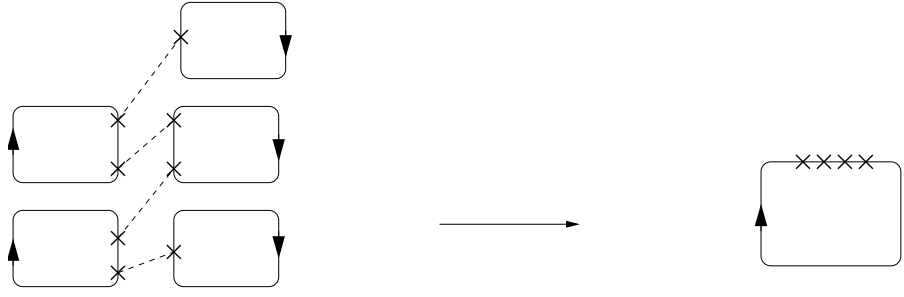} } 
\caption{Diagram for $q_2^+$  }
\label{fig:1m12m2figqiplus}
\end{center}
\end{figure}

Using the diagrammatic method described in section \ref{2mcalculation} 
we  obtain  a general formula
\begin{align}
  & || ( \tr\Phi )^{M_1} (\tr \Phi^2 )^{M_2} ||^2 = \ret
  & (M_2!)^2(M_1!)^2
  2^{2M_2} N^{M_1} \sum_{\{k_i,p_j,q_l^+,q_m^-\}}\frac{N^{2k}
  2^{-k-2q}}{(M_1-2q-p)!}\prod_{i,j,l,m} \frac{ 1}{i^{k_i}
  k_i!}\frac{1}{p_j!} \frac{1}{q_l^+!}\frac{1}{q_m^-!}
  \label{uber12formula}
\end{align}
where the sum is over the sets of non-negative integers
$\{k_i,p_j,q_l^+,q_m^-\}$ satisfying
\begin{itemize} 
\item $k\equiv \sum_i k_i$
\item $p\equiv \sum_i p_i$
\item $q \equiv \sum_i q_i^+ =\sum_i q_i^-$
\end{itemize} 
All the sums above start at $1$. The $k_i$ count diagrams of 
the type encountered in section \ref{2mcalculation}. 
The $p_i$ count diagrams with $i$ 2-cross circles on each side. 
For example, $p_2$ counts diagrams of the type in Figure
 \ref{fig:1m12m2figpi}. 
$q_i^+$ counts diagrams with $i$ 2-cross circles on the left and $(i-1)$
2-cross circles on the right. The diagram in Figure \ref{fig:1m12m2figqiplus}
shows the types of diagrams counted by $q_2^+$. 
$q_i^-$ counts diagrams related to those counted by $q_i^{+}$ by a 
left-right reflection. 
There are also constraints 
\begin{itemize} 
\item $2q+p \leq M_1$
\item $M_2 = \sum_i ik_i + \sum_j jp_j + \sum_l l  ( q_l^+ + q_l^- ) - q $
\end{itemize}

We know that if $L_B \equiv \sum_i ik_i$ and $\l \equiv N^2/2$
\begin{align}
  \sum_{\{k_i\}|\sum_i ik_i = L_B} \l^k L_B! \prod_{i} \frac{ 1}{i^{k_i}
    k_i!} = \frac{(\l+L_B-1)!}{(\l-1)!}
\end{align}
so (\ref{uber12formula}) becomes
\begin{align}
  & || (\tr(\Phi))^{M_1} (\tr(\Phi^2))^{M_2} ||^2 = \ret
  & (M_2!)^2(M_1!)^2
  2^{2M_2} N^{M_1} \sum_{L_B,
    \{p_j,q_l^+,q_m^-\}}\frac{2^{-2q}}{(M_1-2q-p)! L_B!} \frac{(\l+L_B-1)!}{(\l-1)!}\prod_{j,l,m}\frac{1}{p_j!} \frac{1}{q_l^+!}\frac{1}{q_m^-!}
  \label{uber12formulab}
\end{align}
where the sum is over the non-negative integer $L_B$ and the sets of
non-negative integers $\{p_j,q_l^+,q_m^-\}$ satisfying
\begin{itemize}
\item $p\equiv \sum_j p_j$
\item $q \equiv \sum_l q_l^+ =\sum_m q_m^-$
\item $2q+p \leq M_1$
\item $M_2 = \sum_i ik_i + \sum_j jp_j + \sum_l l  ( q_l^+ + q_l^- ) - q $
\end{itemize}

We can simplify this further. Let 
\bea 
&&  Q^{\pm }  = \sum_{i} i q_i^{\pm }  \nnm \\ 
&&  P = \sum_i ip_i \nnm 
\eea 
By using the generating function $ e^{y \over 1 -x } $ we can show that 
the sum over $p_i $ constrained by $p, P $ is given by 
\bea 
\sum_{p_i } { 1 \over \prod p_i! } = { ( P+p -1 ) ! \over p! P ! } \equiv S ( P , p ) 
\eea 
The same sum appears for $ Q^{\pm}$. Hence we can rewrite the norm as 
\bea 
&&  || (\tr(\Phi))^{M_1} (\tr(\Phi^2))^{M_2} ||^2  =  (M_2!)^2(M_1!)^2  2^{2M_2} N^{M_1} \nnm \\ 
&& \sum_{ L_B  , Q^{\pm} , P, p , q } S ( P , p ) S ( Q^{+}  , q )  S ( Q^{-}  , q ) 
\frac{2^{-2q}}{(M_1-2q-p)! L_B!} \frac{(\l+L_B-1)!}{(\l-1)!} \nnm \\ 
\eea 
where $ M_2 = L_B + P +  Q^+ + Q^-  - q $ and $  2q+p \leq M_1$. 

We have checked that the above formula specializes correctly 
to previously derived formulas in the 
cases $M_1 =0 $  ( with $M_2$ arbitrary ) and $M_2=0$ ( with $M_1$ 
arbitrary). In the case $M_1=1$ with  $M_2$ general it gives.  
\begin{align}
   || \tr(\Phi) (\tr(\Phi^2))^{M_2} ||^2 = & (M_2!)^2 2^{2M_2} N
   \sum_{L_B  \leq M_2}\frac{1}{L_B!} \frac{(\l+L_B-1)!}{(\l-1)!} \ret
   =&M_2! 2^{2M_2} N \frac{(\l+M_2)!}{\l!}
 \label{uber12formulagoat}
\end{align}
The case $M_2=2 $ can also be expanded.   
\subsection{More general results}

Some  relevant computations are in Appendix E of \cite{cr}. 
Using these techniques we get, 
for the overlap between an S-giant and multi-KK    
\bea 
&& 
\langle \chi_{[1^L]  } ( \Phi^{ \dagger } )   
 \tr ( \Phi^{c_1} )   \tr ( \Phi^{c_2} )  
 \cdots \tr (  \Phi^{c_k} ) \rangle  \nnm \\ 
&& =  \sum_{R_1, ... R_{k } } 
 g(R_1, R_2 , ..., R_k ; [1^L ] ) f_{[1^{L}]} 
\chi_{R_1} (c_1 ) \chi_{R_2} (c_2 ) ... \chi_{R_k} ( c_k )  \nnm \\ 
&& = (-1)^{c_1 + c_2 ... + c_k - k }   { N! \over (N-L) ! }  \nnm \\ 
&& = (-1)^{ L - k }  { N! \over (N-L) ! }
\eea 
where $ \sum_{i} c_i = L $ by charge conservation. 

For single AdS giant, we have 
\bea 
&& \langle  \chi_{[L]  } ( \Phi^{ \dagger } )    \tr ( \Phi^{c_1} )  
 \tr ( \Phi^{ c_2 } )   ...  \tr (  \Phi^{c_k} ) \rangle  \nnm \\ 
&&    =  \sum_{R_1, ... R_{k } } 
 g(R_1, R_2 , ..., R_k ; [L ] ) f_{[L]} 
\chi_{R_1} (c_1 ) \chi_{R_2} (c_2 ) ... \chi_{R_k} ( c_k )  \nnm \\ 
&& = f_{L} = { ( N+L-1)!  \over (N-1 )! }  \nnm 
\eea

 A special case of interest is
\bea 
 \langle \tr ( \Phi^L ) \chi_R ( \Phi^{\dagger} ) \rangle
\eea 
This is a character of the symmetric $ \chi_R ( \{L \} ) $,
where $ \{ L \} $ is a permutation with a single cycle of length $L$.   
This character is  non-zero only if $R$ is a hook. 
Among the giants, this includes a single sphere giant 
or a single AdS giant, but not two AdS giants or two Sphere giants. 

To compute the above in overlap normalization, we need to look at
correlators of traces.  Using the notation above it is relatively
simple to prove a general formula for the products of traces
\begin{equation}
  \label{eq:generalformula}
  \langle \tr(\Phi^{c_1})\cdots \tr(\Phi^{c_k})\tr(\Phi^{\dagger
    d_1})\cdots \tr(\Phi^{\dagger d_l}) \rangle
\end{equation}
where $\sum_i c_i= \sum_j d_j = n$.  If ${c_i}$ is a cycle of
length $c_i$, e.g. $(12 \dots c_i)$, then
\begin{equation}
  \label{eq:generalformula2}
   \langle \tr(\Phi^{c_1})\cdots \tr(\Phi^{c_k})\tr(\Phi^{\dagger
    d_1})\cdots \tr(\Phi^{\dagger d_l}) \rangle  =
 \sum_R \chi_R \left({c_1}\cdots{c_k}\right) \chi_R
  \left({d_1} \cdots{d_l} \right) f_R
\end{equation}
Now use the same trick on the $\Dim R$ to get
\begin{eqnarray}
  \label{eq:generalformula3}
& &   \langle \tr(\Phi^{c_1})\cdots \tr(\Phi^{c_k})\tr(\Phi^{\dagger
    d_1})\cdots \tr(\Phi^{\dagger d_l}) \rangle \ret
& =&\frac{n!}{ | [ {c_1} \cdots{c_k} ] | }
   \sum_{\s \in [{c_1} \cdots{c_k}
  ]} N^{C\left((  {d_1}\cdots{d_l}     )^{-1}\s\right)}
\end{eqnarray}
It is fairly easy to show that
\begin{equation}
  \label{eq:failruyeasy}
   | [{c_1}  \cdots{c_k} ] |
   = \frac{n!}{1^{l_1}l_1! 2^{l_2}l_2! \cdots m^{l_m}l_m!}
\end{equation}
where $l_1$ of the cycles have length 1, $l_2$ of length 2, etc.  How
can we work this out?  There are a total of $n!$ ways of slotting $n$
things into $n$ boxes.  But some of these configurations will be the
same.  For example, if we mix up the $l_j$ boxes of length $j$ it
won't make any difference, giving us a factor of $l_j!$.  Also in each
box we can cycle round the entries without changing it.  This gives us
$j$ per $l_j$ box.

If we plug this into (\ref{eq:generalformula3}) we get
\begin{eqnarray}
  \label{eq:generalformula4}
& &   \langle \tr(\Phi^{c_1})\cdots \tr(\Phi^{c_k})\tr(\Phi^{\dagger
    d_1})\cdots \tr(\Phi^{\dagger d_l}) \rangle \ret
& =& 1^{l_1}l_1! 2^{l_2}l_2! \cdots m^{l_m}l_m!    \sum_{\s \in [{c_1}
  \cdots {c_k}
  ]} N^{C\left(( {d_1} \cdots {d_l}     )^{-1}\s\right)}
\end{eqnarray}

If $\{c_i\} = \{d_i\}$ (the order doesn't matter) then we can find the
leading term
\begin{eqnarray}
  \label{eq:generalformula5}
& &   \langle \tr(\Phi^{c_1})\cdots \tr(\Phi^{c_k})\tr(\Phi^{\dagger
    c_1})\cdots \tr(\Phi^{\dagger c_k}) \rangle \ret
& =&1^{l_1}l_1! 2^{l_2}l_2! \cdots m^{l_m}l_m!  \left(N^n + O(N^{n-1})\right)
\end{eqnarray}

\section{Conditional probabilities}\label{condprob}  
 Consider operators $ {\cal{O}}_i$ for $ i=1,2,3$ with zero charge. 
 Given the starting state $ \cO_1 $ the joint  probability of 
 getting $ \cO_2 $ and $\cO_3$ as the outgoing states is
 \bea 
P ( \cO_{2} , \cO_3 ) =
{  | \langle \cO_1^{\dagger} \cO_2 \cO_3 \rangle |^2 \over
 \langle \cO_1^{\dagger} \cO_1  \rangle_{G=1}  \langle \cO_2^{\dagger} \cO_2  \rangle
 \langle \cO_3^{\dagger} \cO_3  \rangle }
\eea 
The conditional probability of getting $ \cO_2$ given $ \cO_3$ is
defined as  
\bea 
P ( \cO_2 | \cO_3 ) = { P ( \cO_2 , \cO_3) \over P ( \cO_3 ) } 
\eea 
where the probability of getting $\cO_3$ in the two outgoing states is given 
by 
\bea 
P ( \cO_3 )  =  \sum_i P ( \cO_{i} , \cO_3 ) = \sum_i {  | \langle \cO_1^{\dagger} \cO_i \cO_3 \rangle |^2 \over
 \langle \cO_1^{\dagger} \cO_1  \rangle_{G=1}  \langle \cO_i^{\dagger} \cO_i  \rangle
 \langle \cO_3^{\dagger} \cO_3  \rangle }
\eea 
Similarly 
\bea 
P ( \cO_3 | \cO_2 ) =  { P ( \cO_2 , \cO_3) \over P ( \cO_2 ) } 
\eea 
From these it is clear that the Bayesian rule 
\bea 
{ P ( \cO_3 | \cO_2 ) \over P ( \cO_2 | \cO_3 ) } 
= {  P ( \cO_3 ) \over  P ( \cO_2 ) } 
\eea 
is satisfied.

\section{The Metric,  Euclidean time reversal  and Orientation} \label{euclidconj}


Consider the effect of a change of coordinates $ u = { \bar z }^{-1 }$ 
on the operator $ \partial_z Z  ( z ) $. By the chain rule, 
we get 
\bea 
\partial_u Z ( u ) = - { \bar z }^2 \partial_{\bar z} Z ( \bar z  ) 
\eea 
Note that the effect of this coordinate change is to 
reverse Euclidean time. 

As in the general discussion of the Euclidean adjoint in \cite{Polchinski} (see also section \ref{factprob243}) 
we are supposed to follow the Euclidean time reversal with the usual operation of 
conjugation. In this case, where we are working with complex coordinates, 
we should also complex conjugate the coordinate. This leads to 
\bea 
- z^2 \partial_z Z^*
\eea 
This is exactly what we need to get the desired metric. 
When $Z$ is a matrix, the final complex conjugation on $Z$ is accompanied 
by a matrix transposition. 

Note that the $zw=1$ relation we use in the gluing procedure 
is an orientation preserving map. In the 4d discussion , we 
use an orientation reversing map. The reason why  both 
are acceptable ways of expressing the  Hilbert space inner product
is that the additional coordinate-conjugation of the 2d case 
is an orientation reversing map.

\section{Sphere Factorization}\label{gluesph}

In this section, we will explicitly verify sphere factorization by gluing two $S^2$ correlators to give another
$S^2$ correlator. This allows us to realize, in a very concrete way, CFT factorization.
Denote the two Riemann spheres to be glued by $M$ and $N$. 
$M$ has puncture $P$ located at $z_1=0$ with $z_1$ the local coordinate for a chart containing the puncture. 
$N$ has puncture $Q$ located at $z_2=0$ with $z_2$ the local coordinate for a chart containing the puncture. 
Choose an arbitrary constant $r>1$. 
Assume $z_1$ and $z_2$ are well defined in the disks $|z_1|<r$ and $|z_2|<r$. 
The gluing then has two steps

\begin{itemize}
\item Cut the disks $|z_1|<{1\over r}$ and $|z_2|<{1\over r}$ from $M$ and $N$.

\item Sew $M$ and $N$ together by identifying points on the annulus ${1\over r}<|z_1|<r$ that satisfy
$$ z_1 z_2 =1 $$

\end{itemize}

To apply the CFT factorization equation, we need the inverse of
the product on the space of local operators $\{\cA_i(z,\zb) \}$ 
$$  \cG_{ij} = \braket{i}{j} = \corr{\cA_i^{\dagger \prime} (Q)  \cA_j(P)}_{S^2} $$
which also gives the Hermitian inner product on the set of states, as described in section \ref{sec:innerprod243}. 
Explicitly, we need to evaluate
$$ 
\langle \partial^n Z(P)\partial^m Z^\dagger (Q)\rangle =\lim_{z_1,z_2\to 0}
\langle \partial^n Z(z_1)\partial^m Z^\dagger (z_2)\rangle  $$
First, consider the case that $n\ge m$. 
We perform this calculation in the $z_1$ coordinate. $z_2=0$ corresponds to $z_1=\infty$,
so that the inner product that this correlator computes is the one discussed in section 3.4 of \cite{Ginsparg:1988ui}. 
The simplest case is $m=n=1$. Setting
$$ z_2'={1\over z_2} $$
we have the transformation
$$ \partial Z^\dagger (z_2)\to -(z_2')^2\partial Z^\dagger (z_2')$$
so that
\bea 
\langle \partial Z(P)\partial Z^\dagger (Q)\rangle &&=\lim_{z_1,z_2\to 0}
\langle \partial Z(z_1)\partial Z^\dagger (z_2)\rangle\cr
&&=-\lim_{z_1\to 0}\lim_{z_2'\to\infty} (z_2')^2 \langle \partial Z(z_1)
\partial Z^\dagger (z_2')\rangle\cr
&&=\lim_{z_1\to 0}\lim_{z_2'\to\infty} (z_2')^2 {1\over |z_1-z_2'|^2}\cr 
&&=1 
\eea 
Next, consider $n=2$ and $m=1$
\bea 
\langle \partial^2 Z(P)\partial Z^\dagger (Q)\rangle &&=\lim_{z_1,z_2\to 0}\langle \partial^2 Z(z_1)\partial Z^\dagger (z_2)\rangle\cr 
&&=-\lim_{z_1\to 0}\lim_{z_2'\to\infty} (z_2')^2 \langle \partial^2 Z(z_1)\partial Z^\dagger (z_2')\rangle \cr 
&&=\lim_{z_1\to 0}\lim_{z_2'\to\infty} (z_2')^2 {1\over |z_1-z_2'|^3}\cr 
&&=0 
\eea 
Now set $n=2$ and $m=2$. We have
$$ \partial^2 Z^\dagger (z_2)\to {\partial^2 z_2'\over \partial z_2^2}\partial Z^\dagger (z_2') +\left(
{\partial z_2'\over \partial z_2}\right)^2 \partial^2 Z^\dagger (z_2')
=2(z_2')^3\partial Z^\dagger(z_2') +(z_2')^4 \partial^2 Z^\dagger (z_2')$$
so that
\bea 
\langle \partial^2 Z(P)\partial^2 Z^\dagger (Q)\rangle &&=\lim_{z_1,z_2\to 0}\langle \partial^2 Z(z)\partial^2 
Z^\dagger (z_2)\rangle\cr 
&&=\lim_{z_1\to 0}\lim_{z_2'\to\infty} \left[
2(z_2')^3 \langle \partial^2 Z(z_1)\partial Z^\dagger (z_2')\rangle
+ (z_2')^4 \langle \partial^2 Z(z_1)\partial^2 Z^\dagger (z_2')\rangle\right]\cr 
&& = 2  
\eea
The general result is
\begin{equation}
 \langle \partial^n Z(P)\partial^m Z^\dagger (Q)\rangle = m ((m-1)!)^2 \delta_{m,n}
\end{equation}
Consequently
$$\left[\langle \partial^n Z(P)\partial^m Z^\dagger (Q)
\rangle\right]^{-1} = {1\over m ((m-1)!)^2} \delta_{m,n}$$

This result also follows from the state operator map: using
\bea 
&&  \partial^k Z\leftrightarrow -i (k-1)!\alpha_{-k}  \cr 
&& \partial^k Z^\dagger  \leftrightarrow -i (k-1)!\alpha_{-k}^\dagger \nnm 
\eea 
our general result translates into the identity
\begin{equation}
\langle 0|(i (p-1)!\alpha_p)(-i (k-1)!\alpha_{-k}^{\dagger}  |0\rangle = 
 k\big[ (k-1)!\big]^2\delta_{p,k}
\end{equation}

We are now ready to explicitly verify the
 CFT factorization equation, which reads
\begin{equation}
\langle \partial Z(z_1)\partial Z^\dagger(w_1)\rangle =\sum_{m,n}
\langle \partial Z(z_1)\partial^n Z^\dagger(P)\rangle
\left[\langle \partial^n Z(P)\partial^m Z^\dagger (Q)\rangle\right]^{-1}
\langle \partial^m Z(Q)\partial Z^\dagger (w_1)\rangle 
\end{equation}
Start with the RHS which gives
\bea 
\sum_{m,n}\langle \partial Z(z_1)\partial^n Z^\dagger (P)\rangle
&&\left[\langle \partial^n Z(P)\partial^m Z^\dagger (Q)\rangle\right]^{-1}
\langle \partial^m Z(Q)\partial Z^\dagger (w_1)\rangle \cr 
&&=\sum_{m}\langle \partial Z(z_1)\partial^m Z^\dagger (0)\rangle {1\over m ((m-1)!)^2}
\langle \partial^m Z(0)\partial Z^\dagger (w_1)\rangle\cr 
&&=\sum_m {m!\over z_1^{m+1}}{1\over m ((m-1)!)^2}{m!\over w_1^{m+1}}\cr 
&&={1\over z_1w_1}\sum_m {m\over z_1^m w_1^m } \cr 
&&=-{w_1\over z_1 w_1}
{\partial\over\partial w_1}\sum_m {1\over z_1^m w_1^m } \cr 
&&=-{1\over (1-z_1 w_1)^2} 
\eea 
Now, evaluating the LHS
\bea 
\langle \partial Z(z_1)\partial Z^\dagger (w_1)\rangle &&=-(z')^2\langle \partial Z(z_1)\partial Z^\dagger (z')\rangle\cr 
&&-{(z')^2\over (z'-z_1)^2}\cr 
&&-{1\over w_1^2 \left({1\over w_1}-z_1\right)^2}\cr 
&&=-{1\over (1-z_1 w_1)^2}
\eea 
completing the demonstration.

\section{The Weierstrass elliptic function}\label{appweier}

\subsection{Limits of the Weierstrass elliptic function}

If $x \sim x+ 2T_1 \sim x + 2iT_2$ where $T_1$ and $T_2$ are real then we
can find some limits of the Weierstrass elliptic function.

If $T_2 \to \infty$ then we have
\begin{equation}
  \wp(x) = \left(\frac{\pi}{2T_1}\right)^2 \left(\frac{1}{\sin^2(\pi x/2T_1)}-\frac{1}{3}  \right)
\end{equation}
which for $x = is$ gives
\begin{equation}
  \wp(x) = \left(\frac{\pi}{2T_1}\right)^2 \left( - \frac{1}{\sinh^2(\pi s/2T_1)}-\frac{1}{3}  \right) \label{largeT2limit}
\end{equation}

If $T_2$ is now finite and $x = iT_2$ then we have
\begin{equation}
  \wp(iT_2) = \frac{1}{(iT_2)^2}  +
  \sum_{m,n \in \Z | (m,n) \neq (0,0)} \left\{ \frac{1}{(iT_2+2mT_1+2niT_2)^2}
    -  \frac{1}{(2mT_1+2niT_2)^2} \right\}
\end{equation}
Notice that the first term is the $(m,n) = (0,0)$ term of the first
term in the sum.  Then rewrite the sum for the second term in the sum
\begin{equation}
  \sum_{m,n \in \Z | (m,n) \neq (0,0)} = \sum_{m,n \in \Z | n \neq 0}
  + \sum_{n=0,m \in \Z| m\neq 0}
\end{equation}
so that we get
\begin{align}
  \wp(iT_2)  & = \sum_{m,n \in \Z} \frac{1}{(2mT_1+(2n+1)iT_2)^2}
    -   \sum_{m,n \in \Z |n\neq 0}  \frac{1}{(2mT_1+2niT_2)^2}  - \sum_{m \in \Z |m\neq 0} \frac{1}{(2mT_1)^2}\nn \\
 & =  
  \sum_{m,n \in \Z |n\neq 0}  \frac{1}{(2mT_1+niT_2)^2}
    -  2\sum_{m,n \in \Z |n\neq 0}\frac{1}{(2mT_1+2niT_2)^2}  -\frac{\pi^2}{12T_1^2}\nn \\
 & =\left(\frac{\pi}{2T_1}\right)^2\left( \sum_{n\in \Z |n\neq 0}\left\{ - \textrm{cosech}^2(n\pi
   T_2/2T_1)  +2 \textrm{cosech}^2(n\pi
   T_2/T_1) \right\} -\frac{1}{3}\right) \nn \\
 & =\left(\frac{\pi}{2T_1}\right)^2\left(-4 \sum_{n> 0}\left\{\textrm{coth}(n\pi
   T_2/T_1)\textrm{cosech}(n\pi
   T_2/T_1) \right\} -\frac{1}{3}\right)
\end{align}
which tends \emph{up} to the correct limit (\ref{largeT2limit}) as $T_2 \to
\infty$, i.e. $-\pi^2/(12T_1^2)$.

\subsection{The method of images and the Weierstrass function}

For a torus in $x$ coordinates with $x \sim x+1 \sim x+\tau$ we might
na\"ively try to compute the correlator by the method of images (cf.
\cite{Rabadan:2002wy} where they use this method for correlators on
$S^1$).  This would involve summing the correlators from $x_1$ to each
of the images of $x_2$ on the entire $x$ plane.  If $x= x_1-x_2$ we
would have
\begin{equation}
  \label{naivemoi}
  Z_{T^2}^{-1}\corr{\d Z^\dagger(x_1) \d Z(x_2)}_{G=1,\tau} = -\frac{1}{x^2}  -
  \sum_{m,n \in \Z | (m,n) \neq (0,0)}  \frac{1}{(x+n+m\tau)^2}
\end{equation}

Unfortunately this is divergent.  In order to get a physical quantity
we must regulate this sum by subtracting the divergent part to get the
Weierstrass elliptic function
\begin{align}
  Z_{T^2}^{-1} \corr{\d Z^\dagger(x_1) \d Z(x_2)}_{G=1,\tau} & = -\frac{1}{x^2}  -
  \sum_{m,n \in \Z | (m,n) \neq (0,0)} \left\{ \frac{1}{(x+n+m\tau)^2}
    -  \frac{1}{(n+m\tau)^2} \right\} \nn \\
  & \equiv - \wp(x;\tau).\label{moi}
\end{align}

In our conventions we have for a complex scalar field $Z(x, \ov x)$
\begin{equation}
  Z_{T^2}^{-1} \corr{ Z^\dagger(x_1,\ov x_1)  Z(x_2,\ov x_2)}_{\tau} = G(x_1,\ov x_1;x_2, \ov x_2) =-\ln  \left|\theta_1\left(x_1-x_2 \Big| \tau \right) \right|^2 + \frac{2\pi}{\tau_2}[\text{Im}(x_1-x_2)]^2
\end{equation}
so that
\begin{align}
  Z_{T^2}^{-1} \corr{\d_{x_1} Z^\dagger(x_1) \d_{x_2} Z(x_2)}_{\tau} & = - Z_{T^2}^{-1}
  \d_{x_1}^2\corr{Z^\dagger(x_1) Z(x_2)}_{\tau} \nn \\
  & =  \d_x^2 \left(\log
    \vartheta_{11}\left(x;\tau\right)\right) -
  \frac{2\pi}{\tau_2} \nn \\
  & = -  \wp\left(x;\tau\right)
\label{weierstrasspcorr2}
\end{align}

The divergences in the na\"ive method of images correlator
(\ref{naivemoi}) can be understood to arise because in the mode
decomposition we have included the zero mode.  Removing this zero mode
is equivalent to the regulated correlator in (\ref{moi}).

\section{Windings from torus factorization sums}\label{sec:windings}


In what follows, we assume that the metric on the space of local operators 
has been diagonalized ${\cal G}_{ij}=\langle i|i\rangle\delta_{ij}$.
According to factorization, the two point function of $\partial Z$ on the torus is
\begin{align}
\langle \partial Z^\dagger (p_1)\partial Z(p_2)\rangle_{T^2}= & (q\qb)^{-c/24} \sum_{ij}\sum_{kl} q^{h_j} \ov{q}^{\tilde{h}_j} {\cal G}^{ij}{\cal G}^{kl}
\langle \partial Z^\dagger(p_1){\cal A}'_j(z',\bar{z}'=0){\cal A}_l(z,\bar{z}=0)\rangle_{S^2} \nn \\
& \times \langle {\cal A}_k^\dagger(w,\bar{w}=0){{\cal A}_i^{\dagger \prime}} (w',\bar{w}'=0)\partial Z(p_2)\rangle_{S^2} 
\end{align}
To demonstrate how this sum gives the right correlation function on the torus, it is instructive to analyze a few terms explicitly.

For a correlation function to be non-zero in the free field theory, we need to consider 
the correlator of an even number of fields. One type of term which enters is when 
${\cal A}'_j(z',\bar{z}'=0)={{\cal A}_i^{\dagger \prime}}(w',\bar{w}'=0)=1$ 
and ${\cal A}_l(z,\bar{z}=0)=\partial^l Z (z,\bar{z}=0)$, 
${{\cal A}_k^\dagger}(w,\bar{w}=0)=\partial^k Z^\dagger(w,\bar{w}=0)$. Summing over operators of this type, 
we have (the double sum collapses since 
$\langle \partial^k Z^\dagger (w,\bar{w}=0)\partial^l Z(z,\bar{z}=0)\rangle_{S^2}\propto\delta^{lk}$)
\begin{equation}
(q\qb)^{-c/24}\sum_{l}
\langle \partial Z^\dagger(p_1)\partial^l Z(z,\bar{z}=0)\rangle_{S^2}
{1\over\langle l|l\rangle}\langle \partial^l Z^\dagger(w,\bar{w}=0)\partial Z(p_2)\rangle_{S^2}
\end{equation}
which, explicit evaluation shows, is the correlator to go from $p_1$ to $p_2$ 
passing through the point $z=\bar{z}=0$.
Another type of term which enters is when ${\cal A}_l(z,\bar{z}=0)={\cal A}_k^\dagger (w,\bar{w}=0)=1$ and 
${\cal A}'_j(z',\bar{z}'=0)=\partial^j Z' (z'=\bar{z}'=0)$ and 
${\cal A}_i^{\dagger \prime}(w',\bar{w}'=0)=\partial^i Z^{\dagger \prime} (w',\bar{w}'=0)$. Summing over operators of this type, we have
\begin{equation}
(q\qb)^{-c/24}\sum_{i}q^{h_i} \ov{q}^{\tilde{h}_i}
\langle \partial Z^\dagger (p_1)\partial^i Z' (z'=\bar{z}'=0)\rangle
{1\over\langle i|i\rangle}\langle \partial^i Z^{\dagger \prime}(w',\bar{w}'=0)\partial Z(p_2)\rangle
\end{equation}
which is the correlator to go from $p_1$ to $p_2$ passing through point $z'=\bar{z}'=0$. 

Next, consider the contribution when we sum over the terms 
${\cal A}'_j(z',\bar{z}'=0)=\partial^j Z^{\dagger \prime} (z',\bar{z}'=0)$, 
${\cal A}_i^{\dagger \prime}(w',\bar{w}'=0)=\partial^i Z^{ \prime} (w',\bar{w}'=0)$ and
${\cal A}_l(z,\bar{z}=0)=:\partial^{n_1}Z(z,\bar{z}=0)\partial^{n_2}Z(z,\bar{z}=0):$, 
${\cal A}_k^\dagger(w,\bar{w}=0)=:\partial^{m_1} Z^\dagger(w,\bar{w}=0)\partial^{m_2} Z^\dagger(w,\bar{w}=0):$
with $n_1\ge n_2$ and $m_1\ge m_2$ to avoid over counting. We make use of the fact that
in the free field theory expectations of a product of $2n$ operators factorize into sums over $n$ products of expectations 
of pairs of operators (Wick's theorem). Among the terms that appear, we obtain
\begin{align}
&  {\langle \partial Z^\dagger(p_1)\partial^{n_1} Z(z,\bar{z}=0) \rangle_{S^2} 
\; \langle \partial^{n_1} Z^\dagger(w,\bar{w}=0)\partial^i Z'(w',\bar{w}'=0) \rangle_{S^2}\over
\langle \partial^{n_1} Z^\dagger(w,\bar{w}=0)\partial^{n_1} Z(z,\bar{z}=0) \rangle_{S^2}} & \nn \\
&  \times {1\over \langle \partial^i Z^{\dagger \prime}(z',\bar{z}'=0)\partial^{i} Z'(w',\bar{w}'=0) \rangle_{S^2} } & \nn \\
&  \times { \langle \partial^{i} Z^{\dagger \prime}(z',\bar{z}'=0) \partial^{n_2} Z(z,\bar{z}=0) \rangle_{S^2}\;
 \langle \partial^{n_2} Z^\dagger(w,\bar{w}=0)\partial Z(p_2) \rangle_{S^2}
\over
\langle \partial^{n_2} Z^\dagger (w,\bar{w}=0)\partial^{n_2} Z(z,\bar{z}=0) \rangle_{S^2} } & 
\end{align}
To interpret this expression note that the first factor (after summing on $n_1$) gives the 
amplitude to go from $p_1$ to $w'=\bar{w}'=0$, passing through
$z=\bar{z}=0$; the last factor (after summing on $n_2$) gives the amplitude to go from 
$z'=\bar{z}'=0$ to $p_2$ passing through $z=\bar{z}=0$. Finally, after
summing on $i$ we get the amplitude to go from $p_1$ to $p_2$ along a path with winding number 1. 

Terms for which ${\cal A}'_j(z',\bar{z}'=0)$ has $n$ operators $\partial Z^\dagger$ appearing and
${\cal A}_l(z,\bar{z}=0)$ has $n+1$ operators $\partial Z$ appearing give the amplitudes
with winding number $n$; terms for which ${\cal A}'_j(z',\bar{z}'=0)$ 
has $n+1$ operators $\partial Z$ appearing and 
${\cal A}_l(z,\bar{z}=0)$ has $n$ operators $\partial Z^\dagger$ appearing 
give the amplitudes with winding number $-n$.

\section{Some results with correct normalizations}\label{correctnorm}

Here we give the calculations for Section \ref{sec:correctresults}.

\subsection{Sphere factorization}

We want to work out
\begin{align}
  & P\left(R_1(r=e^x, \O_i), R_2(r=e^x,\O_i) \to R(r=0)\right)\nn \\
  & = \frac{\left|\big\langle R_1^\dagger(r=e^x, \O_i)R_2^\dagger(r=e^x, \O_i)
      R(r=0) \big\rangle \right|^2}{\big\langle
    R_2^\dagger(s=e^x, \O_i')R_1^\dagger(s=e^x, \O_i')R_1(r=e^x, \O_i)R_2(r=e^x, \O_i)\big\rangle\big\langle
    R^\dagger R\big\rangle}
\end{align}
for sphere and AdS giants.

For two sphere giant $[1^{N/2}]$ combining into another sphere
giant $[1^N]$
\begin{align}
  & \frac{\left|\big\langle [1^{N/2}]^\dagger(r=e^x)[1^{N/2}]^\dagger(r=e^x)
      [1^N](r=0) \big\rangle \right|^2}{\big\langle
    [1^{N/2}]^\dagger(s=e^x)[1^{N/2}]^\dagger(s=e^x)[1^{N/2}](r=e^x)[1^{N/2}](r=e^x)\big\rangle\big\langle
    [1^N]^\dagger [1^N]\big\rangle} \nn \\
& =  \frac{g\left([1^{N/2}],
  [1^{N/2}];[1^N]\right)^2f_{[1^N]}^2e^{-4Nx}}{\sum_S g\left([1^{N/2}],
  [1^{N/2}];S\right)^2 f_Se^{-2Nx}(e^x - e^{-x})^{-2N} f_{[1^N]}  } 
\end{align}
where $g\left([1^{N/2}], [1^{N/2}];S\right)$ is a
Littlewood-Richardson coefficient.  In the large $x$ limit we get
\begin{align}
  P([1^{N/2}], [1^{N/2}] \to [1^N]) = \frac{f_{[1^N]}}{\sum_S g\left([1^{N/2}],
  [1^{N/2}];S\right)^2 f_S   } < 1
\end{align}
The fusion of the two vertical Young diagrams gives a sum of 
representations, with column lengths   $  ( N/2 + i , N/2 -i )$. 
Hence the denominator can be written as 
\bea 
\sum_{ i = 0}^{N/2 }  { N! ( N+1)!  \over  ( N/2 -i )! ( N/2 + i + 1 ) ! } 
\eea     
 The probability is less than one because $f_{[1^N]}$ is
included in the sum. 
A similar formula can be written 
for two AdS giants $[N/2]$ combining into another AdS
giant $[N]$. Now the denominator becomes 
\bea 
\sum_{i=0}^{N/2} {  ( 3N/2  + i -1 ) ! ( 3N/2 -i -2 ) ! \over ( N-1) ! ( N-2) ! } 
\eea

\subsection{$G=1$ factorization}

We want to work out
\begin{align}
 & P\left(R(r=e^x, \O_i) \to R_1'(r'=0) R_2(r=0)\right)\nn \\
 & = e^{-2T\Delta_1}\frac{\left|\big\langle
       R^\dagger(r=e^x, \O_i) R_1'(r'=0)R_2(r=0) \big\rangle
    \right|^2}{\big\langle
    R^\dagger (s=e^x, \O_i) R(r=e^x, \O_i)\big\rangle_{G=1}  \big\langle
    R_1^{\dagger}R_1\big\rangle\big\langle R_2^\dagger R_2\big\rangle}
\end{align}
in the large $T$ limit.  Here $\sum_{k \geq 1} k \textrm{cosech}(kT)
\sim 2e^{-T}$.  We will calculate the probability for $R$ at
$r=e^{T/2}$, which will maximize the distance of the insertion of $R$
from the boundaries of the cut $S^4$.

For the transition of an AdS giant $[N]$ into to two smaller AdS giants
$[N/2]$
\begin{align}
 & e^{-2T\Delta_1}  \frac{\left|\big\langle
       [N]^\dagger(r=e^{T/2}) [N/2]'(r'=0)[N/2](r=0) \big\rangle
    \right|^2}{\big\langle
    [N]^\dagger (s=e^{T/2}) [N](r=e^{T/2})\big\rangle_{G=1}  \big\langle
    [N/2]^{\dagger}[N/2]\big\rangle\big\langle [N/2]^\dagger
    [N/2]\big\rangle} \nn \\
& \sim \frac{g([N/2],[N/2];[N])^2 f_{[N]}^2  e^{-TN}
  e^{-2N(T/2)}}{e^{-2N(T/2)}(2e^{-T})^Nf_{[N]}f_{[N/2]}^2} \nn \\
& = \frac{1}{2^N} \frac{f_{[N]}}{f_{[N/2]}^2}  
  \qquad  ~~~~ 
 = \frac{1}{2^N} \frac{(2N-1)!(N-1)!}{\left((3N/2-1)!\right)^2} \nn\\
& = \frac{1}{2^N}\frac{9}{8} \frac{(2N)!N!}{\left((3N/2)!\right)^2}
\sim \frac{1}{2^N}\frac{9}{8} \frac{\sqrt{4\pi N}(2N)^{2N}
  e^{-2N}\sqrt{2\pi N}N^{N} e^{-N}}{3\pi N (3N/2)^{3N}e^{-3N}  } \nn
\\
& = \frac{3}{\sqrt{8}}  \left(\frac{16}{27}\right)^N
\end{align}

For a sphere giant $[1^N]$ evolving into two smaller sphere giants
$[1^{\frac{N}{2}}]$ we get
\begin{align}
& \sim \frac{1}{2^N} \frac{f_{[1^N]}}{f_{[1^{N/2}]}^2} 
 = \frac{1}{2^N} \frac{((N/2)!)^2}{N!} \nn\\
& \sim \frac{1}{2^N}\frac{\pi N (N/2)^N e^{-N}}{\sqrt{2\pi N}N^{N}
  e^{-N}}  = \sqrt{\frac{\pi N}{2}} \frac{1}{2^{2N}}
\end{align}

\subsubsection{Giants to KK gravitons}\label{dualbasisec}

Here we must modify our factorization equations because the trace
basis is not a diagonal basis.  Fortunately there is a dual basis to
the trace basis which we shall call the null basis in line with its
use in \cite{deMelloKoch}.  A fuller explanation of this null basis
will be given in a forthcoming paper.

To start with we will only be concerned with the index structure of
the correlators.
Define a set of elements $\s_i$ in the permutation group $S_n$ where
each $\s_i$ is an element of a different conjugacy class of $S_n$,
$i=1, \dots p(n)$ where $p(n)$ is the number of partitions of $n$.

The \emph{trace basis} is given by the $p(n)$ operators
\begin{equation}
  \tr( \s_i \Phi)=\sum_{R(n)}\chi_R(\s_i) \chi_R(\Phi)
\end{equation}
Define the $p(n)$ elements of the \emph{null basis} by
\begin{equation}
  \xi_i(\Phi):= \frac{|[\s_i]|}{n!} \sum_{R(n)} \frac{1}{f_R}\chi_R(\s_i) \chi_R(\Phi)
\end{equation}
where $|[\s_i]|$ is the size of the conjugacy class of $\s_i$.
This basis is useful because it is dual to the trace basis. The matrix
of correlators of the null basis is the inverse of the matrix of
correlators of the trace basis.  To prove this we work out
\begin{equation}
\langle\xi_i(\Phi^\dagger)\xi_j(\Phi)
   \rangle = \frac{|[\s_i]|}{n!}\frac{|[\s_j]|}{n!} \sum_{R(n)} \frac{1}{f_R}\chi_R(\s_i) \chi_R(\s_j)
\end{equation}
and
\begin{equation}
\langle \tr(\s_j \Phi^\dagger)\tr(\s_k \Phi)
   \rangle =\sum_{S(n)}f_S\chi_S(\s_j) \chi_S(\s_k)
\end{equation}
If we sum $\sum_j$ over conjugacy classes of $S_n$ we get
\begin{align}
 &  \sum_j \langle\xi_i(\Phi^\dagger)\xi_j(\Phi)
   \rangle \langle \tr(\s_j \Phi^\dagger)\tr(\s_k \Phi)
   \rangle \nn  \\
& = \sum_j \sum_R \sum_S \frac{|[\s_i]|}{n!}\frac{|[\s_j]|}{n!} \frac{1}{f_R}\chi_R(\s_i) \chi_R(\s_j) f_S\chi_S(\s_j) \chi_S(\s_k) \nn \\
& = \sum_R  \frac{|[\s_i]|}{n!}\chi_R(\s_i)  \chi_R(\s_k) \nn \\
& = \delta_{ik}
\end{align}
using the orthogonality properties of the characters of $S_n$ (see Appendix \ref{sec:idnotcon}).  The
null basis is dual to the trace basis
\begin{equation}
  \sum_j \langle\xi_i(\Phi^\dagger)\xi_j(\Phi)
   \rangle  \tr(\s_j \Phi) = \xi_i(\Phi)
\end{equation}

Now schematically (dropping spacetime dependence and modular
parameters) the genus 1 factorization we are interested in is
\begin{align}
  \corr{R^\dagger R}_{G=1} & = \sum_{ij} \sum_{kl} \cG^{ij} \cG^{kl}
  \corr{R^\dagger \tr(\s_j \Phi) \tr(\s_l \Phi)}\corr{ \tr(\s_k \Phi^\dagger) \tr(\s_i
  \Phi^\dagger) R} \nn\\
& = \sum_{i} \sum_{k}
  \corr{R^\dagger \tr(\s_i \Phi) \tr(\s_k \Phi)}\corr{ \xi_k(\Phi^\dagger) \xi_i(\Phi^\dagger) R}
\end{align}
using the fact that $ \cG^{ij} = \langle\xi_i^\dagger \xi_j \rangle$.
Thus the probability of a transition to KK gravitons is given by
\begin{equation}
  P\left(R \to \tr(\s_i \Phi),\tr(\s_k \Phi)\right) = \frac{\corr{R^\dagger \tr(\s_i \Phi) \tr(\s_k \Phi)}\corr{ \xi_k(\Phi^\dagger) \xi_i(\Phi^\dagger) R}}{\corr{R^\dagger R}_{G=1}}
\end{equation}

Now we shall do the computation for the transition of an AdS giant to two
Kaluza-Klein gravitons.  We will drop the spacetime dependences and
add them in at the end.  First we must work out the two three point
functions
\begin{equation}
\corr{[N]^\dagger \tr( \Phi^{\frac{N}{2}}) \tr(\Phi^{\frac{N}{2}})} = \sum_{R_1,R_2}
\chi_{R_1}(N/2) \chi_{R_2}(N/2) \corr{[N]^\dagger R_1 R_2}
\end{equation}
where $(N/2)$ is understood to be a cycle of length $N/2$.  Since
$[N]$ can only be made from other single-row representations, the only
representations in the sum that contribute are the AdS giants.  We get
\begin{equation}
\corr{[N]^\dagger \tr(\Phi^{\frac{N}{2}}) \tr(\Phi^{\frac{N}{2}})} = f_{[N]}
\end{equation}
Similarly
\begin{align}
&\corr{ \xi_{(N/2)}(\Phi^\dagger) \xi_{(N/2)}(\Phi^\dagger) R}\nn\\
& =
\sum_{R_1,R_2} \frac{|[(N/2)]|}{(N/2)!}\frac{|[(N/2)]|}{(N/2)!}  \frac{1}{f_{R_1}} \frac{1}{f_{R_2}}
\chi_{R_1}(N/2) \chi_{R_2}(N/2) \corr{[N]^\dagger R_1 R_2} \nn \\
& =
\frac{4}{N^2}  \frac{f_{[N]}}{f_{[N/2]}^2}
\end{align}
where we have used $|[(N/2)]| = (N/2 -1)!$.  Thus, adding back in the
spacetime dependencies, we have
\begin{equation}
   P\left([N](r=e^x)\to \tr(\Phi^{\frac{N}{2}})(r'=0) \tr(\Phi^{\frac{N}{2}})(r=0)\right)
    \sim \frac{4}{N^2}\frac{3}{\sqrt{8}}  \left(\frac{16}{27}\right)^{N}
\end{equation}
and
\begin{equation}
   P\left([1^N](r=e^x)\to \tr(\Phi^{\frac{N}{2}})(r'=0) \tr(\Phi^{\frac{N}{2}})(r=0)\right)
    \sim \frac{4}{N^2}\sqrt{\frac{\pi N}{2}} \left(\frac{1}{2}\right)^{2N}
\end{equation}

\subsection{Higher genus factorization}

For the transition of an AdS giant into $n$ smaller AdS giants, 
using the guess involving $k_n $ from section \ref{sec:correctresults}, we have 
\begin{align}
P([N] \to n \times [N/n])  & =  \frac{1}{k_n^N} \frac{ f_{[N]}
}{ f_{[N/n]}^n }\nn \\
& =  \frac{1}{k_n^N}   \frac{(2N-1)!}{(N-1)!}\left(\frac{(N-1)!}{(N + N/n -1)!}  \right)^n \nn \\
& \sim  \frac{1}{k_n^N}\frac{1}{\sqrt{2}}  \left[\frac{(n+1)}{n}  \right]^{\frac{n}{2}} \frac{(2N)^{2N}e^{-2N} N^{N(n-1)}e^{-N(n-1)}}{\left[\frac{(n+1)}{n}  \right]^{N(n+1)}N^{N(n+1)}  e^{-N(n+1)}}\nn \\
 & =  \frac{1}{\sqrt{2}}  \left[\frac{(n+1)}{n}  \right]^{\frac{n}{2}} \left[\frac{4n^{n+1}}{k_n (n+1)^{n+1}}  \right]^{N}  
\end{align}
in the large $N$ limit.

For the transition of a Schur polynomial operator to
KK gravitons we find in general
\begin{align}
P(R \to \tr(\s_{i_1}\Phi), \dots \tr(\s_{i_n}\Phi))  & = \frac{\corr{R^\dagger \tr(\s_{i_1} \Phi) \cdots \tr(\s_{i_n} \Phi)} \corr{ \xi_{i_n}(\Phi^\dagger) \xi_{i_1}(\Phi^\dagger) R}}{\corr{R^\dagger R}_{G=n-1}} \nn  \\
& = \sum_{R_{1}, \dots R_{n}} \chi_{R_1}(\s_{i_1}) \cdots \chi_{R_n}(\s_{i_n})g(R_1, \dots R_n;R)f_R \nn \\
& \times \sum_{S_{1}, \dots S_{n}} \chi_{S_1}(\s_{i_1}) \cdots \chi_{S_n}(\s_{i_n})g(S_1, \dots S_n;R)f_R \nn \\
& \times \frac{1}{f_{S_1} \cdots f_{S_n}}\frac{|[\s_{i_1}]|}{\Delta_1!} \cdots \frac{|[\s_{i_n}]|}{\Delta_n!} \frac{1}{f_R k_n^{\Delta_R}}
\end{align}

Fortunately this simplifies dramatically if $R$ is an $AdS$ (or
sphere) giant, since, by the Littlewood-Richardson rules, an $AdS$ (or
sphere) giant can only be made from other single row (or column)
representations.  Further if $\s_{i_j}$ are single length $\Delta_j$ cycles
for $j=1, \dots n$ we know that $\chi_{[\Delta_j]}(\Delta_j) = 1$ ($\pm 1$
for sphere) and also that $|[(\Delta_j)]| = (\Delta_j -1)!$.  Thus we get
\begin{align}
P([\Delta_R] \to \tr(\Phi^{\Delta_1}), \dots \tr(\Phi^{\Delta_n}))  & =
 \frac{1}{k_n^{\Delta_R}}  \frac{1}{\Delta_1 \cdots \Delta_n} 
\frac{f_{[\Delta_R]}}{f_{[\Delta_1]} \cdots f_{[\Delta_n]} } \nn \\
P([1^{\Delta_R}] \to \tr(\Phi^{\Delta_1}), \dots \tr(\Phi^{\Delta_n}))  & =
 \frac{1}{k_n^{\Delta_R}}  \frac{1}{\Delta_1 \cdots \Delta_n}
 \frac{f_{[1^{\Delta_R}]}}{f_{[1^{\Delta_1}]}
 \cdots f_{[1^{\Delta_n}]} }
\end{align}

\section{Topology for 5D bulk and 4D boundary}

\subsection{Complements of graph neighborhoods in $B^5$ and  wedge sum of 
spheres  }\label{sec:shortexact}

Take a ball $B^4$ and remove $n$ $B^4_{\circ}$ balls from its
interior. Call the resulting surface $X$.  $X$ is homotopic to an
$n$-wedge of 3-spheres, $\vee_n S^3$, for which we know $H_3(\vee_n
S^3) = \Z^n$ ( page 126 of \cite{hatcher} ).

Now quotient $X$ by the outer $S^3$ boundary of the original $B^4$.
$X/S^3$ is an $S^4$ with $n$ $B^4_{\circ}$ balls removed, $\d^{(i)}
N(V_n, B^5)$, which is homotopic to the complement of a connected
graph.  There is an exact sequence ( page 114 of \cite{hatcher} )
\begin{align}
  \cdots \to H_4(X/S^3) \stackrel{\d}{\to} H_{3}(S^3)
  \stackrel{i_\star}{\to} H_{3}(X) \stackrel{j_\star}{\to} H_3(X/S^3)
  \stackrel{\d}{\to} H_{2}(S^3) \to \cdots
\end{align}
where $i_\star$ is induced from the inclusion map on the chain group
$C_3(S^3) \stackrel{i}{\to}C_3(X)$ and $j_\star$ is induced from the
quotient map $C_3(X) \stackrel{j}{\to}C_3 (X/S^3)$.

We know that $H_4(X/S^3)=\{0\}$ since there are no boundaryless chains
in $C_4(X/S^3)$.  We also have $H_3(A) = H_3(S^3) =\Z$, $H_3(X) =
H_3(\vee_n S^3) = \Z^n$ and $H_2(A) = H_2(S^3) =\{0\}$.  Thus we get a
short exact sequence
\begin{align}
  \cdots \to 0\stackrel{\d}{\to} \Z \stackrel{i_\star}{\to}
  \Z^n\stackrel{j_\star}{\to} H_3(\d^{(i)} N(V_n, B^5) )
  \stackrel{\d}{\to} 0 \to \cdots
\end{align}
Because this is a short exact sequence $i_\star$ is an injection,
$j_\star$ is a surjection, and $\textrm{im}\; i_\star = \textrm{ker}\;
j_\star$.  Hence, by the first isomorphism theorem on the map
$j_\star$, $H_3(\d^{(i)} N(V_n, B^5) ) = \textrm{im}\; j_\star \cong
\Z^n/\textrm{ker}\; j_\star = \Z^n/\Z = \Z^{n-1}$.

\subsection{Cell complexes}\label{sec:celldecompdef}

The easiest way to compute the homology groups of $\S_4(n-1)$ is in
terms of its cell complex decomposition.  A $k$-cell is an \emph{open}
$k$-dimensional ball.  We can build a manifold from cells by starting
with 0-cells, i.e. a set of points, and inductively attaching cells of
higher dimension.  We attach the cells along the boundary of the
cells.  A $k$-cell has boundary $S^{k-1}$.  An attaching map
identifies this boundary with some submanifold of the manifold to
which we are attaching the cell, even if that submanifold is of a
lower dimension.  For example, we can attach a $2$-cell (an open disk)
to a 0-cell by identifying the boundary of the $2$-cell, a circle,
with the 0-cell.  This gives us a cell decomposition of the sphere
$S^2$. A more formal description follows. 

A \emph{cell complex} or \emph{CW complex} is a space $X$ constructed
in the following way
\begin{enumerate}
\item Start with a discrete set $X^0$, the 0-cells of $X$.
\item Inductively, form the $k$-skeleton $X^k$ from $X^{k-1}$ by
  attaching $k$-cells $e^k_\a$. Do this via maps $\varphi_\a :S^{k-1}
  \to X^{k-1}$ where $S^{k-1}$ is the boundary of the $k$-ball.  This
  means that $X^k$ is the quotient space of $X^{k-1}\sqcup B^k_\a$
  ($B^k_\a$ is a closed $k$-ball here) under the identifications $x
  \sim \varphi_\a(x)$ for $ x \in \d B^k_\a$.  The cell $e_\a^k$ is the
  homeomorphic image of the open disk $B_\a^k - \d B^k_\a$ under the
  quotient map.

  This is different to handlebody attachment because the boundary of
  the cell can be attached to any part of $X^{k-1}$, even one of
  dimension lower than $k-1$.
\item $X=X^d$ for some $d$ if $X$ is finite-dimensional.
\end{enumerate}

The cell decomposition is virtually identical to the handlebody
decomposition because a $k$-handle can be viewed as a thickening 
of a $k$-cell.

Once we have a cell decomposition of a manifold we can compute the
homology using a boundary map on the cells.  Let $\{e_\a^k\}$ be the
$k$-cells that we are attaching to the lower-dimensional manifold
$X^{k-1}$.  The cellular boundary map $d_k$ can be computed in terms
of degrees. $d_k(e_\a^k) = \sum_\b d_{\a\b} e_\b^{k-1}$ where
$d_{\a\b}$ is the degree of the map $S_\a^{k-1} \to X^{k-1}\to
S_\b^{k-1}$ that is the composition of the attaching map of $e_\a^k$
with the quotient map collapsing $X^{k-1} - e_\b^{k-1}$ to a point.
If the boundary of $e_\a^k$ is identified with a submanifold of
dimension $k-2$ or lower then $d_k(e_\a^k)=0$. Using this boundary map
we can then compute the homology in the standard way.

We will give a cell decomposition of our spaces so that we can
compute the homology groups.

\subsection{Cell decomposition and homology for the complement of a connected graph}\label{sec:celldecompositioncomplementconnected}

Since $\overline{B^5 \setminus N(V_n, B^5)}$ is homeomorphic to a
thickening of the internal surface $\d^{(i)} N(V_n, B^5) \times
B^1$ it is also of the same homotopy type as the internal surface
itself $\d^{(i)} N(V_n, B^5)$.  The internal surface is in fact a
deformation retract of $\overline{B^5 \setminus N(V_n, B^5)}$ (a
subspace $R$ of a manifold $X$ is a \emph{retract} of $X$ if there is
a continuous map $f: X \to R$ such that $f|_R = \id_R$; if $\id_X$ and
$f$ are homotopic then $R$ is a \emph{deformation retract} of $X$).

The cell decomposition of the $4$-dimensional space $\d^{(i)}
N(V_n,B^5)$ is as follows.
\begin{itemize}
\item Take a single 0-cell.
\item Attach to it $n$ $3$-cells, with a map which identifies the
  boundaries of the $3$-cells ($S^2$s) with the 0-cell.  The space is
  now a \emph{wedge sum} of $n$ spheres $S^3$, i.e. it is the disjoint
  union of $n$ spheres $S^3$ with a point on each sphere identified to
  a single point.  The wedge sum of $n$ spheres is sometimes written
  $\vee_n S^3$.
\item Now wrap a $4$-cell around the $n$ $3$-cells.  The $4$-cell has
  a boundary of $S^3$.  To glue this to the $n$ $S^3$s divide it up
  into $n$ pieces and glue each $n$th around each $3$-sphere
  sequentially.
\end{itemize}

\begin{figure}[h]
\begin{center}
\includegraphics{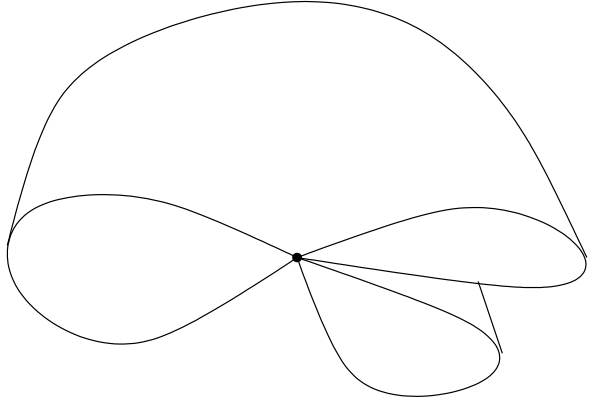}
\caption{The cell decomposition of the pants diagram for $d=3$,
  $n=3$. It has a 0-cell at the center, 3 1-cells attached to the
  0-cell to form the wedge sum of 3 circles $S^1$ and a 2-cell with
  $S^1$ boundary wrapping the 3 circles.}
\label{fig:pants}
\end{center}
\end{figure}

For example for a 3-dimensional bulk we obtain the pants diagram by
taking a point and attaching three open intervals to it.  We get three
circles attached at a single point.  Then we attach an open 2-ball
with its $S^1$ boundary going around each of the three circles,
pinching around the 0-cell.  See Figure \ref{fig:pants}.

Note that for a connected Witten graph $\d^{(i)} N(V_n,B^5)$ is
the same as the $n$-punctured $4$-sphere on which we do our CFT.  The
$n$ spheres $S^3$ are the same as the boundaries of the $4$-balls we
cut out around each puncture.

For the set-up in general dimension see Appendix \ref{celldecompgend}.

Now that we have the cell decomposition of $\d^{(i)} N(V_n,B^5)$ we
can compute the homology groups using our cellular boundary map,
$d_k(e_\a^k) = \sum_\b d_{\a\b} e_\b^{k-1}$ where $d_{\a\b}$ is the
degree of the attaching map.

The homology groups for the complement of a connected Witten graph are
\begin{itemize}
\item $H_4(\d^{(i)} N(V_n,B^5)) = \{0\}$ since the only 4-cell $e^4$
  has a boundary, so $\ker\;d_4 = \{0\}$.
\item $H_3(\d^{(i)} N(V_n,B^5)) = \Z^{n-1}$.  Let $e^3_\a$ for $\a =
  1, \dots n$ be the $n$ 3-cells.  The image of $d_4$ is spanned by
  $d_4 (e^4) = \sum_\a e^3_\a$ since the boundary of $e^4$ wraps the
  3-cells sequentially.  The kernel of $d_3$ is spanned by the $n$
  3-cells $e^3_\a$, since their boundaries are identified to a point.
  Thus $H_3 = \ker \; d_3/\im \; d_4 \cong \Z^{n-1}$.
\item $H_2(\d^{(i)} N(V_n,B^5)) = \{0\}$ since there are no 2-cells.
\item $H_1(\d^{(i)} N(V_n,B^5)) = \{0\}$ since there are no 1-cells.
\item $H_0(\d^{(i)} N(V_n,B^5)) = \Z$ since it is arcwise connected.
\end{itemize}

The only non-trivial homology group, $H_3(\d^{(i)} N(V_n,B^5))$, can
also be computed via a short exact sequence of homology groups (see
Appendix \ref{sec:shortexact}).

These homology groups satisfy the weak Morse inequalities.  If $c_k$
is the number of $k$-cells and $b_k$ is the $k$th Betti number then
\begin{equation}
c_k \geq b_k
\end{equation}
for all $k$.

\subsection{Cell decomposition  and homology 
for the complement of a disconnected  graph}\label{sec:celldecompositioncomplementdisconnected}

Suppose a Witten graph $G$ is composed of $m$ disconnected
components $G = V_{n_1} \sqcup V_{n_2} \sqcup \dots V_{n_m}$.

$\overline{B^5 \setminus N(G, B^5)}$ is homotopic to the $m$ connected
spaces $\d^{(i)} N(V_{n_i},B^5)$ daisy-chained together in a line
by 1-cells. For the cell decomposition
\begin{itemize}
\item Take $m$ 0-cells.
\item Link them in a line by $(m-1)$ 1-cells.  The ends of each 1-cell
  attach to different 0-cells.
\item Attach $n_i$ $3$-cells to each 0-cell for $1\leq i \leq m$
  as above.  We have $m$ wedge sums of $3$-spheres linked together
  in a line by 1-cells.
\item Now wrap a $4$-cell around each collection of $n_i$
  $3$-cells.
\end{itemize}

$\overline{B^5 \setminus N(G, B^5)}$ glued to itself is the same except we
attach a further $(m-1)$ 1-cells, each of which has both ends attached
to the same 0-cell.

$\overline{B^5 \setminus N(G, B^5)}$ has most of the same homology
groups as $\overline{B^5 \setminus N(V_n, B^5)}$ except that now
\begin{equation}
  H_3(\overline{B^5 \setminus N(G, B^5)}) = \Z^{\sum_j n_j - m}
\end{equation}
The first homology group is unchanged because there are no closed
loops from the 1-cells.

$\overline{B^5 \setminus N(G, B^5)}$ glued to itself is a different story.
$H_0$ and $H_i$ for $i >1$ are the same as $\overline{B^5 \setminus N(G,
B^5)}$.

Each of the $(m-1)$ 1-cell loops is a 1-cycle which is not the
boundary of some 2-chain.  Thus it increases the number of free
Abelian generators of $H_1$ by ${m-1}$
\begin{equation}
  H_1((\overline{B^5 \setminus N(G, B^5)})\cup |_{\partial^{(i)} N } 
(\overline{B^5 \setminus N(G, B^5)}))
  = \Z^{m-1}
\end{equation}

\subsection{Cell decomposition and
 homology for $\S_4(G)$}\label{sec:cellboundary}

We want to find the cell decomposition of the 4-dimensional manifold
with genus $G$, $\S_4(G)$.

\begin{figure}[h]
\begin{center}
\includegraphics{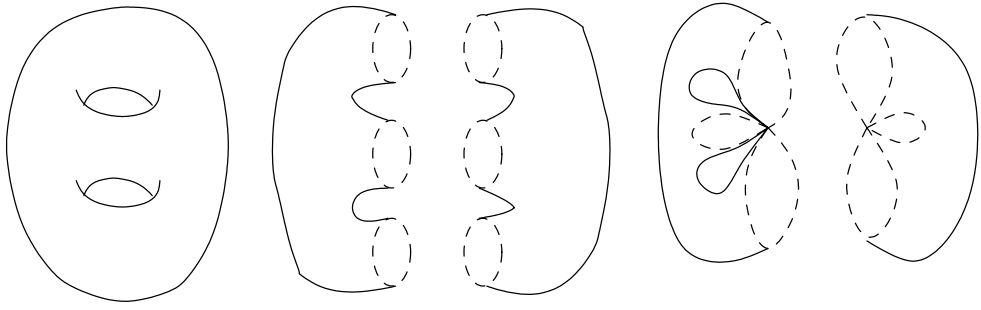}
\caption{$\S_4(G)$, the obvious way to cut it, and the easy way to cut
it for the cell decomposition.}
\label{fig:differentcutting}
\end{center}
\end{figure}
Figure \ref{fig:differentcutting} shows the 2-dimensional analogue of
$\S_4(2)$ and the different ways to cut it up in order to do the cell
decomposition.  The first way, in the middle of Figure
\ref{fig:differentcutting}, cuts $\S_4(G)$ into two $S^4$s with three
holes each (these holes are represented with dotted lines; for
$\S_4(G)$ these holes will be $S^3$-shaped).  It turns out that this
is a tricky way to do the cell decomposition.  It is better to use the
cutting in the final picture of Figure \ref{fig:differentcutting}.
This is homotopically different to the middle cutting because there
are non-trivial 1-cycles in the left-hand piece of the final cutting.

The cell decomposition involves 1 0-cell, $G$ 1-cells, $G+1$ 3-cells
$e^3_\a$ and 2 4-cells, $e^4_L$ representing the left-hand piece of the
last cutting in Figure \ref{fig:differentcutting} and $e^4_R$
representing the right-hand piece.

First attach all the ends of the $G$ 1-cells to the 0-cell so that we
get a wedge of $G$ $S^1$s.

\begin{figure}[h]
\begin{center}
\includegraphics{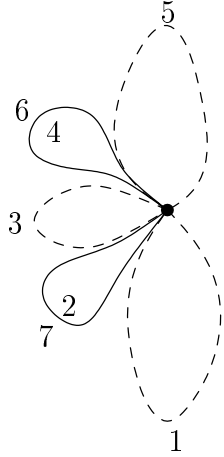}
\caption{The 0-cell, 1-cells and 3-cells of $\S_4(2)$, where the
  0-cell is the blob in the middle, the 1-cells are the thick lines
  and the 3-cells are the dotted lines.}
\label{fig:cellgenus2}
\end{center}
\end{figure}

Next attach the $G+1$ 3-cells $e^3_\a$ to the 0-cell.  The boundary of
the closure of each 3-cell, $S^2$, is identified to the point of the
0-cell so that the 3-cell, an open 3-ball, is closed to become a
sphere $S^3$.  We now have a wedge of $G$ $S^1$s and $G+1$ $S^3$s,
i.e. $G$ $S^1$s and $G+1$ $S^3$s with a point on each of them
identified to the same point. See Figure \ref{fig:cellgenus2} for
$G=2$.

Next we need to describe how to attach the 2 4-cells, $e^4_L$ and
$e^4_R$.

$e^4_L$ attaches to the $G$ 1-cells and the $G+1$ 3-cells.  We need to
specify how the boundary of the closure of $e^4_L$, i.e. an $S^3$, is
mapped to the lower-dimensional cells.  To do this split the 3-sphere
boundary up into $3G+1$ segments.  In terms of coordinates on the
3-sphere we let a $\phi \in [0,2\pi]$ coordinate parameterize the $X_1
- X_2$ plane.  Each segment is defined by $\phi\in
[\frac{2(m-1)\pi}{3G+1}, \frac{2m\pi}{3G+1}]$ for $m = 1, \dots
,3G+1$.  Each segment is like a 3-ball (think segments of the circle
or sphere).  Each of the $G+1$ 3-cells  has a single segment attached
to it and each of the $G$ 1-cells has two segments attached to it on
each side.  The segments are attached in order as indicated in Figure
\ref{fig:cellgenus2}, which generalizes to arbitrary genus.

When a segment is attached to an $S^3$ the boundary of the segment (an
$S^2$) is identified to a point to give $S^3$ (just as when we attach
the 3-cell to the 0-cell).

When we attach a segment to a 1-cell we identify the whole segment
with the 1-ball intersection of the $X^1 - X^2$ plane with the
segment.

$e^4_R$ attaches to the $G+1$ 3-cells.  We split the boundary of the
closure of the $e^4_R$ into $G+1$ pieces and attach them to the
3-cells in order.

From this we can deduce the homology using the boundary operator
$d_k(e_\a^k) = \sum_\b d_{\a\b} e_\b^{k-1}$.

To find $H_4(\S_4(G))$ we use $d_4(e^4_L) = \sum_{\a} e^3_\a$ and
$d_4(e^4_R) = - \sum_\a e^3_\a$ so that $H_4(\S_4(G))$ is generated by
$\S_4(G) = e^4_L + e^4_R$, which has no boundary.  Thus $H_4(\S_4(G))
= \Z$.

All the 3-cells $e^3_\a$ are annihilated by $d_3$ so that
$\textrm{ker}\; d_3$ is spanned by $\{e^3_\a\}$.  The image of $d_4$
is spanned by $\sum_\a e^3_\a$, so $H_3(\S_4(G)) = \Z^{G+1}/\Z =
\Z^{G}$.  Roughly, the $S^3$ cross-sections near the different $G$
holes are not homologous.

There are no 2-cells so $H_2(\S_4(G)) = \{0\}$.

By well-known results about the homology of genus $G$ graphs,
$H_1(\S_4(G)) = \Z^{G}$.

$H_0(\S_4(G)) = \Z$ because the manifold is arcwise-connected.

As expected, Poincar\'e duality holds.
For a general $(d-1)$-dimensional boundary constructed this way see
Appendix \ref{boundarytopd}.

The homology groups above can also be obtained by using 
the Mayer-Vietoris sequence which follows from the 
construction of $ \Sigma_4 (  G ) $ as a union of 
two copies of 
$ S^{4} \setminus \sqcup_{\alpha =1 }^{ G+1 }  ( B^4_{\circ} )_{\alpha} $ 
intersecting over $ \sqcup_{\alpha =1 }^{ G+1  }  ( S^3 )_{\alpha } $.

\subsection{Handlebody decompositions } 

As another way to visualize the topologies involved in 
our discussion, we give their handlebody decompositions. 

\subsubsection{Handlebody decomposition for the complement of a connected graph}

Consider the $n$-valent connected Witten graph $V_n$ with a single
$n$-fold vertex.  To start with we will work with a three-dimensional
bulk because it is easy to visualize.

\begin{figure}[h]
\begin{center}
\includegraphics{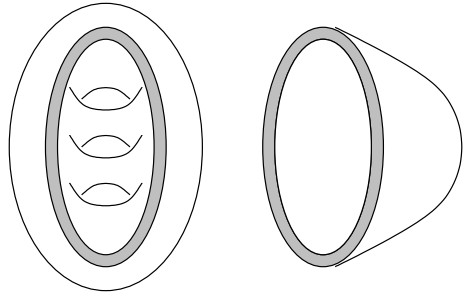}
\caption{The gluing of the 2-handle } \label{fig:glue}
\end{center}
\end{figure}

The handlebody decomposition of $\overline{B^3 \setminus N(V_n,  B^3)}$
is as follows.
\begin{itemize}
\item Start with a 0-handle.
\item Attach $n$ 1-handles to the ball (taking care that the different
  1-handles do not wind around each other).  We now have a filled
  pretzel with $n$ holes, each generated by an $S^1$.
\item Attach a single 2-handle.  Glue it along $S^1
\times B^1$ so that the $S^1$ encircles each of the $n$-holes of the
donut once (see Figure \ref{fig:glue}).
\end{itemize}

This lifts directly to five dimensions for $\overline{B^5 \setminus N(V_n,
 B^5)}$.
\begin{itemize}
\item Start with a 0-handle.
\item Attach $n$ $3$-handles to the ball (taking care to avoid
  non-trivial windings).  The resulting manifold has $n$ holes,
  each generated by an $S^3$.
\item Attach a single $4$-handle.  Glue it along $S^3 \times
  B^1$ so that the $S^3$ encircles each of the $n$ holes once.
\end{itemize}

For the set-up in a general dimension see Appendix
\ref{connhanddecom}.

\subsubsection{Handlebody decomposition 
for the complement of a disconnected graph}

We will describe two ways of providing the handle decomposition of
$\overline{B^5 \setminus N(G, B^5)}$ where $G = V_{n_1} \sqcup V_{n_2}
\sqcup \dots V_{n_m}$.  The first is the one described above: take the
$m$ connected $\overline{B^5 \setminus N(V_{n_i}, B^5)}$ and glue them
together with $(m-1)$ 1-handles.

The second way is similar to that for the connected cases.
\begin{itemize}
\item Start with a 0-handle.
\item Attach $\sum_i n_i$ $3$-handles to the ball (taking care to
  avoid non-trivial windings).
\item For each connected component $V_{n_i}$ attach a $4$-handle.
  Glue it along $S^3 \times B^1$ so that the $S^3$ encircles $n_i$
  holes.  Make sure that the $4$-handles encircle different holes.
\end{itemize}

For $\overline{B^5 \setminus N(G, B^5)}$ glued to a copy of itself use the
same decomposition but add $(m-1)$ 1-handles at the end.


\section{Topology  for gluings in general dimensions}

Here we have a $d$-dimensional bulk and a $(d-1)$-dimensional boundary.

\subsection{Cell decomposition 
and homology 
for the complement of a connected graph}\label{celldecompgend}

The cell decomposition of the $(d-1)$-dimensional space $\d^{(i)}
(N(V_n,B^d))$ is as follows.
\begin{itemize}
\item Take a single 0-cell.
\item Attach to it $n$ $(d-2)$-cells, with a map which sends the boundaries
  of the $(d-2)$-cells ($S^{d-3}$) to the point of the 0-cell.  The
  space is now a \emph{wedge sum} of $n$ spheres $S^{d-2}$, i.e. it is
  the disjoint union of $n$ spheres $S^{d-2}$ with a point on each
  sphere identified to a single point.
\item Now wrap a $(d-1)$-cell around the $n$ $(d-2)$-cells.  The
  $(d-1)$-cell has a boundary of $S^{d-2}$.  To glue this to the $n$
  $S^{d-2}$s divide it up into $n$ pieces and glue each $n$th around
  each $(d-2)$-sphere.
\end{itemize}

Note that for a connected Witten graph $\d^{(i)} (N(V_n,B^d))$ is the
same as the $n$-punctured $(d-1)$-sphere on which we do our CFT.  The
$n$ spheres $S^{d-2}$ are the same as the boundaries of the
$(d-1)$-balls we cut out around each puncture.

The homology groups for a connected Witten graph are
\begin{itemize}
\item $H_0(\d^{(i)} (N(V_n,B^d))) = \Z$ since it is arcwise connected.
\item $H_i(\d^{(i)} (N(V_n,B^d))) = \{0\}$ for $1\leq i \leq d-3$
  since there are no $i$-cells.
\item $H_{d-2}(\d^{(i)} (N(V_n,B^d))) = \Z^{n-1}$ by the same reasoning
  as for $d=5$.
\item $H_{d-1}(\d^{(i)} (N(V_n,B^d))) = \{0\}$ since there are no
  boundaryless $(d-1)$-chains.
\end{itemize}

This gives the  Euler character formula
\begin{align}
 \chi (\d^{(i)} (N(V_n,B^d))) & = \sum_j (-1)^j b_j \nn \\ 
 & = 1 +(-1)^{d-2}(n-1)
\end{align}

\subsection{Cell decomposition and homology 
 for the complement of a disconnected  graph}

Suppose a Witten graph $G$ is composed of $m$ disconnected
components $G = V_{n_1} \sqcup V_{n_2} \sqcup \dots V_{n_m}$.

$\overline{B^d \setminus N(G, B^d)}$ is homotopic to the $m$ connected
spaces $\d^{(i)}(N(V_{n_i},B^d))$ daisy-chained together in a line by
1-cells. For the cell decomposition
\begin{itemize}
\item Take $m$ 0-cells.
\item Link them in a line by $(m-1)$ 1-cells.  The ends of each 1-cell
  attach to different 0-cells.
\item Attach $n_i$ $(d-2)$-cells to each 0-cell for $1\leq i \leq m$
  as above.  We have $m$ wedge sums of $(d-2)$-spheres linked together
  in a line by 1-cells.
\item Now wrap a $(d-1)$-cell around each collection of $n_i$
  $(d-2)$-cells.
\end{itemize}

$\overline{B^d \setminus N(G, B^d)}$ glued to itself is the same except we
attach a further $(m-1)$ 1-cells, each of which has both ends attached
to the same 0-cell.

$\overline{B^d \setminus N(G, B^d)}$ has most of the same homology groups
except that
\begin{equation}
  H_{d-2}(\overline{B^d \setminus N(G, B^d)}) = \Z^{\sum_j n_j - m}
\end{equation}
For $d>3$ the first homology group is unchanged because there are no
closed loops from the 1-cells.

$\overline{B^d \setminus N(G, B^d)}$ glued to itself is a different story.
$H_0$ and $H_i$ for $i >1$ are the same as $\overline{B^d \setminus N(G,
B^d)}$.

Each of the $(m-1)$ 1-cell loops is a 1-cycle which is not the
boundary of some 2-chain.  Thus it increases the number of free
Abelian generators of $H_1$ by ${m-1}$.  So for $d > 3$ we have
\begin{equation}
  H_1((\overline{B^d \setminus N(G, B^d)}) \cup|_{\partial^{(i)} N } 
(\overline{B^d \setminus N(G, B^d)}))
  = \Z^{m-1}
\end{equation}
and for $d=3$
\begin{equation}
  H_1((\overline{B^3 \setminus N(G, B^3)})  \cup|_{\partial^{(i)} N } 
  (\overline{B^3 \setminus N(G, B^3)}))
  = \Z^{\sum_j n_j-1}
\end{equation}

\subsection{General genus boundaries}\label{boundarytopd}

Here we work with a bulk of dimension $d$ and a boundary of dimension
$d-1$.

The homology groups for a boundary of genus $n-1$ for $d > 3$ are
\begin{itemize}
\item $H_0(\S_{d-1}(n-1)) = \Z$ since the manifold is arcwise-connected.
\item $H_1(\S_{d-1}(n-1)) = \Z^{n-1}$ from the topology of the graph.
\item $H_i(\S_{d-1}(n-1)) = \{0\}$ for $2 \leq i \leq d-3$.
\item $H_{d-2}(\S_{d-1}(n-1)) = \Z^{n-1}$ because the $S^{d-2}$
  cross-sections near the different $(n-1)$ holes are not homologous.
  This also follows from Poincar\'e duality for a closed, compact,
  oriented surface.
\item $H_{d-1}(\S_{d-1}(n-1))= \Z$ since the manifold itself has no
  boundary and is arcwise-connected.
\end{itemize}

\subsection{Handlebody decompositions } 

\subsubsection{Handlebody decomposition for 
the complement of a connected graph}\label{connhanddecom}

Here we do a handlebody decomposition of $\overline{B^d \setminus N(V_n,
 B^d)}$.
\begin{itemize}
\item Start with a 0-handle.
\item Attach $n$ $(d-2)$-handles to the ball (taking care to avoid
  non-trivial windings).  The resulting manifold has $n$ holes,
  each generated by an $S^{d-2}$.
\item Attach a single $(d-1)$-handle.  Glue it along $S^{d-2} \times
  B^1$ so that the $S^{d-2}$ encircles each of the $n$ holes once.
\end{itemize}

From the CFT point of view this set-up corresponds to an $S^{d-1}$
with $n$ operator insertions.  To perform the gluing we cut out from
the $S^{d-1}$ a $B^{d-1}_{\circ}$ hole around each operator insertion.
This gives us $n$ disconnected $S^{d-2}$ boundaries.  Each $S^{d-2}$
boundary corresponds to the $S^{d-2}$ that generates each hole in the
handlebody decomposition above of the $\overline{B^d \setminus N(V_n,
B^d)}$ bulk.  The $(d-1)$-handle above then makes sure that the $n$
holes meet up inside the bulk.

\subsubsection{Handlebody decomposition for the 
complement of a disconnected graph}
We will describe two ways of providing the handle decomposition of
$\overline{B^d \setminus N(G, B^d)}$ where $G = V_{n_1} \sqcup V_{n_2}
\sqcup \dots V_{n_m}$.  The first is the one described above: take the
$m$ connected $\overline{B^d \setminus N(V_{n_i}, B^d)}$ and glue them together with
$(m-1)$ 1-handles.

The second way is similar to that for the connected cases.
\begin{itemize}
\item Start with a 0-handle.
\item Attach $\sum_i n_i$ $(d-2)$-handles to the ball (taking care to
  avoid non-trivial windings).
\item For each connected component $V_{n_i}$ attach a $(d-1)$-handle.
  Glue it along $S^{d-2} \times B^1$ so that the $S^{d-2}$ encircles
  $n_i$ holes.  Make sure that the $(d-1)$-handles encircle different
  holes.
\end{itemize}

For $\overline{B^d \setminus N(G, B^d)}$ glued to a copy of itself use the
same decomposition but add $(m-1)$ 1-handles at the end.

\section{Identities, notation and conventions}\label{sec:idnotcon}
We define
\begin{equation}
   \tr (\s \Phi) = \sum_{i_1, i_2, \dots i_n} \Phi^{i_1}_{i_{\s(1)}} \Phi^{i_2}_{i_{\s(2)}} \cdots  \Phi^{i_n}_{i_{\s(n)}}
\end{equation}
The Schur polynomials are defined as a sum of these trace operators
over the elements $\s$ of $S_n$, weighted by the characters of $\s$ in
the representation $R$ of $S_n$,
\begin{equation}
  \chi_R(\Phi) = \frac{1}{n!}\sum_{\s \in S_n}  \chi_R(\s) \tr (\s \Phi)
\end{equation}
A representation $R$ of $S_n$ can be written as a Young diagram with $n$ boxes, with which we also associate a representation $R$ of the unitary group\footnote{This arises because $U(N)$ and $S_n$ have a commuting action on $V^{\otimes n}$ where $V$ is the fundamental representation of $U(N)$.  If the Schur polynomial takes an element of $U(N)$ as its argument it is the character of that element in the irreducible representation $R$. }.  We can reverse the relation between traces and Schur polynomials
\begin{equation}
  \tr(\s \Phi) = \sum_{R(n)}  \chi_R(\s) \chi_R(\Phi)
\end{equation}
where we sum over representations $R$ of $S_n$ with Young diagrams of $n$ boxes.   To do this we have used the orthogonality relation for two elements $\s,\tau\in S_n$
\begin{equation}
  \label{eq:sumRorthogonal}
    \sum_{R(n)} \chi_R (\s) \chi_R (\tau) =\frac{n!}{|[\sigma]|} \delta_{\tau\in [\s]}
\end{equation}
where we have summed over representations of $S_n$.  We also have another orthogonality relation for two representations $R,S$ of $S_n$
\begin{equation}
  \label{eq:sumsigmaorthogonal}
    \sum_{\s\in S_n} \chi_R (\s) \chi_S (\s) =  n! \delta_{RS}
\end{equation}

\begin{table}[!h]
\centering
\begin{tabular}{|l| p{10cm}|}
 \hline
$R$ & a representation of the symmetric group and the unitary group; as an operator we mean $\chi_R(\d Z)$ for the 2d theory and $\chi_R(\Phi)$ for the 4d theory  \\ \hline
$R^\dagger$ & as an operator $\chi_R(\d Z^\dagger)$ for the 2d theory and
  $\chi_R(\Phi^\dagger)$ for the 4d theory \\ \hline
$[L]$ & the Young diagram with a single row of length $L$; for $L \sim N$, $[L]$ corresponds to the $AdS$ giant graviton\\ \hline
$[1^L]$ & the Young diagram with a single column of length $L$; for $L \sim N$, $[1^L]$ corresponds to the sphere giant graviton\\ \hline
$f_R$ & the combinatorial coefficient that appears in the two-point
 function of the Schur polynomials.  It is computed by $f_R =
 \frac{\Dim_R \Delta_R!}{d_R}$ where $\Dim_R$ is the dimension of the $U(N)$ representation $R$ and $d_R$
is the dimension of the symmetric group $S_{\Delta_R}$ representation $R$.  A useful identity is $f_R = \prod_{i,j} (N-i+j)$ where we sum over the boxes of the Young diagram for $R$, $i$ labeling the rows and $j$ the columns.\\ \hline
$g(R_1, R_2, \dots, R_n;R)$ & the Littlewood-Richardson ( LR )  coefficient,
which counts the number of times the representation $R$ appears in the
tensor product of $R_1, \dots, R_n$ \\ \hline
$|[\s]|$ & the size of the conjugacy class $[\s]$ of $\s$, an element of the symmetric group \\ \hline
\end{tabular}
\caption{representation theory notation}
\end{table}
\begin{table}[!h]
\centering
\begin{tabular}{|l| p{10cm}|}
\hline
$B^k$ & the closed $k$-ball \\ \hline
$B^k_{\circ}$ & the open $k$-ball \\ \hline
$V_n$ & the Witten graph obtained by joining $n$ points on the
boundary to a single vertex in the bulk \\ \hline
$\sqcup V_{n_i}$ & a disjoint union of graphs $V_{n_i}$ \\ \hline
$N ( G,B^5 )$ & the neighborhood of the graph $G$ in $B^5$.  Formally $N ( G,B^5 )= \{x \in B^5 : || G -x || \leq
  \epsilon \}$ where we are using the metric of $\R^5$, not Euclidean
  $AdS$. \\ \hline
$\d^{(i)} N (  G ,  B^5 )$ & the interior boundary of the
  neighborhood of the graph $G$.  Formally $\d^{(i)} N (  G ,  B^5 )= \{x \in B^5 : || G -x || =
  \epsilon \}$. \\ \hline
$ \Sigma_4 ( G ) $ &  the 4-dimensional analog of 
a genus $G$ surface in two dimensions. It can be obtained by taking two copies of 
$S^4$ with  $G+1$ non-intersecting balls removed, and gluing the two
along the $S^3$ boundaries.  \\ \hline
$\vee_n S^k$ & the wedge sum of $n$ $k$-spheres. It is the disjoint
  union of $n$ spheres $S^k$ with a point on each sphere identified to
  a single point. \\ \hline
\end{tabular}
\caption{topology notation}
\end{table}

\end{appendix}


\newpage

\begin{thebibliography}{99}

\bibitem{malda} 
  J.~M.~Maldacena,
  ``The large N limit of superconformal field theories and supergravity,''
  Adv.\ Theor.\ Math.\ Phys.\  {\bf 2} (1998) 231
  [Int.\ J.\ Theor.\ Phys.\  {\bf 38} (1999) 1113]
  [arXiv:hep-th/9711200].

\bibitem{gkp}
  S.~S.~Gubser, I.~R.~Klebanov and A.~M.~Polyakov,
  ``Gauge theory correlators from non-critical string theory,''
  Phys.\ Lett.\ B {\bf 428} (1998) 105
  [arXiv:hep-th/9802109].

\bibitem{witten} 
  E.~Witten,
  ``Anti-de Sitter space and holography,''
  Adv.\ Theor.\ Math.\ Phys.\  {\bf 2} (1998) 253
  [arXiv:hep-th/9802150].


\bibitem{mst}
  J.~McGreevy, L.~Susskind and N.~Toumbas,
  ``Invasion of the giant gravitons from anti-de Sitter space,''
  JHEP {\bf 0006} (2000) 008
  [arXiv:hep-th/0003075].

\bibitem{GMT}
  M.~T.~Grisaru, R.~C.~Myers and O.~Tafjord,
  ``SUSY and Goliath,''
  JHEP {\bf 0008}, 040 (2000)
  [arXiv:hep-th/0008015].

\bibitem{HHI}
  A.~Hashimoto, S.~Hirano and N.~Itzhaki,
  ``Large branes in AdS and their field theory dual,''
  JHEP {\bf 0008}, 051 (2000)
  [arXiv:hep-th/0008016].


\bibitem{bbns}
  V.~Balasubramanian, M.~Berkooz, A.~Naqvi and M.~J.~Strassler,
  ``Giant gravitons in conformal field theory,''
  JHEP {\bf 0204} (2002) 034
  [arXiv:hep-th/0107119].


\bibitem{cjr}
  S.~Corley, A.~Jevicki and S.~Ramgoolam,
  ``Exact correlators of giant gravitons from dual N = 4 SYM theory,''
  Adv.\ Theor.\ Math.\ Phys.\  {\bf 5} (2002) 809
  [arXiv:hep-th/0111222].

\bibitem{dms}
  A.~Dhar, G.~Mandal and M.~Smedback,
  ``From gravitons to giants,''
  JHEP {\bf 0603} (2006) 031
  [arXiv:hep-th/0512312].


\bibitem{llm}
  H.~Lin, O.~Lunin and J.~M.~Maldacena,
  ``Bubbling AdS space and 1/2 BPS geometries,''
  JHEP {\bf 0410} (2004) 025
  [arXiv:hep-th/0409174].

\bibitem{fmmr}
D.~Z.~Freedman, S.~D.~Mathur, A.~Matusis and L.~Rastelli,
  ``Correlation functions in the CFT($d$)/AdS($d+1$) correspondence,''
  Nucl.\ Phys.\ B {\bf 546} (1999) 96
  [arXiv:hep-th/9804058].

\bibitem{lmrs}
  S.~M.~Lee, S.~Minwalla, M.~Rangamani and N.~Seiberg,
   ``Three-point functions of chiral operators in D = 4, N = 4 SYM at  large
  N,''
  Adv.\ Theor.\ Math.\ Phys.\  {\bf 2} (1998) 697
  [arXiv:hep-th/9806074].

\bibitem{bmn}
  D.~Berenstein, J.~M.~Maldacena and H.~Nastase,
  ``Strings in flat space and pp waves from N=4 Super Yang Mills,''
  AIP Conf.\ Proc.\  {\bf 646}, 3 (2003).

\bibitem{cfh}
  N.~R.~Constable, D.~Z.~Freedman, M.~Headrick, S.~Minwalla, L.~Motl, A.~Postnikov and W.~Skiba,
  ``PP-wave string interactions from perturbative Yang-Mills theory,''
  JHEP {\bf 0207}, 017 (2002)
  [arXiv:hep-th/0205089].

\bibitem{kpss}
  C.~Kristjansen, J.~Plefka, G.~W.~Semenoff and M.~Staudacher,
  ``A new double-scaling limit of N = 4 super Yang-Mills theory and PP-wave
  strings,''
  Nucl.\ Phys.\ B {\bf 643}, 3 (2002)
  [arXiv:hep-th/0205033].

\bibitem{sv}
  M.~Spradlin and A.~Volovich,
  ``Superstring interactions in a pp-wave background,''
  Phys.\ Rev.\ D {\bf 66}, 086004 (2002)
  [arXiv:hep-th/0204146].

\bibitem{cr} 
  S.~Corley and S.~Ramgoolam,
   ``Finite factorization equations and sum rules for BPS correlators in  N = 4
  SYM theory,''
  Nucl.\ Phys.\ B {\bf 641} (2002) 131
  [arXiv:hep-th/0205221].

\bibitem{bdhm}
  T.~Banks, M.~R.~Douglas, G.~T.~Horowitz and E.~J.~Martinec,
  ``AdS dynamics from conformal field theory,''
  arXiv:hep-th/9808016.

\bibitem{Berenstein:2005fa}
  D.~Berenstein, D.~H.~Correa and S.~E.~Vazquez,
  ``Quantizing open spin chains with variable length: An example from giant
  gravitons,''
  Phys.\ Rev.\ Lett.\  {\bf 95}, 191601 (2005)
  [arXiv:hep-th/0502172].

\bibitem{Berenstein:2006qk}
  D.~Berenstein, D.~H.~Correa and S.~E.~Vazquez,
  ``A study of open strings ending on giant gravitons, spin chains and
  integrability,''
  JHEP {\bf 0609}, 065 (2006)
  [arXiv:hep-th/0604123].

\bibitem{Berenstein}
  D.~Berenstein,
  ``A toy model for the AdS/CFT correspondence,''
  JHEP {\bf 0407}, 018 (2004)
  [arXiv:hep-th/0403110].

\bibitem{deMelloKoch}
  R.~de Mello Koch and R.~Gwyn,
  ``Giant graviton correlators from dual SU(N) super Yang-Mills theory,''
  JHEP {\bf 0411} (2004) 081
  [arXiv:hep-th/0410236].

\bibitem{taktak}
  H.~Takayanagi and T.~Takayanagi,
  ``Notes on giant gravitons on pp-waves,''
  JHEP {\bf 0212} (2002) 018
  [arXiv:hep-th/0209160].

\bibitem{okuyama}
  K.~Okuyama,
  ``1/2 BPS correlator and free fermion,''
  JHEP {\bf 0601} (2006) 021
  [arXiv:hep-th/0511064].

\bibitem{taktsu}
  Y.~Takayama and A.~Tsuchiya,
   ``Complex matrix model and fermion phase space for bubbling AdS
  geometries,''
  JHEP {\bf 0510} (2005) 004
  [arXiv:hep-th/0507070].

\bibitem{intril}
  K.~A.~Intriligator,
  Nucl.\ Phys.\ B {\bf 551} (1999) 575
  [arXiv:hep-th/9811047].

\bibitem{ehsw1}
  B.~U.~Eden, P.~S.~Howe, A.~Pickering, E.~Sokatchev and P.~C.~West,
  ``Four-point functions in N = 2 superconformal field theories,''
  Nucl.\ Phys.\ B {\bf 581} (2000) 523
  [arXiv:hep-th/0001138].

\bibitem{ehsw2}
  B.~U.~Eden, P.~S.~Howe, E.~Sokatchev and P.~C.~West,
  ``Extremal and next-to-extremal n-point correlators in four-dimensional
  SCFT,''
  Phys.\ Lett.\ B {\bf 494} (2000) 141
  [arXiv:hep-th/0004102].

\bibitem{balbabel1}
  V.~Balasubramanian, V.~Jejjala and J.~Simon,
  ``The library of Babel,''
  Int.\ J.\ Mod.\ Phys.\ D {\bf 14}, 2181 (2005)
  [arXiv:hep-th/0505123].

\bibitem{balbabel} 
  V.~Balasubramanian, J.~de Boer, V.~Jejjala and J.~Simon,
  ``The library of Babel: On the origin of gravitational thermodynamics,''
  JHEP {\bf 0512}, 006 (2005)
  [arXiv:hep-th/0508023].

\bibitem{ost}
  K.~Osterwalder and R.~Schrader,
  Commun.\ Math.\ Phys.\  {\bf 31} (1973) 83.

\bibitem{pst} 
  J.~Polchinski, L.~Susskind and N.~Toumbas,
  ``Negative energy, superluminosity and holography,''
  Phys.\ Rev.\ D {\bf 60} (1999) 084006
  [arXiv:hep-th/9903228].

\bibitem{susskindflat}
  L.~Susskind,
  ``Holography in the flat space limit,''
  arXiv:hep-th/9901079.

\bibitem{polchinskiflat}
  J.~Polchinski,
  ``S-matrices from AdS spacetime,''
  arXiv:hep-th/9901076.

\bibitem{giddingsflat}
  S.~B.~Giddings,
  ``The boundary S-matrix and the AdS to CFT dictionary,''
  Phys.\ Rev.\ Lett.\  {\bf 83}, 2707 (1999)
  [arXiv:hep-th/9903048].

\bibitem{Ginsparg:1988ui}
  P.~H.~Ginsparg,
  ``Applied Conformal Field Theory,''
  arXiv:hep-th/9108028.

\bibitem{Polchinski}
  J.~Polchinski,
  ``String theory. Vol. 1: An introduction to the bosonic string, '' CUP (1998).

\bibitem{Vafa:1987ea}
  C.~Vafa,
  ``Conformal Theories and Punctured Surfaces,''
  Phys.\ Lett.\ B {\bf 199} (1987) 195.

\bibitem{Sonoda:1988mf}
  H.~Sonoda,
  ``Sewing Conformal Field Theories,''
  Nucl.\ Phys.\ B {\bf 311} (1988) 401.

\bibitem{Sonoda:1988fq}
  H.~Sonoda,
  ``Sewing Conformal Field Theories. 2,''
  Nucl.\ Phys.\ B {\bf 311} (1988) 417.

\bibitem{Sonoda:1987ra}
  H.~Sonoda,
  ``Functional Determinants on Punctured Riemann Surfaces and their Applications to String Theory,''
  Nucl.\ Phys.\ B {\bf 294}, 157 (1987).

\bibitem{ginsmoore} 
  P.~H.~Ginsparg and G.~W.~Moore,
  ``Lectures on 2-D gravity and 2-D string theory,''
  arXiv:hep-th/9304011.

\bibitem{wit2dgrav} 
  E.~Witten,
  ``On the Structure of the Topological Phase of Two-Dimensional Gravity,''
  Nucl.\ Phys.\ B {\bf 340} (1990) 281.

\bibitem{dijkhou}
  R.~Dijkgraaf,
  ``Fields, strings and duality,''
  arXiv:hep-th/9703136.

\bibitem{Brigante:2005bq}
  M.~Brigante, G.~Festuccia and H.~Liu,
   ``Inheritance principle and non-renormalization theorems at finite
  temperature,''
  Phys.\ Lett.\ B {\bf 638} (2006) 538
  [arXiv:hep-th/0509117].

\bibitem{Birrell:1982ix}
  N.~D.~Birrell and P.~C.~W.~Davies,
 ``Quantum Fields In Curved Space,'' CUP (1982).

\bibitem{amp}
  O.~Aharony, J.~Marsano, S.~Minwalla, K.~Papadodimas and M.~Van Raamsdonk,
  ``The Hagedorn / deconfinement phase transition in weakly coupled large N
  gauge theories,''
  Adv.\ Theor.\ Math.\ Phys.\  {\bf 8}, 603 (2004)
  [arXiv:hep-th/0310285].

\bibitem{kmm}
  J.~Kinney, J.~M.~Maldacena, S.~Minwalla and S.~Raju,
  ``An index for 4 dimensional super conformal theories,''
  arXiv:hep-th/0510251.

\bibitem{gibbhawkfest} 
 G. Gibbons, ``Euclidean quantum gravity: the view from 2002'' 
 in ``The Future of Theoretical Physics and Cosmology,'' edited by 
 G.Gibbons, E.Shellard, S.Rankin CUP (2003)

\bibitem{wittenpt}
  E.~Witten,
  ``Anti-de Sitter space, thermal phase transition, and confinement in  gauge
  theories,''
  Adv.\ Theor.\ Math.\ Phys.\  {\bf 2}, 505 (1998)
  [arXiv:hep-th/9803131].

\bibitem{gs} 
  R. Gompf, A. Stipsicz, ``4-manifolds and Kirby calculus,'' 
  Graduate Studies in Mathematics, Volume 20.
  AMS, Providence, Rhode Island. 


\bibitem{dowgar}
  H.~F.~Dowker and R.~S.~Garcia,
  ``A handlebody calculus for topology change,''
  Class.\ Quant.\ Grav.\  {\bf 15} (1998) 1859
  [arXiv:gr-qc/9711042].

\bibitem{hartn} 
  S.~A.~Hartnoll,
  ``Compactification, topology change and surgery theory,''
  Class.\ Quant.\ Grav.\  {\bf 20} (2003) 3093
  [arXiv:hep-th/0302072].

\bibitem{horo} 
  G.~T.~Horowitz,
  ``Topology change in classical and quantum gravity,''
  Class.\ Quant.\ Grav.\  {\bf 8}, 587 (1991).

\bibitem{maldainflat}
  J.~M.~Maldacena,
   ``Non-Gaussian features of primordial fluctuations in single field
  inflationary models,''
  JHEP {\bf 0305} (2003) 013
  [arXiv:astro-ph/0210603].

\bibitem{harthawk}
  J.~B.~Hartle and S.~W.~Hawking,
  ``Wave Function Of The Universe,''
  Phys.\ Rev.\ D {\bf 28} (1983) 2960.

\bibitem{fs} 
  B.~Freivogel and L.~Susskind,
  ``A framework for the landscape,''
  Phys.\ Rev.\ D {\bf 70} (2004) 126007
  [arXiv:hep-th/0408133].

\bibitem{bousso}
  R.~Bousso,
  ``Holographic probabilities in eternal inflation,''
  arXiv:hep-th/0605263.

\bibitem{page} 
  D.~N.~Page,
  ``Predictions and tests of multiverse theories,''
  arXiv:hep-th/0610101.

\bibitem{vilenk} 
  J.~Garriga, D.~Schwartz-Perlov, A.~Vilenkin and S.~Winitzki,
  ``Probabilities in the inflationary multiverse,''
  JCAP {\bf 0601} (2006) 017
  [arXiv:hep-th/0509184].

\bibitem{fssc} 
  B.~Freivogel, Y.~Sekino, L.~Susskind and C.~P.~Yeh,
  ``A holographic framework for eternal inflation,''
  Phys.\ Rev.\ D {\bf 74} (2006) 086003
  [arXiv:hep-th/0606204].

\bibitem{horava} 
  P.~Horava and P.~G.~Shepard,
  ``Topology changing transitions in bubbling geometries,''
  JHEP {\bf 0502} (2005) 063
  [arXiv:hep-th/0502127].

\bibitem{lpstu}
  D.~A.~Lowe, J.~Polchinski, L.~Susskind, L.~Thorlacius and J.~Uglum,
  ``Black hole complementarity versus locality,''
  Phys.\ Rev.\ D {\bf 52} (1995) 6997
  [arXiv:hep-th/9506138].

\bibitem{gmh}
  S.~B.~Giddings, D.~Marolf and J.~B.~Hartle,
  ``Observables in effective gravity,''
  Phys.\ Rev.\ D {\bf 74} (2006) 064018
  [arXiv:hep-th/0512200].

\bibitem{freidel}
  A.~Baratin and L.~Freidel,
  ``Hidden quantum gravity in 3d Feynman diagrams,''
  arXiv:gr-qc/0604016.

\bibitem{maldaloop}
  J.~M.~Maldacena,
  ``Wilson loops in large N field theories,''
  Phys.\ Rev.\ Lett.\  {\bf 80} (1998) 4859
  [arXiv:hep-th/9803002].

\bibitem{drukgro}
  N.~Drukker, D.~J.~Gross and H.~Ooguri,
  ``Wilson loops and minimal surfaces,''
  Phys.\ Rev.\ D {\bf 60} (1999) 125006
  [arXiv:hep-th/9904191].

\bibitem{okusem}
  K.~Okuyama and G.~W.~Semenoff,
  ``Wilson loops in N = 4 SYM and fermion droplets,''
  JHEP {\bf 0606} (2006) 057
  [arXiv:hep-th/0604209].

\bibitem{chen} 
J.Q. Chen, Group Representation theory for Physicists,
 World Scientific, 1987, Chapter 7. 

\bibitem{partensky} 
A. Partensky, J. Math. Phys. {\bf 13 } , 1972 , 1503. 

\bibitem{sramthesis} 
S. Ramgoolam, Thesis, Yale 1995. 

\bibitem{Rabadan:2002wy}
  R.~Rabadan and F.~Zamora,
  ``Dilaton tadpoles and D-brane interactions in compact spaces,''
  JHEP {\bf 0212} (2002) 052
  [arXiv:hep-th/0207178].

\bibitem{hatcher}
  Allen Hatcher, ``Algebraic Topology'', CUP (2002). \\
  \url{http://www.math.cornell.edu/~hatcher/}












\end{thebibliography}
\end{document}